\documentclass[twocolumn, trackchanges]{aastex63}
\usepackage{graphicx}
\usepackage{float}
\usepackage{amsmath,amssymb}
\usepackage{xcolor}
\usepackage{multirow} 

\usepackage{colortbl}

\def\review{\textcolor{black}}
\def\reviewt{}

\def\paltas{\texttt{paltas\:}}
\def\thetaE{$\theta_{\rm E}\:$}
\def\gammalens{$\gamma_{\rm lens}\:$}
\def\ho{$H_{0}\:$}
\received{October 23, 2025}
\revised{July 10, 2026}
\accepted{}
\shorttitle{LSST Lensed AGN Model Accuracy}
\shortauthors{Padma Venkatraman et al.}

\begin{document}

\title{Lens Model Accuracy in the Expected LSST Lensed AGN Sample}


\correspondingauthor{Padma Venkatraman}
\email{vpadma23@slac.stanford.edu}

\author[0000-0001-8638-2780]{Padmavathi Venkatraman}
\affiliation{Department of Astronomy, University of Illinois Urbana-Champaign, Urbana, IL 61801, USA}
\affiliation{Kavli Institute for Particle Astrophysics and Cosmology, Department of Physics, Stanford University, Stanford, CA 94309, USA}

\author[0000-0001-5717-2688]{Sydney Erickson}
\affiliation{Kavli Institute for Particle Astrophysics and Cosmology, Department of Physics, Stanford University, Stanford, CA 94309, USA}
\affiliation{SLAC National Accelerator Laboratory, 2575 Sand Hill Road, Menlo Park, CA 94025, USA}

\author[0000-0002-0113-5770]{Phil Marshall}
\affiliation{Kavli Institute for Particle Astrophysics and Cosmology, Department of Physics, Stanford University, Stanford, CA 94309, USA}
\affiliation{SLAC National Accelerator Laboratory, 2575 Sand Hill Road, Menlo Park, CA 94025, USA}

\author[0000-0001-7051-497X]{Martin Millon} \affiliation{Institute for Particle Physics and Astrophysics, ETH Zurich,
Wolfgang-Pauli-Strasse 27, CH-8093 Zurich, Switzerland}
\affiliation{Kavli Institute for Particle Astrophysics and Cosmology, Department of Physics, Stanford University, Stanford, CA 94309, USA}

\author[0009-0002-8896-6100]{Philip Holloway} \affiliation{Department of Physics, University of Oxford, Keble Road, Oxford, UK}\affiliation{Institute of Cosmology and Gravitation, University of Portsmouth, Portsmouth, UK}

\author[0000-0003-3195-5507
]{Simon Birrer} \affiliation{Department of Physics, Stony Brook University, NY, U.S.A.}

\author[0000-0002-4773-1463]{Steven Dillmann}\affiliation{Kavli Institute for Particle Astrophysics and Cosmology, Department of Physics, Stanford University, Stanford, CA 94309, USA}\affiliation{Stanford Artificial Intelligence Laboratory, Department of Computer Science, Stanford University, Stanford, CA 94305, USA}\affiliation{Institute of Computational and Mathematical Engineering, Stanford University, Stanford, CA 94305, USA}

\author[0000-0001-7113-0599]{Xiangyu Huang} \affiliation{Department of Physics, Stony Brook University, NY, U.S.A.}

\author[0000-0002-5386-7076]{Sreevani Jaragula}\affiliation{Fermi National Accelerator Laboratory, Kirk Road and Pine Street, Batavia IL 60510-5011, USA}

\author{Ralf Kaehler}\affiliation{Kavli Institute for Particle Astrophysics and Cosmology, Department of Physics, Stanford University, Stanford, CA 94309, USA}\affiliation{SLAC National Accelerator Laboratory, 2575 Sand Hill Road, Menlo Park, CA 94025, USA}

\author[0000-0001-5512-2716]{Narayan Khadka} \affiliation{Department of Physics, Stony Brook University, NY, U.S.A.} \affiliation{Kavli Institute for Particle Astrophysics and Cosmology, Department of Physics, Stanford University, Stanford, CA 94309, USA}\affiliation{SLAC National Accelerator Laboratory, 2575 Sand Hill Road, Menlo Park, CA 94025, USA}

\author{Grzegorz Madejski} \affiliation{Kavli Institute for Particle Astrophysics and Cosmology, Department of Physics, Stanford University, Stanford, CA 94309, USA}
\affiliation{SLAC National Accelerator Laboratory, 2575 Sand Hill Road, Menlo Park, CA 94025, USA}

\author[0000-0002-9436-8871]{Ayan Mitra}
\affiliation{Center for AstroPhysical Surveys, National Center for Supercomputing Applications, University of Illinois Urbana-Champaign, Urbana, IL, 61801, USA}
\affiliation{Department of Astronomy, University of Illinois Urbana-Champaign, Urbana, IL 61801, USA}

\author[0000-0002-2234-749X
]{Kevil Reil}\affiliation{SLAC National Accelerator Laboratory, 2575 Sand Hill Road, Menlo Park, CA 94025, USA}

\author[0000-0001-5326-3486]{Aaron Roodman} \affiliation{Kavli Institute for Particle Astrophysics and Cosmology, Department of Physics, Stanford University, Stanford, CA 94309, USA}
\affiliation{SLAC National Accelerator Laboratory, 2575 Sand Hill Road, Menlo Park, CA 94025, USA}

\author{the LSST Dark Energy Science Collaboration}

\begin{abstract}
Strong gravitational lensing of active galactic nuclei (AGN) enables measurements of cosmological parameters through time-delay cosmography (TDC). With data from the upcoming LSST survey, we anticipate using a sample of $\mathcal{O}(1000)$ lensed AGN for TDC. To prepare for this dataset and enable this measurement, we construct and analyze a realistic mock sample of 1300 systems drawn from the OM10 (Oguri \& Marshall 2010) catalog of simulated lenses with AGN sources at $z<3.1$ in order to test a key aspect of the analysis pipeline, that of the lens modeling. We realize the lenses as power law elliptical mass distributions and simulate 5-year LSST $i$-band coadd images. From every image, we infer the lens mass model parameters using neural posterior estimation (NPE). 
Focusing on the key model parameters, \thetaE (the Einstein Radius) and \gammalens (the projected mass density profile slope), with consistent mass-light ellipticity correlations in test and training data, we recover \thetaE with less than 1\% bias \review{on average}, 6.5\% precision per lens and \gammalens with less than 3\% bias \review{on average}, 8\% precision per lens. We find that lens light subtraction prior to modeling is only useful when applied to data sampled from the training prior. If emulated deconvolution is applied to the data prior to modeling, precision improves across all parameters by a factor of 2. Finally, we combine the inferred lens mass models using Bayesian Hierarchical Inference to recover the global properties of the lens sample with less than 1\% bias under consistent train-test mass-light correlations. 
\end{abstract}




\section{Introduction}
\label{sec:intro}


Understanding the \reviewt{physical origin of the} accelerating expansion of our Universe has been a long-standing challenge in cosmology. The two \reviewt{most precise} probes \reviewt{inferring} the current expansion rate of the Universe (Hubble Constant, $H_0$) from the early universe, using cosmic microwave background anisotropies \cite[e.g.][]{planck2020}, and late universe, using Type Ia SNe \reviewt{anchored by the distance ladder} \citep[e.g.][]{Riess_2022} \reviewt{are increasingly precise yet discrepant}, leading to a 5$\sigma$ difference on $H_0$, known as the Hubble tension \citep[e.g.][]{Verde2019, abdalla2022}. There is also growing evidence that the \reviewt{evolution of dark energy is not constant in time, challenging $\rm \Lambda CDM$} \citep[e.g.][]{desi_2025, des_combine_probes}. The discrepancies could point to new physics or be a result of systematic errors in use of the aforementioned probes. Here, strong lensing time-delay cosmography (TDC) provides an independent competitive measurement on the current expansion rate of the universe and evolution of dark energy over cosmic time \citep[e.g.][]{Treu_2022, Birrer2024, investigating_de_constraint}.


TDC can be applied to any strong lensing system with a time-variable, bright, point-like source, situated behind a massive deflector, both lying along the line of sight of the observer. It uses the source's light travel time around the massive deflector to constrain angular diameter distances \citep{refsdal1964}.

\ho has been measured to \reviewt{4.6\% precision with 8 lensed} active galactic nuclei (AGN) \citep{wong2020, millon2020, tdcosmo_2025} \reviewt{under $\rm \Lambda CDM$}. \review{Lenses which do not have measurable time-delays can also }be used to increase the precision of cosmological inferences made using time-delay lenses \citep[e.g.][]{tdcosmo4, tdcosmo_2025}, \review{by constraining the lens mass models, and using these within the context of hierarchically constructing an informative population model for deflector galaxies.} Surveys such as the Legacy Survey of Space and Time (LSST) \citep{lsst}, are expected to increase our cosmology grade time-delay sample size from $\sim40$ lensed AGN to $\mathcal{O}(100)$ lensed AGN with measured time delays \citep{taak_treu} and $\mathcal{O}(1000)$ lensed AGN in total \citep{om10, yue_2022, abe_2025}. In particular, lensed AGN brighter than 22\,mag with variability detectable in LSST will be able to contribute to TDC cosmology constraints \citep{taak_treu}. 


Modeling the deflector galaxy's gravitational potential is a key component of the TDC analysis pipeline. We anticipate a ``Gold'' sample of lensed AGN in LSST that will consist of $\mathcal{O}(100)$ systems with all data required for high precision lens modeling including space-based imaging. The remaining  $\mathcal{O}(1000)$ lensed AGN, constituting a ``Silver'' sample, will only have LSST imaging. We posit that combining the population constraints from this larger Silver sample with the Gold sample will improve the cosmological constraining power through an improved understanding of the mass distributions of deflector galaxies and a finer sampling of the distance-redshift relation. Within this framework, the inferred population-level distribution of lens mass model parameters from ground-based data becomes a key benchmark for our pipeline.
\begin{figure*}
    \centering
    \includegraphics[width=0.8\textwidth]{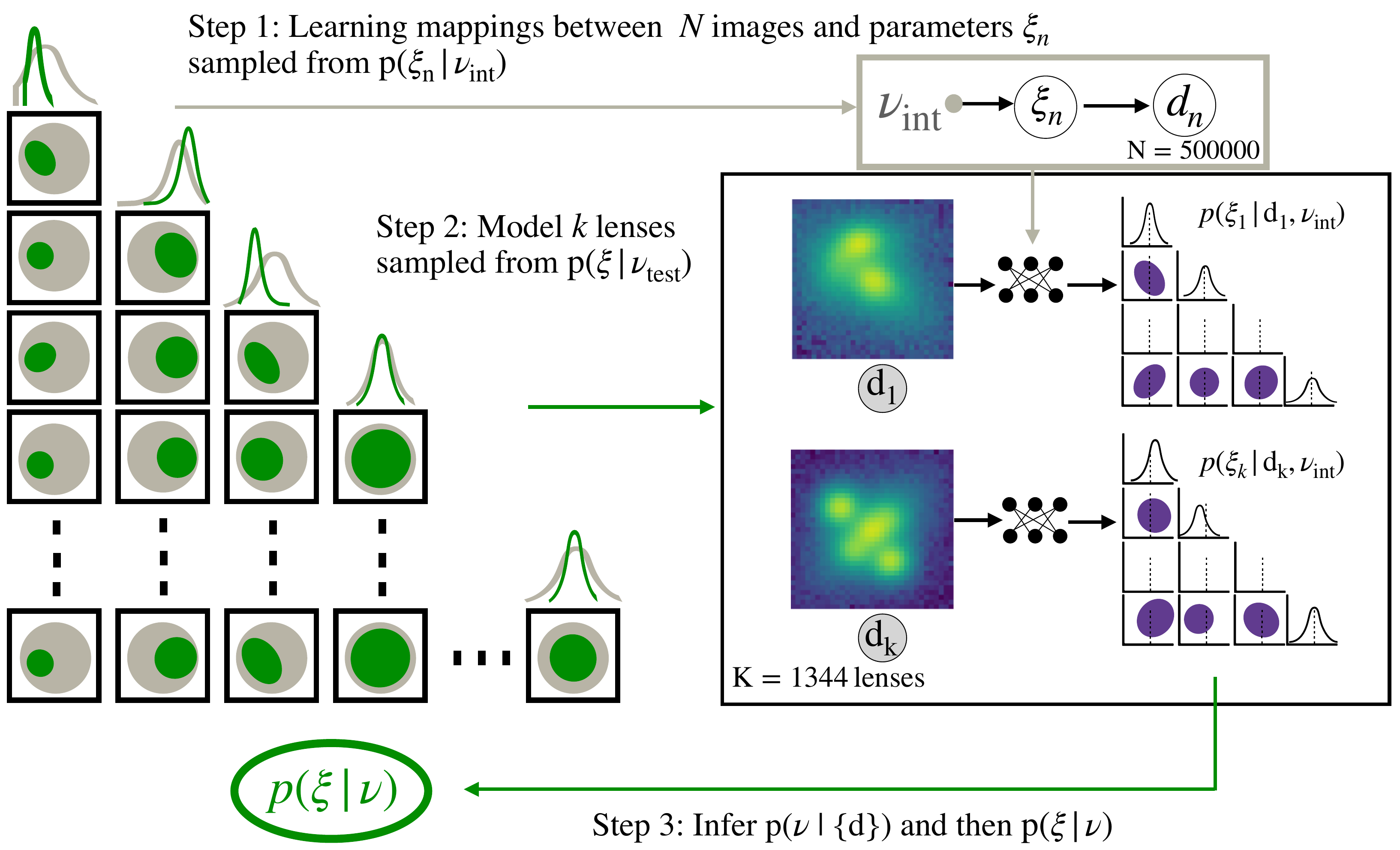}
    \caption{Probabilistic Graphical Model (PGM) depicting the levels of inference in this project. We assert broad priors on lensing parameters $\rm \nu_{int}$ (shown in gray) and sample \review{$\rm N=5\times10^5$} lens systems with model parameters $\xi_{\review{n}}$. This is used to optimize model weights that we apply to the test image ($d_k$) for each of the \review{$K$} lensed AGN systems. From the \review{test} data, we recover lens model parameters $\rm p(\xi_k|d_k, \nu_{int})$ (shown in purple contours), and infer the lens population model $\rm p(\nu|\{d\})$ (shown as overlaid green contours).}
    \label{fig:mot_HI}
\end{figure*}
In this work we carry out LSST-scale modeling of ground-based imaging of simulated lensed AGN. Constructing mass models given an image of a strongly lensed system is a computationally expensive task. Current state of the art forward modeling techniques take $10^4 - 10^6$ CPU hours \citep{shajib_2020}, and several investigator hours (i.e. researcher analysis time), for a single lens system \citep{ding2021time}. With further automation as shown in \citet{time, strides_schmidt} and \citet{dolphin}, this is brought down to days, and in the best cases, hours. In the best-case scenario, GPU-accelerated models have reached modeling times down to 5 GPU hours per lens \citep[e.g.][]{desi_foundry,galan2022}. 

In order to scale up to LSST, and model thousands of lenses within a reasonable analysis time, we turn to deep learning methods to infer lens mass models, as first introduced by \cite{hezaveh2017, npe_uncertainties}. Previously, such machine learning methods have been validated on large samples of simulated space-based imaging data of lensed AGN 
in \cite{Park_2021}, and applied to real HST data of lensed AGN taken from STRIDES \citep{strides_schmidt} in \cite{erickson2024}, achieving comparable accuracy to forward modeling. Similarly, they have been validated against simulated ground-based data of galaxy-galaxy lenses (G-G lenses) \citep[e.g.][]{lsst_euclid, Wagner_Carena_2021,poh2022}, and tested on samples of space-based and ground-based imaging of G-G lenses \citep[e.g.][]{schuldt2021holismokes,gawade2024,lemon,holismokes10, lemon_euclid}.
In this work, we verify the use of machine learning based lens modeling on a large realistic sample of ground-based mock LSST imaging of lensed AGN.

We use a Simulation Based Inference (SBI) technique called Neural Posterior Estimation (NPE) \citep{npe_loss,Cranmer_2020} to \review{infer} the lens mass models. Using open-source software \paltas \citep{paltas}\footnote{https://github.com/swagnercarena/paltas}, we produce approximate multivariate Gaussian posterior probability distributions (PDF) describing the lens model parameters learned from $5 \times 10^5$ mappings of images to lens model parameters. Once \review{we have posterior distributions for the mass model parameters of all \reviewt{the} lens systems,} we combine them in a Hierarchical Bayesian Inference (HBI) framework to learn the properties of the \reviewt{underlying} sample. This process is outlined in Figure \ref{fig:mot_HI}.


We aim to answer the following questions:
\begin{itemize}
    \setlength\itemsep{0em}
    \setlength\parskip{0em}
    \item How accurately can we measure the key lens model parameters \thetaE (the Einstein Radius) and \gammalens (the density profile slope) from the LSST 5-year coadd imaging alone? 
    \item To what extent does this performance depend on the subtraction of the lens galaxy light, and the AGN point image light? How does deconvolution applied to the data prior to modeling improve the accuracy of measured lens models? 
    \item How do distribution shifts in learned and latent parameters impact the inference of lens mass model parameters?
    \item How does the presence of mass-light ellipticity correlations in the training data impact the accuracy of inferred key lens model parameters, specifically \gammalens?
    \item \review{How accurately can we recover the underlying parameter distribution that describes the population of key lens mass model parameters?}

\end{itemize}

This paper is organized as follows. 
Section \ref{sec:data} presents the expected LSST lensed AGN catalog constituting our sample, lens modeling choices, and image data specifications. Section \ref{sec:method} describes the NPE method used to model lenses followed by the HBI inference pipeline used to infer the population. Section \ref{sec:results} presents the lens mass model parameter recovery results, on individual lenses and on the population-level parameters. 
Section \ref{sec:discussion} presents a discussion of the method and outlines the future investigation required. Section \ref{sec:conclusions} presents our key conclusions. 


\vspace{-0.42cm}

\section{Data}\label{sec:data}
To create a mock dataset of lensed AGN expected from LSST, we start from the forecast lensed quasar catalog from \citet[][hereafter OM10]{om10}, which contains lens systems with a deflector galaxy and background AGN. We detail the profile assumptions and modifications we make to the existing catalog in Section \ref{data:om10_lens}. We then perform a matching with the \texttt{cosmoDC2} \citep{cosmodc2} mock galaxy catalog to assign an AGN host galaxy to each system, as described in Section \ref{data:cosmodc2}. The end result is an updated catalog of 1344 lensed AGN. From this catalog, we create simulated LSST images with  multiple ``preparations,'' as detailed in Section \ref{data:3_preps}.


\subsection{OM10 Lens Systems}\label{data:om10_lens}
We start with the catalog of forecast strongly lensed AGN produced in OM10 \footnote{https://github.com/drphilmarshall/OM10}. The OM10 catalog assumes a singular isothermal ellipsoid (SIE) model for the lens mass distribution. It also contains AGN apparent brightness, K-correction and redshift information. First, we make some cuts on the catalog to select systems observable by LSST. We require the image separation of the lensed systems to be greater 0.7 arcsec. This is a little less than the expected median point spread function (PSF) FWHM, or image quality, in the LSST coadd images. In Section \ref{data:lsst}, we revisit the image-separation cut after applying a realistic PSF to images of our modified lens models. We also require the brightness of third image of a quadruply imaged lensed AGN (or second image of a doubly \review{imaged} lensed AGN) to have $i_3 < 23.3$ \citep[to ensure that image is detected at 10-sigma significance, on average, in every visit image, ][]{om10}. This results in 2818 lensed AGN systems selected from a large sample of 15658 lensed AGN systems. In the following sections, we lay out the changes we made to the lens mass, lens light, and AGN host galaxy light profiles to make our mock lens systems more realistic. \review{Changing the lens model, adding AGN host galaxy information and applying a realistic observation model change the number of lenses in our sample, bringing the sample down to 1344 lenses. This is detailed in Section \ref{data:lsst}.}

\subsubsection{Lens Mass Model}\label{data:lens_mass}
In OM10, the lens mass model is a singular isothermal ellipsoid (SIE). We re-parameterize the deflector galaxy as a power-law elliptical mass distribution (PEMD) \citep{Barkana_1998}. While SIE assumes a fixed projected radial density slope, PEMD allows the slope to vary. The lens modeling process is an inference of the convergence $\kappa$, or normalized surface mass density, of the deflector galaxy in a lens system using lensing observables. The lens model is combined with lensed source position to predict the Fermat potential, which is the quantity of interest for cosmology. Using the PEMD mass model, we can parameterize convergence as:
\begin{equation}\label{eqn:lens_model}
    \rm \kappa(x, y) = \frac{3 - \gamma_{\rm lens}}{2} \left(\frac{\theta_E}{\sqrt{q_{m}x_{lens, M}^2 + y_{lens, M}^2/q_{m}}}\right)^{\gamma_{\rm lens}-1}.
\end{equation}
where $\rm \gamma_{lens}$ is the negative logarithmic projected mass density power-law slope and $\rm \theta_E$ is the Einstein radius -- the radius within which the deflector galaxy's integrated convergence is unity (assuming no ellipticity), $\rm x_{lens, M}$ and $\rm y_{lens, M}$ are defined in a coordinate system aligned with the major and minor axis of the lens \review{mass}, and $\rm q_{m}$ is the \review{lens mass} minor/major axis ratio.
\begin{figure*}
    \centering
    \includegraphics[width=1\textwidth]{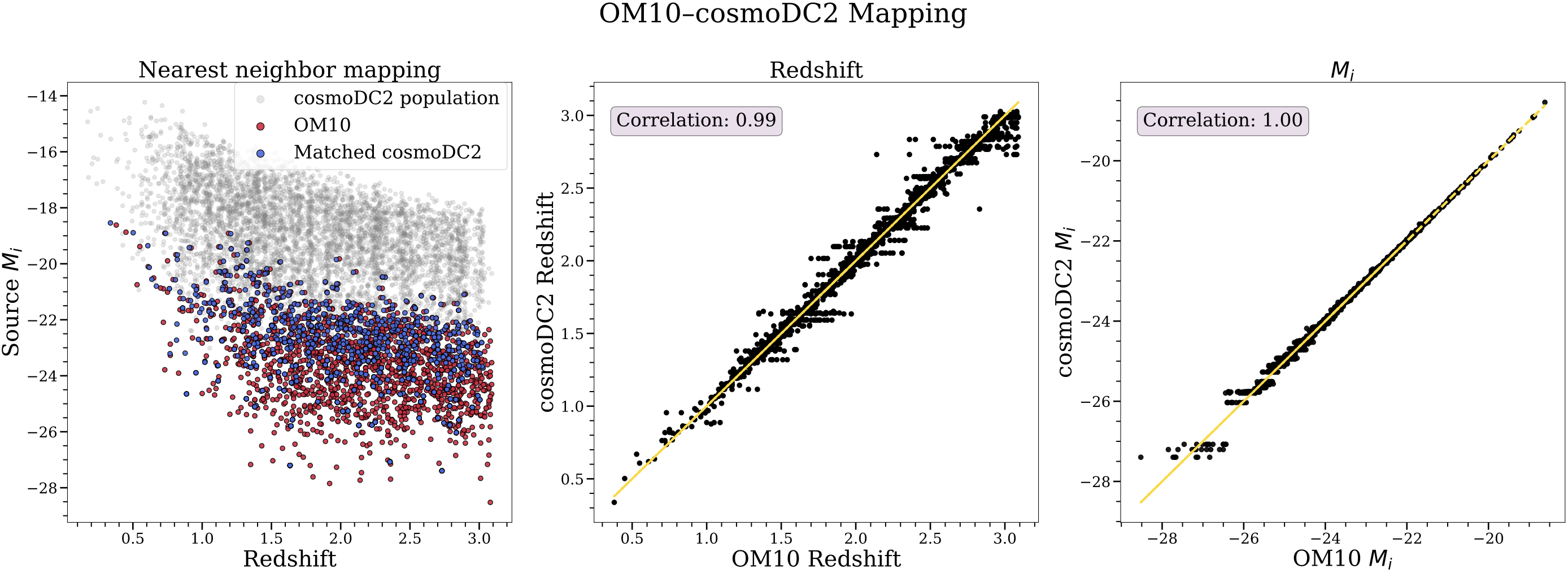}
    \caption{\textbf{OM10-cosmoDC2 mapping.} We ``paint on" host galaxies from cosmoDC2 catalog \review{(outlined in Section \ref{data:cosmodc2})} to systems in the OM10 catalog \review{(outlined in Section \ref{data:om10_lens})}. Nearest neighbor mapping: To obtain a mapping between the two catalogs, we map AGN in OM10 to AGN in cosmoDC2 using their redshift and absolute magnitude and use their corresponding host galaxy. \textbf{First panel:} The red OM10 systems are mapped to the blue cosmoDC2 systems. The blue points are a subset of the gray points. \textbf{Second panel:} We demonstrate the scatter of the OM10-cosmoDC2 system mapping along AGN redshift axis. \textbf{Third panel:} We demonstrate the scatter of the OM10-cosmoDC2 system mapping along AGN $i$-band absolute brightness (AB mag) axis.}
    \label{fig:mapping}
\end{figure*}
Instead of learning the lens $\rm q_{m}$ and orientation angle $\rm \phi_m$, we parameterize the two-component ellipticity of the lens mass as:
\begin{equation}
\begin{split}
\rm e_1 = \frac{1 - q_m}{1 + q_m}\cos(2\phi_m).\\    
\rm e_2 = \frac{1 - q_m}{1 + q_m}\sin(2\phi_m).
\end{split}
\end{equation}
This is because inference of periodic parameters, like $\rm \phi_m$, is unstable.

We include external shear coming from the environment of the deflector and fit for these parameters. These parameters add an additional component to the deflection angle, but do not change the convergence $\kappa$. We also change the parameterization of external shear from strength ($\rm \gamma_{ext}$) and orientation ($\phi_{\gamma})$ to $\gamma_1$ and $\gamma_2$ where:
\begin{equation}
\begin{split}
\gamma_1 = \rm \gamma_{ext}\cos(2\phi_{\gamma}). \\
\gamma_2 = \rm \gamma_{ext}\sin(2\phi_{\gamma}).
\end{split}
\end{equation}

We obtain the following mass model parameters from OM10: $\rm \theta_E$, $\rm q_m$, $\rm \phi_m$, $\gamma$, $\phi_\gamma$, $\rm x$, $\rm y$. We randomly assign each deflector a density profile \review{slope} sampled from a Gaussian distribution centered on 2, with a scatter of 0.16 ($\gamma_{\rm lens} \sim $ N(2, 0.16)) because massive galaxies are approximately isothermal ($\gamma_{\rm lens} = 2$) with values ranging from 1.6 to 2.4 as shown in \citep{auger}. 


PEMD is often adopted in the context of TDC \citep{suyu_gamma}, as it allows flexibility in jointly constraining the density slope and \ho. When there is no arc information the density slope is degenerate with \ho. This is a manifestation of the mass-sheet degeneracy (MSD) detailed in \cite{msd}. This is constrained by kinematics information about the lensing galaxy or by knowing lensed AGN host galaxy properties such as brightness and size. \citep{tdcosmo4, Treu_2022}.


\begin{figure*}[t]    
\includegraphics[width=0.5\textwidth]{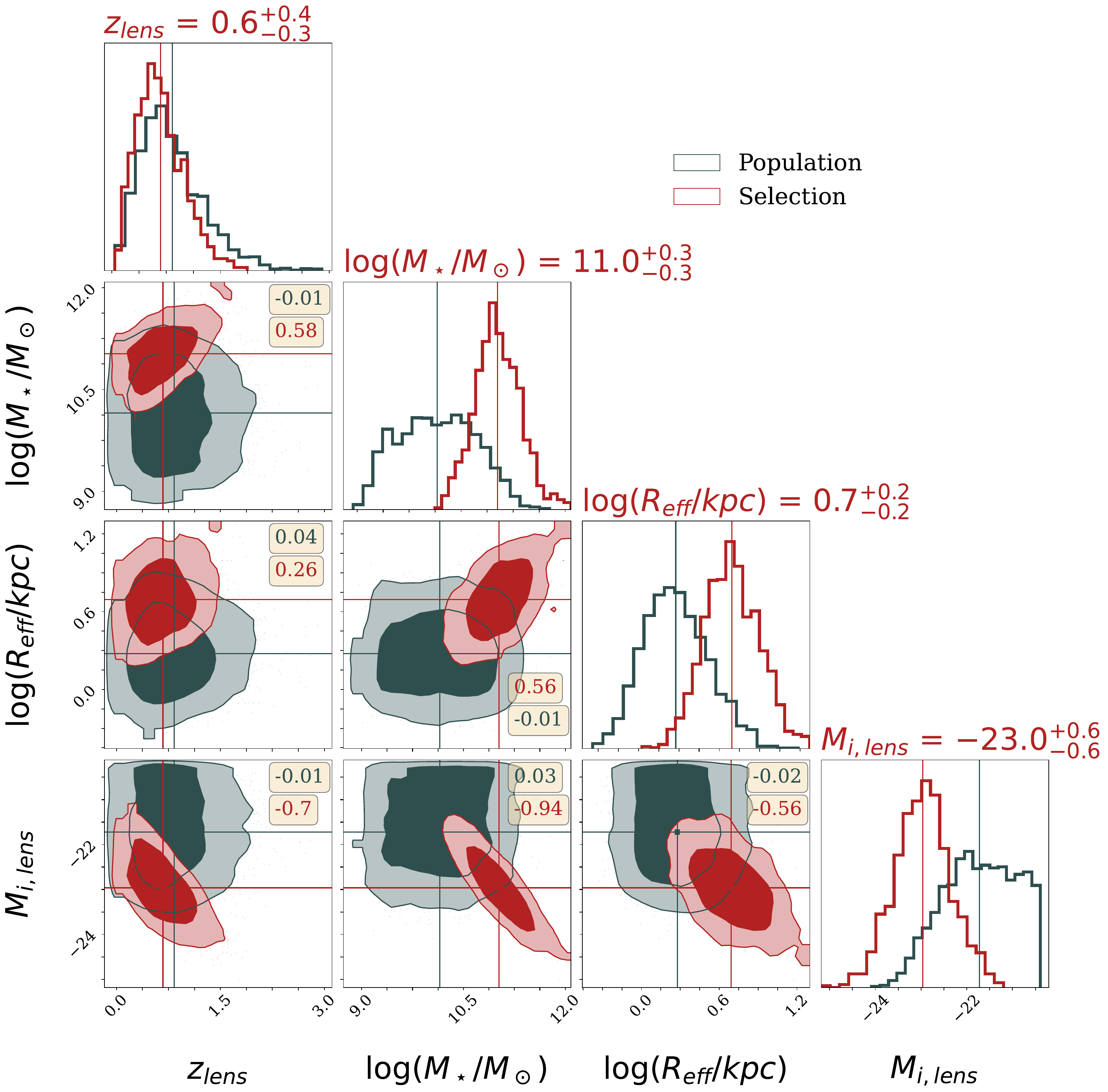}
\includegraphics[width=0.5\textwidth]{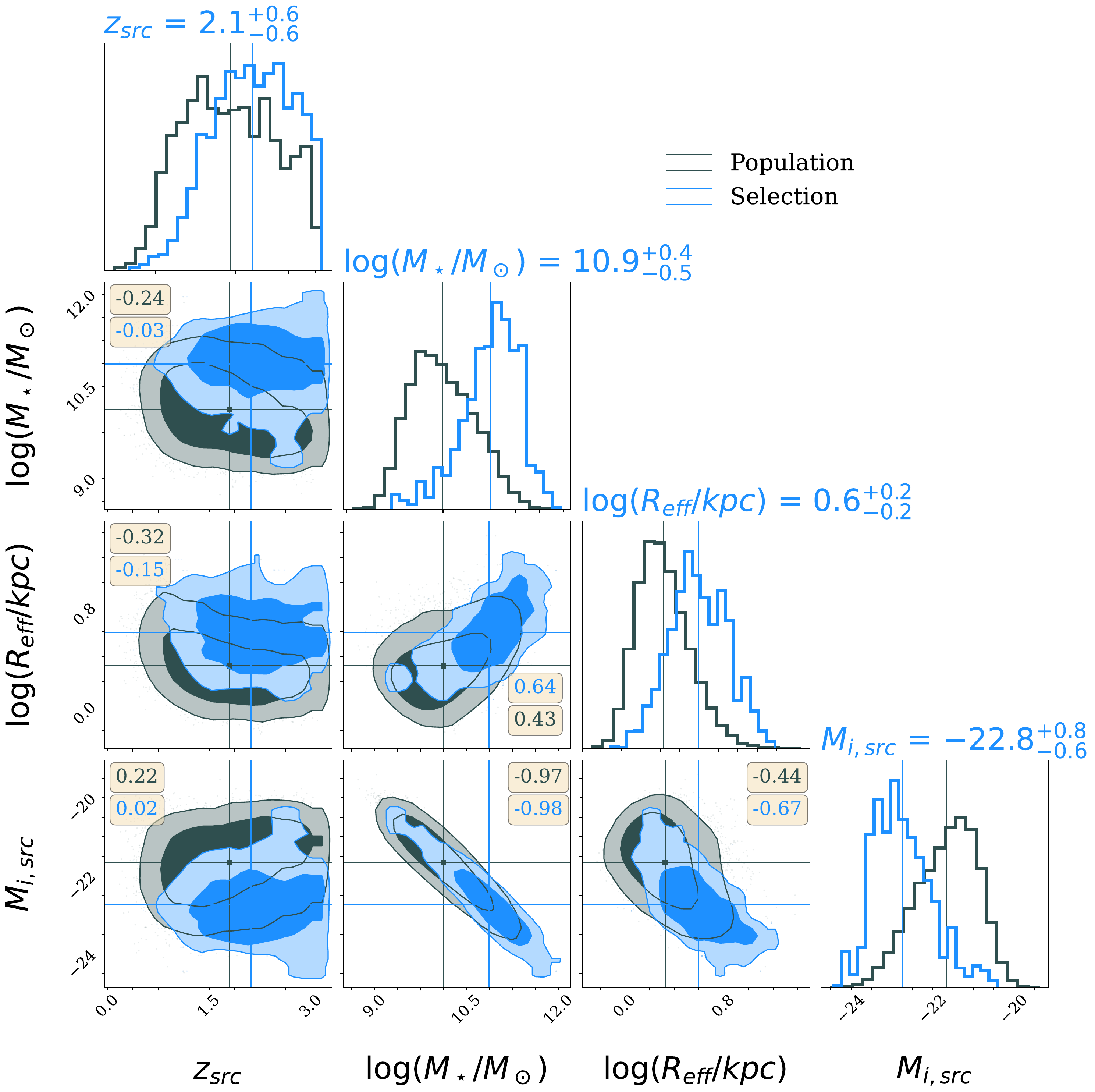}
    \caption{\textbf{Deflector Galaxy and AGN Host Galaxy Selection. }We display the following four galaxy properties: Redshift ($z$), stellar mass ($\log(M_*/M_{\odot})$), effective size ($\log(R_{\rm eff}/\rm kpc)$) and absolute brightness ($M_{i}$). \textbf{Left -- } In gray, we show properties of a sample of randomly selected galaxies from a cone. In red, we show massive elliptical galaxies that constitute the lenses in our sample. The selection of galaxies is intrinsic: i.e more massive galaxies act as lenses. \textbf{Right --} In gray, we show the population of galaxies that contain an AGN. In blue, we show the AGN containing galaxies in our sample that are lensed by a massive foreground galaxy. The selection on AGN host galaxies in this case arises from cuts placed to ensure that these systems are detectable in LSST. Thus, this selection of AGN host galaxies in blue is observationally imposed, rather than intrinsic.}
    \label{fig:om10_select}
\end{figure*}

\subsubsection{Lens Light Model}\label{data:lens_light}
We model the deflector's surface brightness as an elliptical S\'ersic profile with $\rm n_{sersic} = 4$ since a \review{large} fraction of massive elliptical galaxies exhibit this profile \citep{sersic_massive_elliptical}. This is expressed as:

\begin{equation}\label{eqn:sersic}
\begin{split}
    \rm I_*(x, y) = I_*\exp
    \left[-b\left(\left[\frac{R}{R_{eff}}\right]^{\frac{1}{\rm n_{sersic}}} - 1 \right)\right], \\
    \rm where \;R = \sqrt{qx^2 + y^2/q}.
\end{split}
\end{equation}
where $\rm I_*$ is the initial amplitude, $\rm R$ is the elliptical radius, $\rm b$ scales the equation so that $\rm R_{eff}$ is the effective radius of the light profile, $\rm q$ is the axis ratio of the lens light and $\rm x,y$ are the galaxy position. 

For the light profile of deflector galaxies in our catalog, we start with the spherical light profile in OM10, then introduce ellipticity. We do this by mapping OM10 deflectors to galaxies sampled from a cone in the synthetic sky catalog, \texttt{cosmoDC2}, produced by LSST-DESC \citep[][hereafter \texttt{cosmoDC2}]{cosmodc2}. First, we select galaxies that have a bulge-to-total luminosity ratio $\rm B/T > 0.9$ - this selects for bulge-dominated galaxies and can be used to select elliptical galaxies. \review{For a given galaxy in OM10 with mass axis ratio $\rm q_m$, we find the nearest neighbor in the \texttt{cosmoDC2} catalog by mapping absolute magnitude, redshift and $\rm q_m$ (from OM10) to light axis ratio $\rm q_l$ (available in \texttt{cosmoDC2}). This matching process results in a small scatter in the mass and light axis ratio (shown in Figure \ref{fig:lens_light_impact}). We force the mass and light profiles to be perfectly aligned ($\rm \phi_l = \phi_m$).} We show the lens sample properties in Figure \ref{fig:om10_select}, left panel.

\subsubsection{Extended Source Light Model: Painting on AGN Host Galaxies}\label{data:cosmodc2}
Source light consists of light from the AGN and the AGN's host galaxy. The AGN is modeled as a point source. For the AGN host galaxy, we assume that the majority of its light comes from the bulge component of the galaxy.

In the OM10 catalog, we need to assign each lens systems with realistic AGN host galaxy properties. To do this, we map AGN properties (i.e. AGN redshift and brightness) provided in the OM10 catalog to AGN in \texttt{cosmoDC2}, as shown in Figure \ref{fig:mapping}, and use their corresponding host galaxies which are decomposed into bulge and disk components. The quasar-host relation in \texttt{cosmoDC2} is implemented as follows (reproduced for completion): the mass of the black hole ($M_{\rm BH}$) at the center of a galaxy and the Eddington rate ($\lambda_{\rm edd}$) are related to the bulge mass of the galaxy as follows (Equation 15 and 16, Section 5.4.2, \citet{cosmodc2}): 
\begin{equation}
\begin{split}
    M_{\rm BH} = 0.0049M_{\rm bulge}(M_{\rm bulge}/M_0)^\alpha, \\
    P(\lambda_{\rm edd}|z) = 0.00071\frac{(1+z)}{(1+z_0)^{\gamma_z}}\lambda_{\rm edd}^{\gamma_e}.
\end{split}
\end{equation}
where $M_0=10^{11}M_{\odot}, \alpha=0.15, \gamma_z = 3.47, \gamma_e = -0.65, z_0=0.6$, and different power-law relations are empirically obtained to account for variation in $\lambda_{\rm edd}$ as a function of redshift. 
The $i$-band magnitude ($M_i$) of the AGN is then assigned using the empirical relation between $M_i$, $\lambda_{\rm edd}$ and $M_{BH}$ presented in Figure 15 of \cite{MacLeod_2010}. 

\review{To paint on the AGN host galaxy light, we only include light from the AGN host galaxy bulge, as this is the dominant component. We model the surface brightness profile of the host galaxy bulge as an elliptical S\'ersic profile. We obtain the parameters corresponding to an elliptical S\'ersic profile (given in Equation \ref{eqn:sersic}), axis ratio of the galaxy bulge ($\rm q$), the effective radius of the galaxy bulge ($\rm R_{eff}$) and $\rm n_{sersic} = 4$, from cosmoDC2. We show the AGN host galaxy sample properties in Figure \ref{fig:om10_select}, right panel.}


\subsection{LSST Images}\label{data:lsst} 
We \reviewt{start with the lens model catalog} described in Section \ref{data:om10_lens} and produce \reviewt{corresponding} LSST-like images. \reviewt{To simulate the whole catalog of lenses, we provide lens models and LSST observation specifications shown in  Table \ref{Tab:specs} to \texttt{paltas} \citep{paltas} (described further in Section \ref{method:NPE}), which builds on \texttt{lenstronomy} \citep{Birrer_2018, Birrer2021}. \texttt{lenstronomy} computes the lensing image positions and convergence following our model assumptions.} All images have 33$\times$33 pixels to cover 6.6 arcsec on a side, given the LSST pixel scale of 0.2 arcsec. All images are simulated in LSST $i$-band, and we defer multi-band modeling to future work (Section \ref{disc:future}). All noise properties and zeropoints pertain to 5-year coadd $i$-band imaging, shown in Table \ref{Tab:specs}.

\begin{deluxetable}{c|c}
\tablecaption{Assumed LSST Camera $i$-band image specifications}
\tablehead{
\colhead{Property} & \colhead{Value}
}
\startdata
Pixel scale (arcsec/pixel) & 0.2 \\  
CCD gain (electron/ADU) & 2.3 \\
Instrumental noise/exposure & 9 \\
$i$-band instrumental zeropoint & 28.17 \\
Sky brightness (mag/arcsec$^2$) & 20.46\\
Exposure time (s) & 15 \\
Number of Exposures & 150 [5 year-coadd] \\
\review{Median PSF FWHM (arcsec)} & \review{0.85} \\
\enddata
\tablecomments{We describe each property in Section \ref{data:lsst}. Values are taken from dp0.2 \footnote{https://smtn-002.lsst.io/}.\label{Tab:specs}}
\end{deluxetable}
\begin{figure*}
    \centering
    \includegraphics[width=1\textwidth]{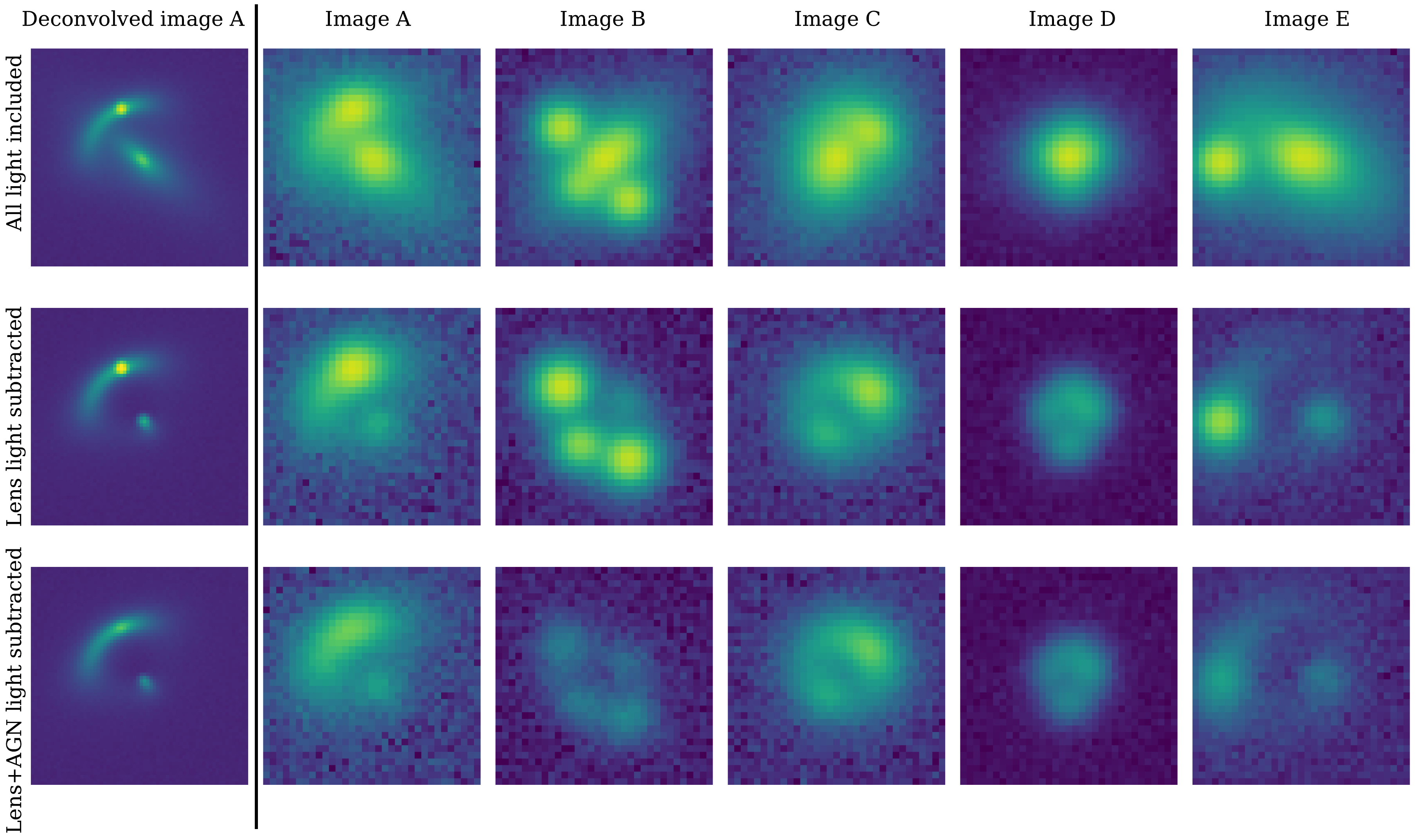}
    \caption{5 simulated LSST lensed \reviewt{AGN} systems in the $i$-band. We test our inference pipeline on 3 different preparations of the data. Subtraction of different light components have been shown to improve the performance of the modeling pipeline in lens parameter recovery. We discuss this further in Section \ref{data:3_preps}. In the first column (to the left of the black line), we show the deconvolved version of the second column of images (to the right of the black line). Deconvolution also helps in the lens modeling, detailed also in Section \ref{data:3_preps}. \textbf{Top row:} include lens light, AGN light and host galaxy light. \textbf{Middle row:} Lens light subtracted. Poisson noise due to lens light remains. \textbf{Bottom row:} Lens light and AGN light subtracted. Poisson noise due to all light remains. The background Gaussian noise also remains constant across 3 different preparations. }
    \label{fig:gallery}
\end{figure*}

Ground-based images are heavily impacted by the atmosphere which blurs the light we receive on the telescope and this effect is captured in the PSF. \review{The PSF varies across the focal plane; we extract a grid of PSF kernels at different positions from Data Preview 0.2 and assign each simulated lens a different kernel. The median measured PSF-FWHM, when we fit a gaussian profile to the entire sample of PSF grids, is 0.85 arcsec.} 

\review{We realize LSST lensed AGN images with properties from our modified catalog described in Section \ref{data:om10_lens}, which result in different multiplicity of images, image separations and magnifications from the original OM10 catalog. We then eliminate systems with: (i) undetectable images, (ii) only a single image or (iii) unresolved images. This brings our sample from 1557 lenses down to 1344 lenses.}





\subsection{Image Preparations}\label{data:3_preps}
In order to understand how lens modeling performance depends on the preparation of image data, we simulate the catalog data under three different preparations (see Image \ref{fig:gallery})\review{: all light included, lens light subtracted and lens and AGN light subtracted}. We use \texttt{lenstronomy}'s Image Simulation module\footnote{\href{https://github.com/lenstronomy/lenstronomy/tree/main/lenstronomy/ImSim}{lenstronomy/ImSim GitHub link}}, under the hood of \texttt{paltas}, to make these.

First, for our fiducial test set \review{-- all light included}, we simulate images that include all light from all components of a lensed AGN system: deflector, lensed AGN, and lensed AGN host galaxy. 

In a second test set \review{-- lens light subtracted}, we emulate perfect lens light subtraction, including Poisson noise from all components but omitting the surface brightness of the deflector itself. De-blending and removing the lens light makes the point sources and arcs more visible, which can help in inferring the structure of the lensing system. We are motivated by previous works that indicate that removing lens light is beneficial for lens parameter recovery \citep[e.g.][]{lsst_euclid, madireddy2019}. 

In a third test set, we assume subtraction of both the lens light and the AGN point source light. This last preparation aims to highlight information coming from the lensed arc of the AGN host galaxy which constrains model parameters like the density slope of the deflector, $\gamma_{\rm lens}$ \citep{gamma_lens_degeneracy}. AGN point source light can remove arc information, since a) blending between the AGN point source and the lens reduces the contrast of the lensed arcs, and b) the high Poisson noise due to the bright AGN lowers the SNR of fainter arcs. 

In all three preparations described above, the background Gaussian noise and Poisson noise due to flux from the deflector, AGN, and AGN host galaxy remain in the image, as it would in real data. These different preparations are illustrated in Figure \ref{fig:gallery}. 

Beyond the three preparations, we are also interested in the impact that image data quality has on the the NPE modeling of lensed AGN images. We expect to see that with better data quality, all features of the system are more visible and separable leading to better modeling results. One of the key parameters of interest is the projected mass density slope \gammalens, and this parameter can be constrained with high-quality image data with sharp arc features. Obtaining higher-resolution images of a system we have ground based data for is possible through dedicated follow-up through a space-based telescope, or through deconvolution. In this project, we emulate the deconvolution of our LSST images, and study an additional three preparations (deconvolved-fiducial, deconvolved-lens light subtracted and deconvolved-lens and AGN light subtracted) shown in Column 1, Figure \ref{fig:gallery}. \review{We do this as technology for deconvolving data at LSST scale is not yet available, and preparing deconvolved training data is also challenging (further elaborated in Section \ref{disc:lensing_info}). In practice, we will combine dithered exposures across 5 years to reconstruct a higher resolution image of the lens, using STARRED \citep{starred}. This would require generating individual exposures which is beyond the scope of this paper. For consistency, we emulate STARRED-like deconvolution of images, which produces a known Gaussian PSF with 2 pixel FWHM, at 0.1"/pixel (subsampled by a factor of 2 compared to the original data).}

\section{Method}\label{sec:method}
Our goal is to \review{infer} lens mass models from simulated ground-based images of lensed AGN, and then recover sample properties of the lenses. We are primarily interested in learning the following lens mass model parameters used in Equation \ref{eqn:lens_model}: Einstein radius ($\theta_E$), density profile slope ($\gamma$), external shear parameters ($\gamma_1$ \& $\gamma_2$), mass ellipticity parameters ($e_1$ \& $e_2$), lens mass center ($\rm x_{lens, M}$ \& $\rm y_{lens, M}$). When only the arc information is retained in the images with sufficient signal (in cases where the lensed AGN host galaxy is visible after lens and AGN subtraction), we also learn AGN host galaxy parameters: apparent magnitude ($\rm m_i$) and effective radius or half light radius ($\rm R_{eff}$).

We use a SBI method called Neural Posterior Estimation (NPE) to \review{infer} individual lens mass model posterior PDFs from LSST images, outlined in Section \ref{method:NPE}. We use Hierarchical Bayesian Inference (HBI) to learn the underlying distribution the lenses were sampled from, detailed in Section \ref{method:HBI}. Code used in this work is publicly available on GitHub\footnote{\href{https://github.com/padma18-vb/lsst_lensed_agn_modeling}{\texttt{lsst\_lensed\_agn\_modeling}}} with datasets archived on Zenodo \citep{venkatraman_zenodo}.


\begin{deluxetable*}{c|c}
\tablecaption{Training Distribution}
\tablehead{
\colhead{Parameter} & \colhead{Distribution}
}
\startdata
\textbf{Lens Mass Model: PEMD + External Shear} & \\
\hline
Einstein Radius (") & $\theta_E \sim \mathcal{N}_{tr}(\sigma_{low} = -0.5, \sigma_{high} = \infty, \mu = 0.8, \sigma = 1$) \\
Density Profile Slope & $\rm \gamma_{lens} \sim \mathcal{N}(\mu = 2, \sigma = 0.2)$ \\
External Shear & $\gamma_{1,2} \sim \mathcal{N}(\mu = 0, \sigma = 0.1)$ \\
Mass Ellipticity ($\rm e_{mass}$) & $e_{1,2} \sim \mathcal{N}(\mu = 0, \sigma = 0.2)$\\
Lens Mass position (")($\rm xy_{mass}$)& $\rm x_{lens, M},y_{lens, M} \sim \mathcal{N}(\mu = 0, \sigma = 0.06) $ \\
\hline
\textbf{Lens Light Model: Single Sersic Ellipse} & \\
\hline
Light Ellipticity ($\rm e_{light}$) & $\rm e_{1,2} \sim e_{mass} + 0.05 * N(0, 1)$ \\
Lens Light position & $x_{\rm lens, L},y_{\rm 
lens, L} \sim \rm xy_{mass} + 0.001*N(0, 1)$\\
Sersic Index & $\rm n_{lens} \sim \mathcal{N}(\mu=4,\sigma=0.005)$ \\
Effective Radius (") & $\rm R_{lens} \sim \mathcal{N}_{tr}(\sigma_{low}=-0.5, \sigma_{high} = \infty, 
\mu=0.7, \sigma=1)$ \\
Apparent Magnitude (AB mag) & $\rm m_i \sim \mathcal{N}(\mu=20.5,\sigma=2)$\\
\hline
\textbf{Extended Source Light Model: Single Sersic Ellipse} & \\
\hline
Light Ellipticity ($\rm e_{src}$) & $e_{1,2} \sim \mathcal{N}(0, 0.1)$ \\
Source Light position (") & $x_{\rm src},y_{\rm src} \sim \mathcal{N}(0, 0.4)$\\
Sersic Index & $n_{\rm src} \sim \mathcal{N}(\mu=4,\sigma=0.001)$ \\
Effective Radius(") & $\rm R_{eff,\,src} \sim \mathcal{N}_{tr}(\sigma_{low}=-0.5, \sigma_{high} = \infty, 
\mu=0.7, \sigma=1)$ \\
Apparent Magnitude (AB mag) & $\rm m_i \sim \mathcal{N}(\mu=24,\sigma=2)$
\enddata
\tablecomments{This table contains the parameter distributions we sampled from to generate the $5\times10^5$ training images. All parameters are described in Section \ref{data:om10_lens}. $x_{\rm lens, M}$ refers to the lens mass x-position, and $x_{\rm lens, L}$ is the lens light x-position, and same notation follows for y-position. $\gamma_1 , \gamma_2$ refers to the Cartesian shear components and $\rm e_1, e_2$ refer to Cartesian ellipticity components introduced in Section \ref{data:lens_mass}. $\mathcal{N}$ is the Normal distribution and $\mathcal{N}_{tr}$ is the Truncated Normal distribution truncated at $\rm \sigma_{low}$ and $\rm \sigma_{high}$.\label{Tab:train}}
\end{deluxetable*}

\subsection{Individual Lens Mass Models: Neural Posterior Estimation}\label{method:NPE}
In this section, we outline how we use NPE to obtain the posterior PDF, represented as $\rm p(\xi_k|d_k)$, for the PEMD lens mass model parameters $\rm \xi_k$, given an image of the lensed AGN, $\rm d_k$. 

Deep convolutional neural networks (CNNs) can be used to map images to features. NPE guides the CNN output features to approximately describe a posterior PDF for the parameters listed above given an input image. \review{During training, an image $\rm d_n$ is mapped to an output $\rm q_\phi(\xi_n | {d_n}, \review{\nu_{int}})$. Here, $\phi$ represents the model weights \citep{npe_loss, Cranmer_2020}, and $\rm q_\phi(\xi_n | {d_n}, \review{\nu_{int}})$ represents a multivariate Gaussian with a full covariance matrix over model parameters $\rm \xi_n$, conditioned on the image $\rm d_n$ and the hyperparameters governing the prior PDF assumed during training, $\rm \nu_{int}$}. For each image $\rm d_n$, the network returns parameter means ($\mu$) and the Cholesky factor \review{($C$)} of the covariance matrix (\review{$C$}: lower triangular matrix such that \review{$C C^T = \Sigma$}) defining the posterior PDF.

The NPE output $\rm q_\phi(\xi_n | {d_n}, \review{\nu_{int}})$ is an approximation of the true lens model posterior PDF, $\rm \review{p}(\xi_n | {d_n}, \review{\nu_{int}})$. The multivariate Gaussian posterior distribution $\rm q_\phi$ is shifted, made narrow or broader along different parameter axes in order to minimize the Kullback-Leibler (KL) divergence between the approximate posterior ($\rm q_\phi$) and the true posterior (\review{$\rm p$}). As shown in \cite{npe_loss}, this corresponds to minimizing the loss below:

\begin{equation}\label{eqn:npe_loss}
L(\phi) = -\sum_{\review{n}=1}^{N}\log{\rm q_{\phi}(\xi_{\review{n}, \review{truth}}|d_{\review{n}}, \review{\nu_{int}}).}
\end{equation}

In the limit of infinite training data, $N\rightarrow\infty$, given an expressive network and sufficiently flexible functional form of $\rm q_\phi$ \citep{npe_loss},
$\rm q_\phi(\xi_n | {d_n}, \review{\nu_{int}} ) \approx \rm \review{p}(\xi_n | {d_n}, \review{\nu_{int}})$.

\subsection{Network Training}\label{method:train}
In the following sections, we elaborate on the training data and procedure used to construct individual lens mass model posteriors using NPE. 

\subsubsection{Training Data}\label{method:train_data}
To learn the posterior PDF of lens mass model parameters from an image, the estimator $q_\phi$ requires multiple and repeated exposures to image-parameter mappings. We sample from $\rm N = 5\times10^5$ lensing configurations from the broad training prior ($\rm \xi_{\review{n}} \sim p(\xi_{\review{n}}| \nu_{int})$) and simulate an image, $\rm d_{\review{n}}$ for each configuration. Here, $\rm \nu_{int}$ refers to hyperparameters governing the broad training prior, and $\rm p(\xi_{\review{n}}|\nu_{int})$ is the training distribution. We present this distribution for each of the model parameters and for latent parameters that we do not infer (such as lens apparent magnitude) in Table \ref{Tab:train}. These image-parameter mappings ($\rm d_{\review{n}}; \xi_{\review{n}}$) becomes our training data.

We use \texttt{paltas} to rapidly sample the training prior, reject parameter configurations that do not form detectable lensing systems and simulate the training images. \texttt{paltas} uses \texttt{lenstronomy} simulation code to make the images, so our train and test images are produced in a consistent way.

Choice of the training prior is important because the quality of lens parameter inference relies on what the network sees during training. The training prior distribution is chosen to be wider than the test distribution for all learned and latent parameters. For certain parameters, the training distribution is truncated at limits beyond which the system is unphysical. \review{The only physical correlation we include in our training data is between the ellipticity of the lens mass and lens light as shown Table \ref{Tab:train}.} Further discussion on training distribution realism and including physical correlations between parameters in the training data is available in Section \ref{disc:training_data}.
We generate and \review{store training images} corresponding to the image preparations described in Section \ref{data:3_preps} \review{using CPU and memory resources at the National Energy Research Scientific Computing Center (NERSC).}

\subsubsection{Training Procedure}\label{method:train_proc}
We use a CNN based on xResNet-34 architecture \citep{resnet34} to model image data and extract lensing parameters. We start with a learning rate of $10^{-4}$ and use an exponential decay schedule. We apply random rotations to the images. \review{We use a fixed seed for all training runs.} We train for 200 epochs, revisiting the data with batches of 1024 randomly sampled images. The network training loss shown in \review{Equation \ref{eqn:npe_loss}} is optimized using the gradient-descent optimizer Adam \citep{adamopt}. We use early stopping, defined as terminating training if the validation loss does not improve for 10 consecutive epochs, to select the model weights. We find that this consistently yields more calibrated and robust results compared to picking model weights with the lowest validation loss. \review{All training hyperparameters can be found in the training configuration files stored in the code repository}.

\review{The paltas network training, built on \texttt{tensorflow} \citep{tensorflow}, is run using GPU resources at NERSC which is approximately 10x faster than CPU based training.}

\subsection{Sample-level property inference: Hierarchical Bayesian Inference}\label{method:HBI}
\review{Applying our trained network to the expected LSST lensed AGN sample of lenses produces lens mass models for each system, from which we can hierachically infer sample-level properties.} Sample information can be folded into hierarchical analysis to infer cosmology parameters as shown in TDCOSMO-IV \citep{tdcosmo4}.

We are interested in inferring the \review{underlying} distribution of lensing parameters i.e. $\rm p(\xi_k | \nu)$, \review{where $\nu$ refers to the hyperparameters that govern the true underlying distribution, also referred to as the conditional PDF or cPDF.} Given a sample of LSST images, $\{d\}$, we can infer the hyperparameters, i.e. $\rm p(\nu | \{d\})$, using hierarchical Bayesian inference \citep[e.g.][]{erickson2024}.

To infer properties of $\nu$, we start with Bayes' Theorem. We assume each lens provides an independent constraint, and thus multiply the likelihood from each modeled lens $k$:

\begin{equation}
    \rm p(\nu | \{d\}) = p (\nu)\prod_k\frac{p(d_k| \nu)}{p(\{d\})}.
\end{equation}

We elaborate this expression to include lens parameter information pertaining to each system i.e. $\xi_k$, and then marginalize over $\xi_k$, as shown:
\begin{equation}\label{eqn:pop_model}
    \rm p(\nu | \{d\}) = p (\nu)\prod_k\int d\xi_k\frac{p(d_k| \xi_k, 
    \nu)p(\xi_k| \nu) }{p(\{d\})}.
\end{equation}

We have access to the lens model posterior, $\rm p(\xi_k | d_k, \nu_{int})$, for each lens $k$, from the NPE modeling. We note that the lens parameters $\xi_k$, is conditioned on both the image, $\rm d_k$, and the previously assumed model during training $\rm \nu_{int}$ (as detailed in Section \ref{method:NPE}). As the inferred lens models are not conditioned on the actual population model $\nu$, we need to rewrite the posterior using the following rearrangement:

\begin{equation}\label{eqn:rewrite}
    \rm p(\xi_k | d_k, \nu_{int}) = \frac{p(d_k | \xi_k, \nu_{int})p(\xi_k|\nu_{int})}{p(d_k|\nu_{int})}.
\end{equation}

Rearranging Equation \ref{eqn:rewrite}, and inserting it into Equation \ref{eqn:pop_model}, we find that:

\begin{equation}\label{eqn:p_nu_d}
\begin{split}
    \rm p(\nu | \{d\}) = p (\nu) 
    \prod_k\int d\xi_k \ p(\xi_k | d_k, \nu_{int}) \\
    \rm \times \frac{p(d_k|\nu_{int})}{p(\{d\})} \frac{p(\xi_k| \nu) }{p(\xi_k|\nu_{int})}.
\end{split}
\end{equation}

In Equation \ref{eqn:p_nu_d}, the last term in the integrand, $\rm \frac{p(\xi_k|\nu)}{p(\xi_k|\nu_{int})}$, serves to divide out the interim prior and replace it with the cPDF, which can be thought of as the prior we would like to have assigned in the first place. Formalism for this is also presented in \citep[e.g.][]{Wagner_Carena_2021, foreman_mackey_exo}. We infer $\nu$ by adapting on scripts in \texttt{lens-npe} \citep{lens-npe} which builds on the MCMC package \texttt{emcee} \review{\citep{emcee}}. \review{Our set of 16 inferred $\nu$ parameters are: $\rm \mathcal{M}(\theta_E), \mathcal{M}(\gamma_{lens}),\mathcal{M}(e_{1\&2}), \mathcal{M}(\gamma_{1\&2}), \mathcal{M}(x\,\&\,y_{lens, M})$, $\rm \Sigma(\theta_E), \Sigma(\gamma_{lens}),\Sigma(e_{1\&2}), \Sigma(\gamma_{1\&2}), \Sigma(x\,\&\,y_{lens, M})$.}

We can now return to \reviewt{the} distribution of lensing parameters given that we know the true underlying distribution, i.e. $\rm p(\xi_k | \nu)$, the cPDF. We model the cPDF as a multivariate Gaussian distribution with a diagonal covariance matrix. We start with a simple cPDF to study the available precision of the hyperparameters under such a simple assumption, as well as any bias incurred by it. In future work, we will investigate more expressive cPDFs and hyperparameters governing the cPDF. 

In order to infer cosmological parameters such as $H_0$ correctly, we must infer $H_0$ and $\nu$ jointly as is done in \citet{tdcosmo4}. Biases on particular parameters like the population mean of the density profile slope will translate to a bias on Fermat potential, and thus $H_0$. 

\begin{figure*}
    \centering
    \includegraphics[width=0.8\textwidth]{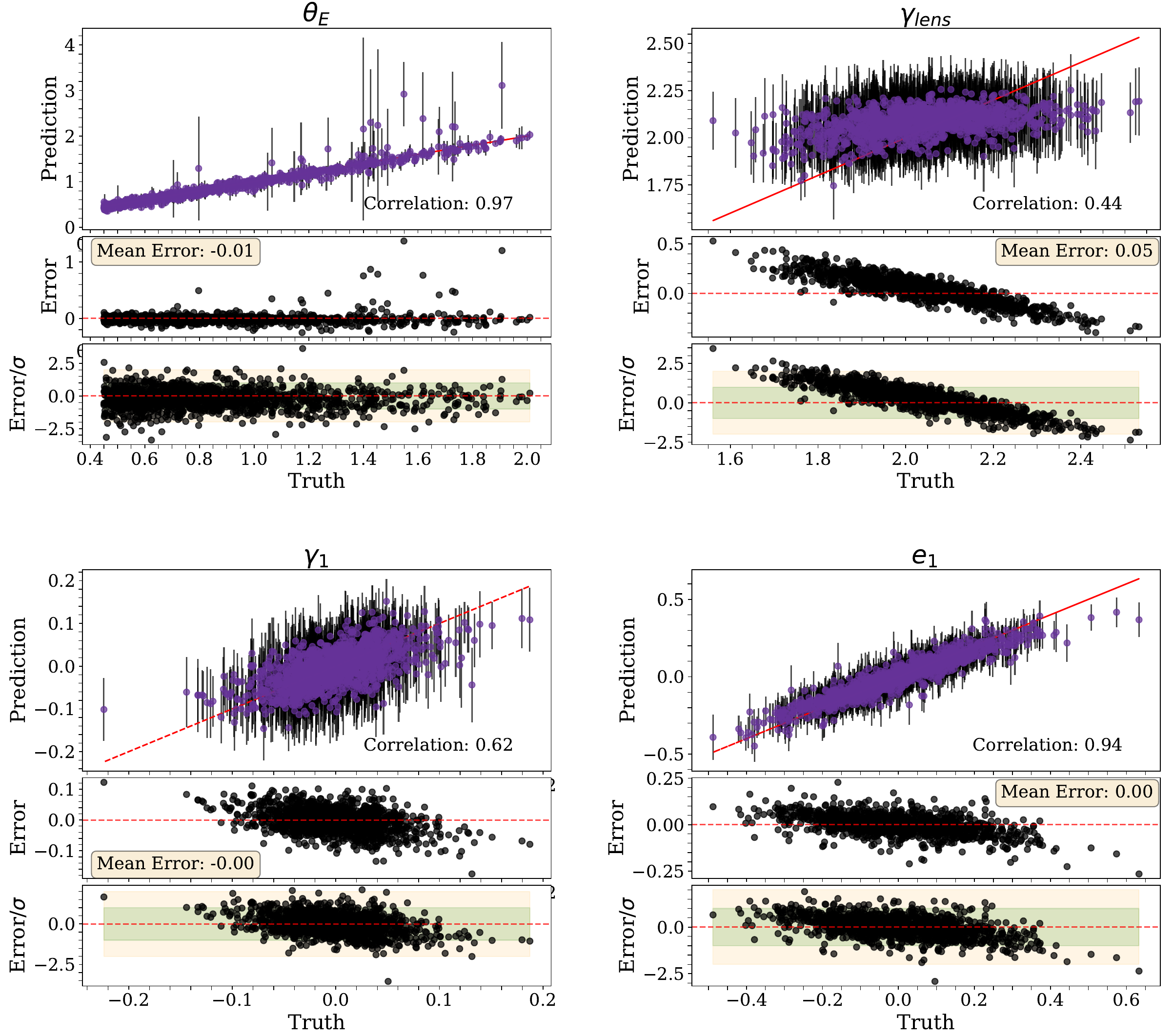}
    \caption{Comparison of 1344     predicted Gaussian posteriors (obtained as described in Section \ref{method:NPE}) for fiducial data (refer: Section \ref{data:lsst}) against ground truth. Purple points correspond to the mean of the posteriors, and black lines correspond to the posterior $1 \,\sigma$ width or standard deviation. In the bottom most panel of each parameter subplot, the green shaded region represents the $1\,\sigma$ area and the yellow shaded region represents the $2\,\sigma$ region.}
    \label{fig:all_interim_results}
\end{figure*}


\begin{deluxetable*}{|c||c|cccccccc|}
\tablecaption{Individual lens mass model posterior results on fiducial test set}
\tablehead{
\colhead{Preparation} & \colhead{Metric} & \colhead{$\theta_E$} & \colhead{$\gamma_{\rm lens}$} & \colhead{$e_{1}$} & \colhead{$e_{2}$} & \colhead{$\gamma_{1}$} & \colhead{$\gamma_{2}$} & \colhead{$x_{\rm lens}$} & \colhead{$y_{\rm lens}$} \\
\hline
 &  & (arcsec) &  &  &  &  &  & (arcsec) & (arcsec)
}
\startdata
All Light [Fiducial] & ME & -0.01 &  0.05 &  0.   & -0.   & -0.   & -0.   & -0.   & -0. \\  
& MAE & 0.02 & 0.1  & 0.03 & 0.03 & 0.02 & 0.02 & 0.01 & 0.01 \\
 & Precision & 0.05 & 0.16 & 0.09 & 0.09 & 0.06 & 0.06 & 0.02 & 0.02 \\
\hline
Lens Light Subtracted & ME & -0.01 &  0.07 &  0.   & -0.   &  0.01 & -0.   & -0.   & -0. \\  
& MAE & 0.02 & 0.1  & 0.06 & 0.06 & 0.03 & 0.03 & 0.02 & 0.02 \\
 & Precision & 0.04 & 0.16 & 0.13 & 0.13 & 0.07 & 0.06 & 0.04 & 0.04 \\
\enddata
\tablecomments{Fiducial experiment results, outlined in Section \ref{res:LSST}. ME = Average deviation of posterior means from truth, indicator of bias. MAE = Median absolute error of posterior means from truth, indicator of scatter. Precision = Median uncertainty of posteriors.\label{Tab:fiducial}}
\end{deluxetable*}

\section{Experiments and Results}\label{sec:results}

In this section, we present lens modeling results we obtain from the NPE method and population-level model inference. 

For all tests, including the fiducial (Section \ref{res:LSST}), we show results with and without lens light subtraction (details on image preparation in Section \ref{data:3_preps}). After running the fiducial, we construct experiments to investigate the  performance drivers. First, we test the input images to the method. We investigate the selection of systems with bright host galaxies (Section \ref{res:bhg}), and the resolution of the images (Section \ref{res:DECONV}). Second, we investigate our method. We try removing mass-light correlations in training data (Section \ref{res:corr}), and removing distribution shift between the test and training data (Section \ref{res:dist_shift}). \review{\textit{Note: }Throughout this work, we use `distribution shift' to refer to a difference in the mean, variance, and/or skewness of one or more parameters between the training and test distributions.}
We are interested in recovery of $\theta_E$ and \gammalens, which are, respectively, the easiest and most difficult parameter to measure from imaging alone \citep{erickson2024}. We present NPE recovery for all experiments for these key parameters in Table \ref{Tab:optimize_data_and_method}. We present the error in population-level parameter recovery for all parameters in Table \ref{Tab:pop_model}, and the recovery with uncertainties in Table \ref{Tab:pop_model_full}. Further, we present NPE posterior recovery of all parameters in the Appendix \ref{app:full} in Table \ref{Tab:optimize_data_and_method_full}.


\begin{figure}
    \includegraphics[width=0.47\textwidth]{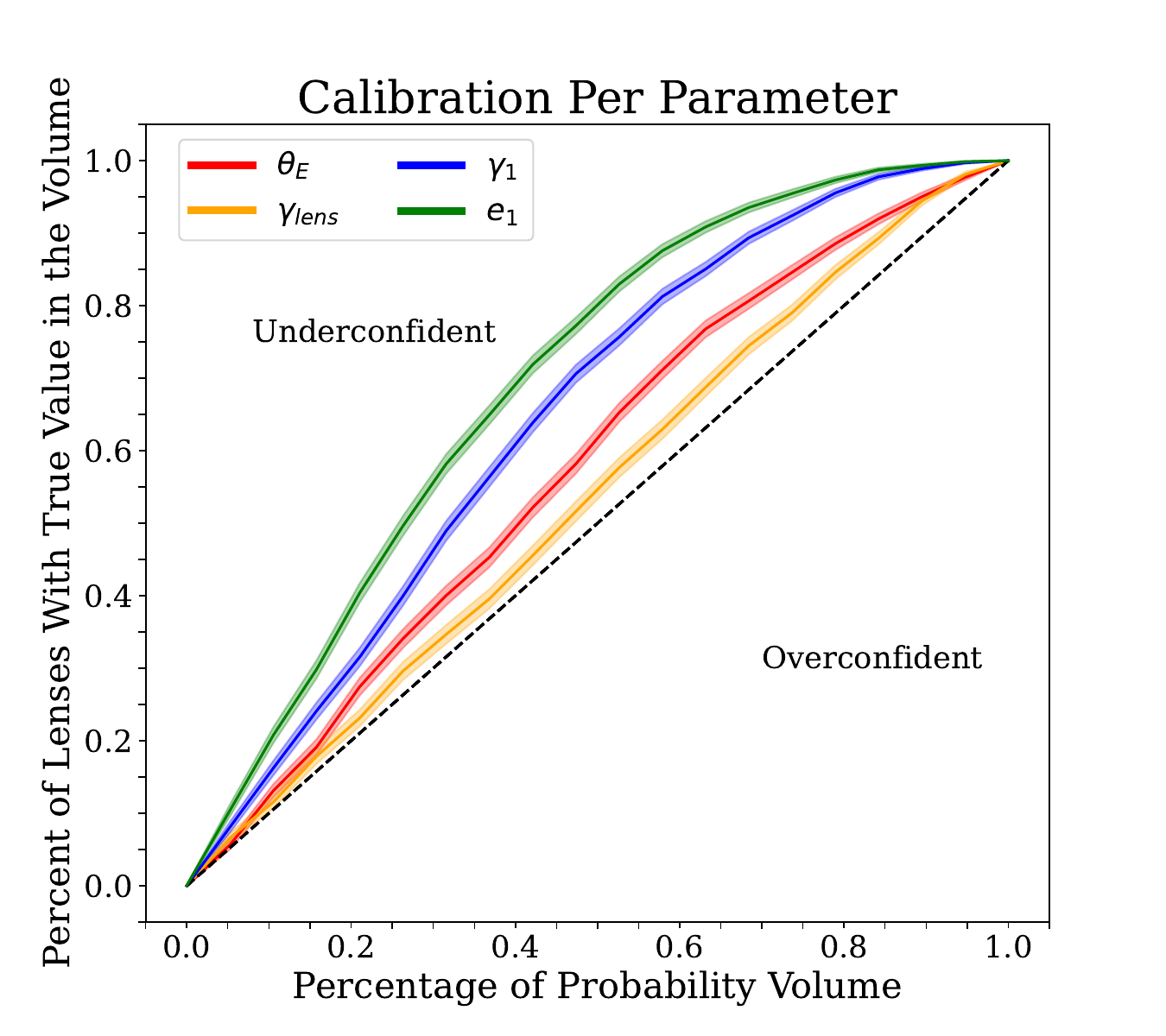}
    \caption{Calibration for 4 parameters in the fiducial case (Section \ref{res:LSST}.)}
    \label{fig:calibration_all}
\end{figure}


\begin{figure*}
    \centering 
    \includegraphics[width=0.7\textwidth]{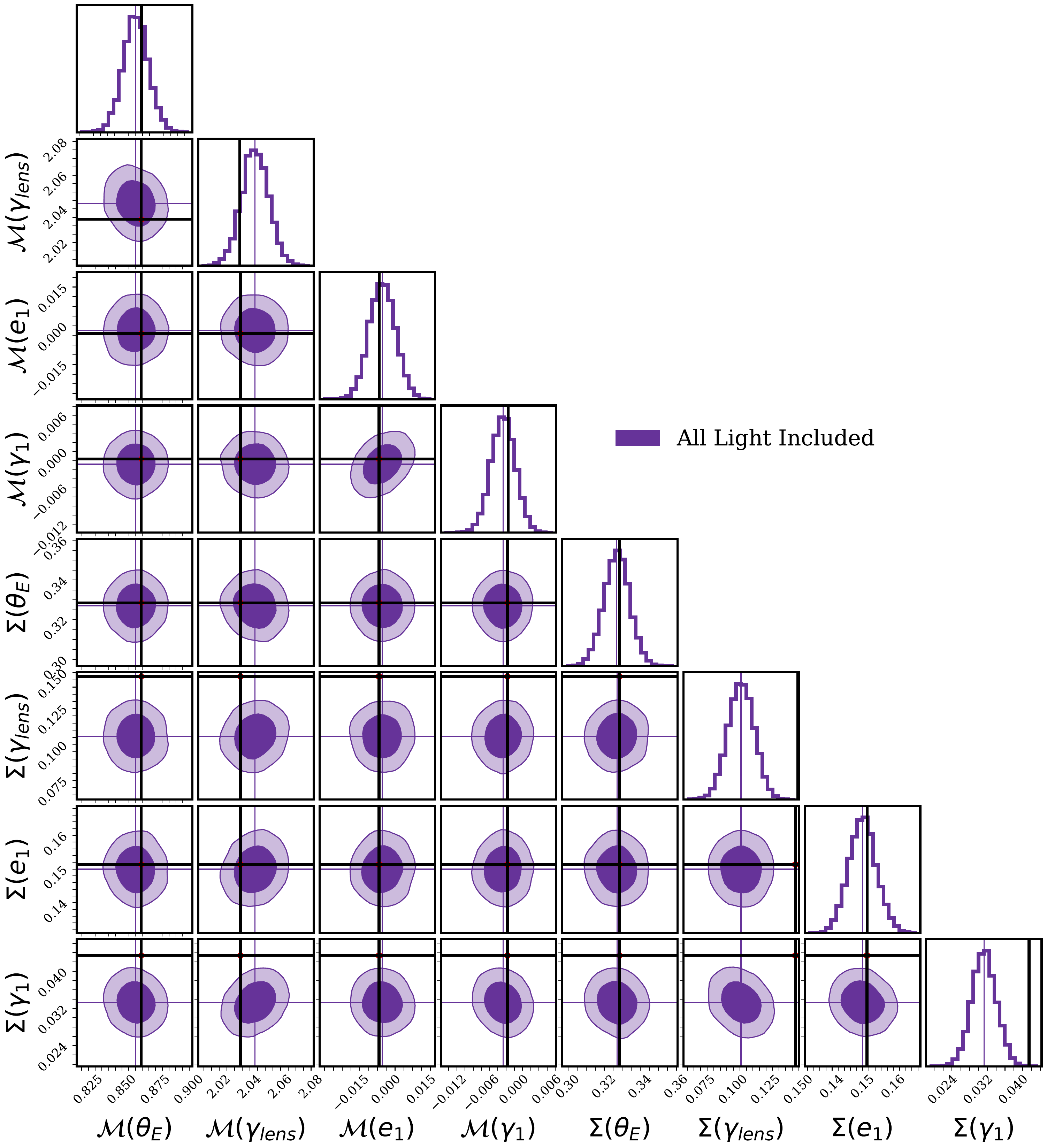}
    \caption{Population level model recovery for $\theta_E, \gamma_{\rm lens}, e_{1}$ and $\gamma_{1}$ using posteriors recovered from the fiducial test set. The black solid line is the ground truth of the test set, and the purple lines indicate the mean value of the converged chains. Note that we have removed 71 objects from our sample that had wider posteriors than the training distribution along one or more parameter axes as this breaks the HBI framework (refer Section \ref{method:HBI}). We show results for $e_1$ and $\gamma_1$ only, as $e_2$ and $\gamma_2$ reproduce the same result. We note that the population model learns a relationship between ellipticity and shear encoded into the test. The population mean of $\rm \gamma_{lens}$ is well constrained by the fiducial test set, and we go on to demonstrate how this can change with different model assumptions or data quality in Figure \ref{fig:pl_slope_pop} and Figure \ref{fig:lens_light_impact}.}
    \label{fig:HI_model}
\end{figure*}



\subsection{Fiducial Results}\label{res:LSST}
\review{In this section, we present NPE lens modeling results from LSST i-band 5-year coadd images (descriptions for catalog in Section \ref{data:om10_lens} and data preparation in Section \ref{data:3_preps}).} These data are selected as the baseline, since they will not require any additional processing beyond survey data products. First, we assess the inference of mass model posteriors for individual lens parameters. We use the following metrics: 
\begin{enumerate}
    \item Mean error (ME): ME for a given parameter is the average difference between the mean of the posterior and the ground truth value for that parameter across all lenses. ME tracks bias by testing for systematically low or high predictions. 
    \item Median absolute error (MAE): MAE for a given parameter is the median absolute difference between the mean of the posterior and the ground truth value for that parameter across all lenses. MAE tracks accuracy as lower values indicate that predictions are clustered more tightly around the truth. 
    \item Precision: Precision of a given parameter is the median uncertainty or 1-$\sigma$ posterior width  across all lenses. 
\end{enumerate}

For parameters $\rm \theta_E$ and $\rm \gamma_{lens}$, we also present the relative precision\footnote{Defined as: $\rm precision/lens \,\,(\%)$ = $\rm median (\sigma/truth) \times 100$} and relative mean error percentage\footnote{Defined as: $\rm mean \, error (\%)$ = $\rm mean (error/truth) \times 100$} to better compare between different experiments.

Finally, we are interested in the uncertainty \textit{calibration} of the posteriors. We test whether the predicted uncertainties of the approximate posteriors are consistent by verifying that x\% of posterior volume contains the truth x\% of the time. If less than x\% of the truth is contained in x\% of the posterior volume, this volume is too large and the predicted uncertainties are too big, the network is under-confident. Correspondingly, if more than x\% of the truth is in x\% of the volume, the predicted uncertainties are too small, the network is over-confident. This helps assess calibration over the entire test set. This formalism was introduced in \citep{Park_2021}.

We present recovery of $\rm \theta_E, \gamma_{lens}, \gamma_1, e_1$ in Figure \ref{fig:all_interim_results}. We present recovery of all model parameters in Table \ref{Tab:fiducial}. First we present results when all light components are included in the image. On the individual mass models, we are able to recover key mass model parameter $\theta_E$ with $6.5\%$ precision per lens, and \review{$-1\%$ bias on average}. This corresponds to a median precision of 0.05 and average error of -0.01 overall on the entire test set as shown in Table \ref{Tab:fiducial}. We recover the density profile slope $\gamma_{\rm lens}$ with $8\%$ precision per lens and $3\%$ \review{bias on average}. This corresponds to 0.16 median precision and 0.05 mean error on the whole test set as shown in Table \ref{Tab:fiducial}. 

\review{We also note that the network picks up on additional information not present in the training data: the posterior PDFs across lenses contain strong correlation between learned ellipticity and external shear. Ellipticity and shear have the same effect in changing the appearance of the lens: trying to fit each of their contribution to the lens model results in a shear-ellipticity degeneracy \citep[e.g.][]{shear_ellipticity, fleury_shear_ellipticity}, and traditional forward modeling tools also produce covariate posteriors in shear-ellipticity space.}

We show the calibration for individual model parameters in Figure \ref{fig:calibration_all}. In the fiducial case, we see that the network calibration is under-confident, which means that we overestimate the uncertainties. \review{In this case, for 71 lenses out of the whole sample,} we infer \review{NPE posterior distributions that are wider than the prior distribution}. We discard these ill-constrained samples before inferring the population-level parameters.

Next, we apply a neural network trained on lens light subtracted images, to the lens light subtracted dataset (described in Section \ref{data:3_preps}). We present the NPE posterior metrics in Table \ref{Tab:fiducial}. We find that lens light subtraction has little impact on recovery of $\theta_E$. We find that it worsens recovery of ellipticity parameters by a  factor of 2, compared to the case where all light components are included. We hypothesize that this is because the network has been trained under mass-traces-light framework, seen in the gray scatter in Figure \ref{fig:lens_light_impact}. Subtracting lens light removes this information. Finally, we see that with lens light subtraction, bias on $\rm \gamma_{lens}$ increases from the fiducial case \review{from 3\% to 4\%}. We discuss the reason for this in Section \ref{disc:gamma_lens_bias}.

In addition to evaluating performance on individual lens models, we assess recovery of the population model. As described in Section \ref{method:HBI}, we run a hierarchical inference to infer the population distribution over lens parameters. This results in a posterior over the hyperparameters, $\rm p(\nu|{d})$. Our inferred $\rm p(\nu|{d})$ for the fiducial test set is shown in Figure \ref{fig:HI_model}. \reviewt{The difference between the inferred hyperparameters and their ground truth values are shown in Table \ref{Tab:pop_model}}. Overall, we find that the inferred population-level model is within 1-$\sigma$ of the truth for population mean parameter $\mathcal{M}(\theta_E), \mathcal{M}(\gamma_{\rm lens}),\mathcal{M}(e_{1\&2}), \mathcal{M}(\gamma_{1\&2})$ as shown in Figure \ref{fig:HI_model}. \reviewt{Complete recovery of population mean and uncertainty is presented in Table \ref{Tab:pop_model_full}.} We are unable to learn intrinsic scatter for \review{$\rm \gamma_{lens}$ and shear parameters $\rm \gamma_{1/2}$, quantified by $\Sigma(\gamma_{\rm lens})$ and $\Sigma(\gamma_{1/2})$ respectively}. Under-confident NPE posteriors (seen in Figure \ref{fig:calibration_all}) yield a population model with low intrinsic scatter. We discuss this further in Section \ref{disc:hypermodel}.

We also tried modeling the posterior PDF for a given lens as diagonal multivariate Gaussian distribution. Given that many uncertainties are correlated, we find that this produces worse results (higher average error on parameters like \gammalens) consistently than using a full covariance matrix, as expected.


\subsection{Bright AGN Host Galaxies}\label{res:bhg}

We are interested in how the presence of a lensed arc impacts the NPE lens modeling performance, and subsequent population inference. To test this factor, we isolate the signal from the lensed arcs by removing both the lens light and the AGN light (more details about this preparation in Section \ref{data:3_preps}). In some systems, the AGN host galaxy is faint, and this leaves no signal above the noise floor. We only test on systems with a bright enough AGN host galaxy to produce arc light above the noise floor. We define signal as the fraction of pixels above the noise level: $S = (N > \sigma)/N_{tot} $ where $\sigma$ is the background noise level and N is the number of pixels, similar to the lensing information criterion introduced in \cite{Tan_2024}. Our selection criteria for bright hosts is images with $S > 0.5$, i.e more than half the pixels contain signal higher than noise -- this results in 757 systems. We re-train the network to operate on arc-light only, and present results on the NPE posteriors in Table \ref{Tab:optimize_data_and_method}. While average error and MAE for $\theta_E$ and $\rm \gamma_{lens}$ remain similar to the fiducial test, we see that systems with bright host galaxies provides better precision in all light included, and lens light subtracted cases. \review{If we subtract lens and AGN light, so that only the lensed host galaxy arcs remain in the image, we obtain the highest precision on $\theta_E$ and $\rm \gamma_{lens}$ (in comparison to all other experiments in this work, on LSST-quality imaging data, and using the fiducial test set).} This shows that most of the information on these parameters is retained in the arcs and points to the need for future modeling techniques that isolate the arc before modeling \review{ground based image data}.

We find that population recovery of $\mathcal{M}(\theta_E), \mathcal{M}(e_1)$ improve on the selection of bright host galaxies, as seen in Table \ref{Tab:pop_model}. 


\subsection{Image Deconvolution}\label{res:DECONV}
Here we explore how parameter recovery is affected by data quality, in particular whether the network performs better on deconvolved images. \review{We train and test on emulated deconvolved data, image preparation and deconvolution processed detailed in Section \ref{data:3_preps}.} As shown in Table \ref{Tab:optimize_data_and_method}, we find that this improves parameter recovery to 3\% precision per lens on \thetaE (compared to 6.5\% precision reported on simulated LSST imaging) and 5\% precision per lens on \gammalens (compared to 8\% on simulated LSST imaging). In the most optimal case, where we emulate deconvolution and isolate the arcs in the images via lens and AGN light subtraction, we obtain no bias and 4\% precision per lens on $\rm \gamma_{lens}$. Unlike $\theta_E$, that sees marginal improvement in parameter recovery between LSST and emulated deconvolved imaging, performance on $\rm \gamma_{lens}$ improves substantially. We find that calibration for all parameters is still under-confident but improved from the fiducial case, showing that better resolution also improves calibration. When only lens light is subtracted, we find that the average error on \gammalens is 0.05, which is still high. We discuss this in Section \ref{disc:lens_light_subtraction}.

Regarding the population parameter recovery, we find that the uncertainty on all inferred parameters, $\mathcal{M}(\xi)$ and $\Sigma(\xi)$, is lower. The emulated deconvolved data has better constraining power than simulated ground-based imaging. Despite lower uncertainties, we still see that bias on inferred $\mathcal{M}(\theta_E)$ is higher compared to the Fiducial case, as seen in Table \ref{Tab:pop_model}. We hypothesize that this could be because with higher data resolution, the inference is more sensitive to the mis-specified functional form. We present a more in-depth discussion of this in Sections \ref{disc:lensing_info} and  \ref{disc:hypermodel}.




\setlength{\tabcolsep}{2pt} 

\begin{deluxetable}{|c|c||c|c|c|}
\tablecaption{Optimizing for data contained in images: Individual lens mass model posterior results}
\tablehead{
\colhead{Experiment} &
\colhead{Preparation} & \colhead{Metric} & \colhead{$\theta_E$} & \colhead{$\gamma_{\rm lens}$} \\
\hline
\cline{3-5}
 &  &  & (arcsec) & 
}
\startdata
\multirow{6}{*}{Fiducial} & \multirow{3}{*}{ALL} & ME & -0.01 &  0.05 \\  
& & MAE & 0.02 & 0.10 \\
& & Precision & 0.05 & 0.16 \\
\cline{3-5}
& \multirow{3}{*}{LLS} & ME & -0.01 &  0.07 \\ 
& & MAE & 0.02 & 0.10 \\
& & Precision & 0.04 & 0.16 \\
\hline
\hline
\multicolumn{5}{c}{\textbf{Experiments optimizing the data}} \\
\hline
\hline
\multirow{9}{*}{Bright Host Galaxies} & \multirow{3}{*}{ALL} & ME & -0.00 &  0.04 \\  
& & MAE & 0.02 & 0.10 \\
& & Precision & 0.04 & 0.16 \\
\cline{3-5}
& \multirow{3}{*}{LLS} & ME & -0.01 &  0.06 \\  
& & MAE & 0.02 & 0.09 \\
& & Precision & 0.04 & 0.15 \\
\cline{3-5}
& \multirow{3}{*}{LALS} & ME & -0.01 &  0.06 \\  
& & MAE & 0.02 & 0.10 \\
& & Precision & 0.03 & 0.14 \\
\hline
\multirow{9}{*}{Deconvolved} & \multirow{3}{*}{ALL} & ME & 0.01 &  0.03 \\  
& & MAE & 0.01 & 0.07 \\
& & Precision & 0.02 & 0.12 \\
\cline{3-5}
& \multirow{3}{*}{LLS} & ME & -0.00 &  0.05 \\
& & MAE & 0.01 & 0.05 \\
& & Precision & 0.02 & 0.08 \\
\cline{3-5}
& \multirow{3}{*}{LALS}  & ME & -0.01 &  0.02 \\  
& & MAE & 0.01 & 0.03 \\
& & Precision & 0.01 & 0.05 \\
\hline
\hline
\multicolumn{5}{c}{\textbf{Experiments modifying the modeling method}} \\
\hline
\hline
\multirow{6}{*}{\shortstack{No mass light \\ correlations in \\training distribution}} & \multirow{3}{*}{ALL} & ME & -0.02 &  0.06 \\  
& & MAE & 0.03 & 0.10 \\
& & Precision & 0.05 & 0.16 \\
\cline{3-5}
& \multirow{3}{*}{LLS} & ME & -0.01 &  0.06 \\ 
& & MAE & 0.02 & 0.09 \\
& & Precision & 0.04 & 0.16 \\
\hline
\multirow{6}{*}{No Distribution Shift} & \multirow{3}{*}{ALL} & ME & 0.01 &  0.01 \\  
& & MAE & 0.04 & 0.11 \\
& & Precision & 0.06 & 0.16 \\
\cline{3-5}
& \multirow{3}{*}{LLS} & ME & -0.00 &  0.01 \\ 
& & MAE & 0.03 & 0.09 \\
& & Precision & 0.05 & 0.14 \\
\hline
\hline
\enddata
\tablecomments{Notes: Details about every experiment is available in Section \ref{sec:results}. Details above every preparation is available in Section \ref{data:3_preps}. ALL = All Light Included, 
LLS = Lens Light Subtracted, 
LALS = Lens and AGN Light Subtracted. Metrics are further defined in Section \ref{sec:results}.\label{Tab:optimize_data_and_method}}
\end{deluxetable}

\begin{deluxetable*}{|c|c||cccccccc|}
\tablecaption{Error in Population Model Parameter Recovery}
\tablehead{
\colhead{Experiment} &
\colhead{Preparation} &
\colhead{$\mathcal{M}({\theta_E})$} & 
\colhead{$\Sigma({\theta_E})$} & \colhead{$\mathcal{M}({\gamma_{\rm lens})}$} & 
\colhead{$\Sigma({\gamma_{\rm lens}})$} & 
\colhead{$\mathcal{M}(e_{1\&2})$} & 
\colhead{$\Sigma(e_{1\&2})$} & 
\colhead{$\mathcal{M}(\gamma_{1\&2})$} & 
\colhead{$\Sigma(\gamma_{1\&2})$}}
\startdata
Fiducial & ALL & \textbf{-0.004} & \textbf{-0.001} & 0.01 & -0.041 & -0.001 & -0.01 &\textbf{ 0.001 }& \textbf{-0.001} \\  
 & LLS & -0.01 & -0.002 & 0.10 & -0.004 & 0.01 & 0.001 & 0.02 & 0.03 \\
\hline
Bright Host Galaxies & ALL & 0.07 & 0.01 & -0.05 & -0.05 & \textbf{0} & -0.01 & -0.01 & -0.01 \\ 
 & LLS & 0.06 & 0.01 & 0.04 & -0.03 & 0.01 & \textbf{0} & 0.01 & 0.01 \\ 
 & LALS & 0.05 & -0.3 & 0.04 & -0.01 & 0.01 & 0.05 & 0.02 & -0.11 \\ 
\hline
Emulated Deconvolved & ALL & 0.01 & -0.27 & 0.02 & \textbf{0.002} & -0.002 & 0.07 & 0.01 & -0.11  \\ 
 & LLS & -0.002 & -0.29 & 0.04 & 0.02 & -0.001 & 0.09 & -0.01 & -0.103 \\ 
 & LALS &  -0.01 & -0.3 & \textbf{0.009} & 0.01 & -0.002 & 0.1 & -0.01 & -0.1\\ 
\hline
\hline
W/o mass/light correlations & ALL & -0.03 & -0.01 & 0.1 & -0.15 & 0.01 & 0.01 & 0.02 & 0.03  \\ 
& LLS & -0.01 & -0.02 & 0.08 & -0.02 & 0.01 & 0.002 & 0.01 & 0.02 \\ 
\hline
\enddata
\tablecomments{Table contains the difference between inferred population mean or scatter and fiducial test distribution mean and scatter. \label{Tab:pop_model}}
\end{deluxetable*}

\begin{figure*}
    \centering
    \includegraphics[width=0.85\textwidth]{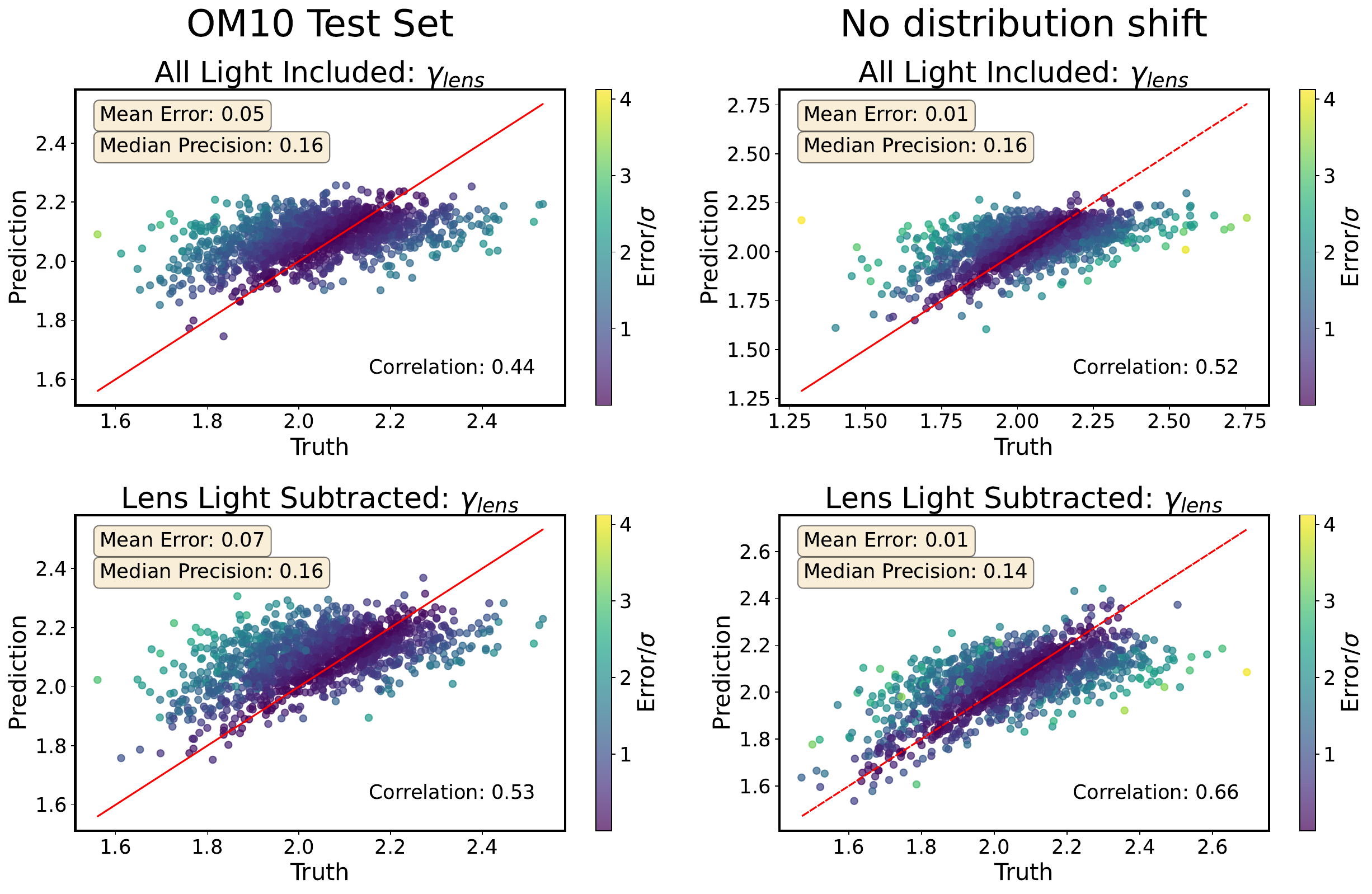}
    \caption{Top panels show recovery of $\gamma_{\rm lens}$ in presence and absence of distribution shift, when all light is included. We can see that the bias on $\gamma_{\rm lens}$ is significantly lower in the absence of distribution shift. The bottom left and right panels show the same when lens light is subtracted from the LSST quality images. We can see that in this case, not only is $\gamma_{\rm lens}$ inference less biased, it is also more accurate in the absence of distribution shift. Error is defined as predicted posterior mean - truth. The red line indicates perfect recovery.}
    \label{fig:nolens_valid_interim}
\end{figure*}

\subsection{Removing Mass-Light Correlations in the Training Data}\label{res:corr}

We hypothesize that one of the main factors affecting performance is the correlation of mass and light encoded into the training set. In our baseline set-up, the network is trained on lenses where ellipticity of the mass and light profiles are correlated (shown in Table \ref{Tab:train}). In this experiment, we regenerate training data without the ellipticity mass-light correlations, and re-train the NPE modeler. We show posteriors generated from applying this model to the fiducial dataset, where mass and light ellipticity are correlated, in Table \ref{Tab:optimize_data_and_method}. We see that bias on $\theta_E$ increases from 1\% to 2.5\%. As $\rm \gamma_{lens}$ is a difficult parameter to measure from images, we do not find any major impact on $\rm \gamma_{lens}$ from using a less informative training prior. However, we note that ellipticity bias and precision is impacted heavily, shown in Table \ref{Tab:optimize_data_and_method_full}. Ellipticity bias is doubled, and precision is halved. On the contrary, calibration on ellipticity parameters is improved, while other parameters have the same calibration as the fiducial case. We note that this degradation in NPE posterior performance is later folded into the hierarchical analysis of population-level parameters, and this affects $\rm \mathcal{M}(\gamma_{lens})$, $\Sigma(e_1)$ and $\Sigma(e_2)$ inference. We discuss the reason for this, and techniques to mitigate this effect in Section \ref{disc:gamma_lens_bias} and Section \ref{disc:hypermodel}. Overall, we find that performance is better when the network is trained with mass-light correlations.


\subsection{Removing distribution shift between test and training data}\label{res:dist_shift}
The network learns to produce lens model posteriors, given an image, based on what it sees from the training distribution. Any distribution shift between the test and training distribution can lead to biased inference \citep{dist_shift}. If the test set lies in an undersampled part of parameter space, posteriors produced will be less constraining. Fiducial results are produced from a test set shifted from the training distribution across multiple dimensions (test distribution shown in green and train distribution shown in gray in Figure \ref{fig:cPDF_fiducial}). To test if distribution shift is the driver of bias in our results, we apply the network on a test set that is sampled from the training distribution. This is referred to as a `held-out test set' going forward. Results are presented in Table \ref{Tab:optimize_data_and_method}. We show that there is still a residual bias, on parameters like $\theta_E$ and $\rm \gamma_{lens}$, and thus this is a function of misrepresented posterior PDF functional form or lack of expressivity of the network, and not just due to the distribution shift. We also find that the scatter of predictions around the true value is larger in the held-out test set. While the fiducial test has a complex selection function that makes all the lenses detectable and informative, the held-out test set lenses are completely randomly selected, and thus consist of lenses with lower lensing information. Further, the held-out test set contains some lenses that are in poorly sampled regions of the training data. This could add to the amplified MAE for all parameters seen in Table \ref{Tab:optimize_data_and_method}. Finally, we see that the calibration is perfect on all parameters in this experiment. This indicates that the under-confidence is largely a function of distribution shift, and it is insufficient to assess a network just using its calibration.

We do not propagate held-out test set posteriors into a hierarchical inference, since running the inference on a test set that matches the training set distribution would violate one of our key assumptions: that the test distribution is narrower than the training distribution (refer to Section \ref{method:HBI}).

\begin{figure*}
    \centering
    \includegraphics[width=1\textwidth]{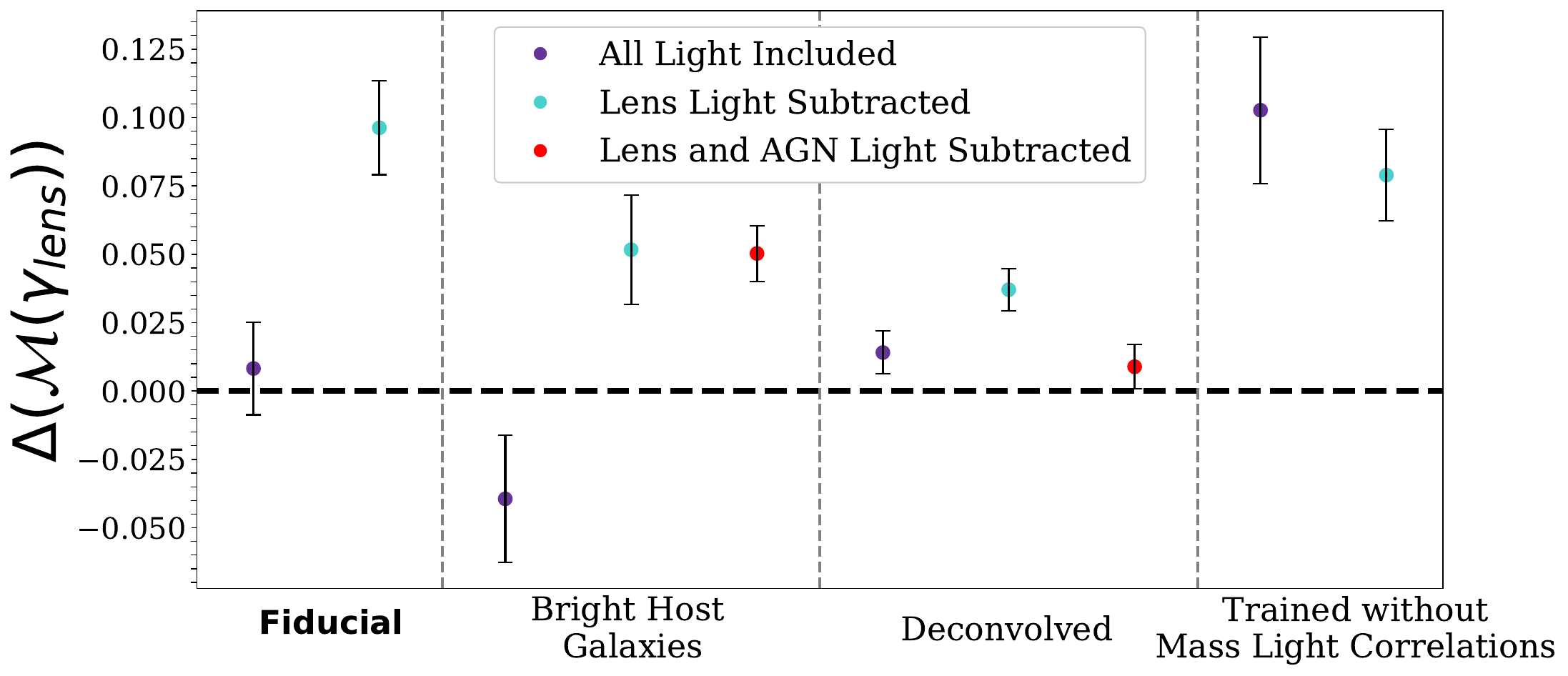}
    \caption{Recovery of $\mathcal{M}(\gamma_{\rm lens})$ under various experiments. Error bars indicate 2-$\sigma$ uncertainty on the hierarchically inferred population mean. Note that the fiducial preparation captures the truth within one $\sigma$, and is lower than the average error on the individual lens mass model posteriors reported in Table \ref{Tab:fiducial}.}
    \label{fig:pl_slope_pop}
\end{figure*}

\begin{figure*}
    \centering 
    \includegraphics[width=0.8\textwidth]{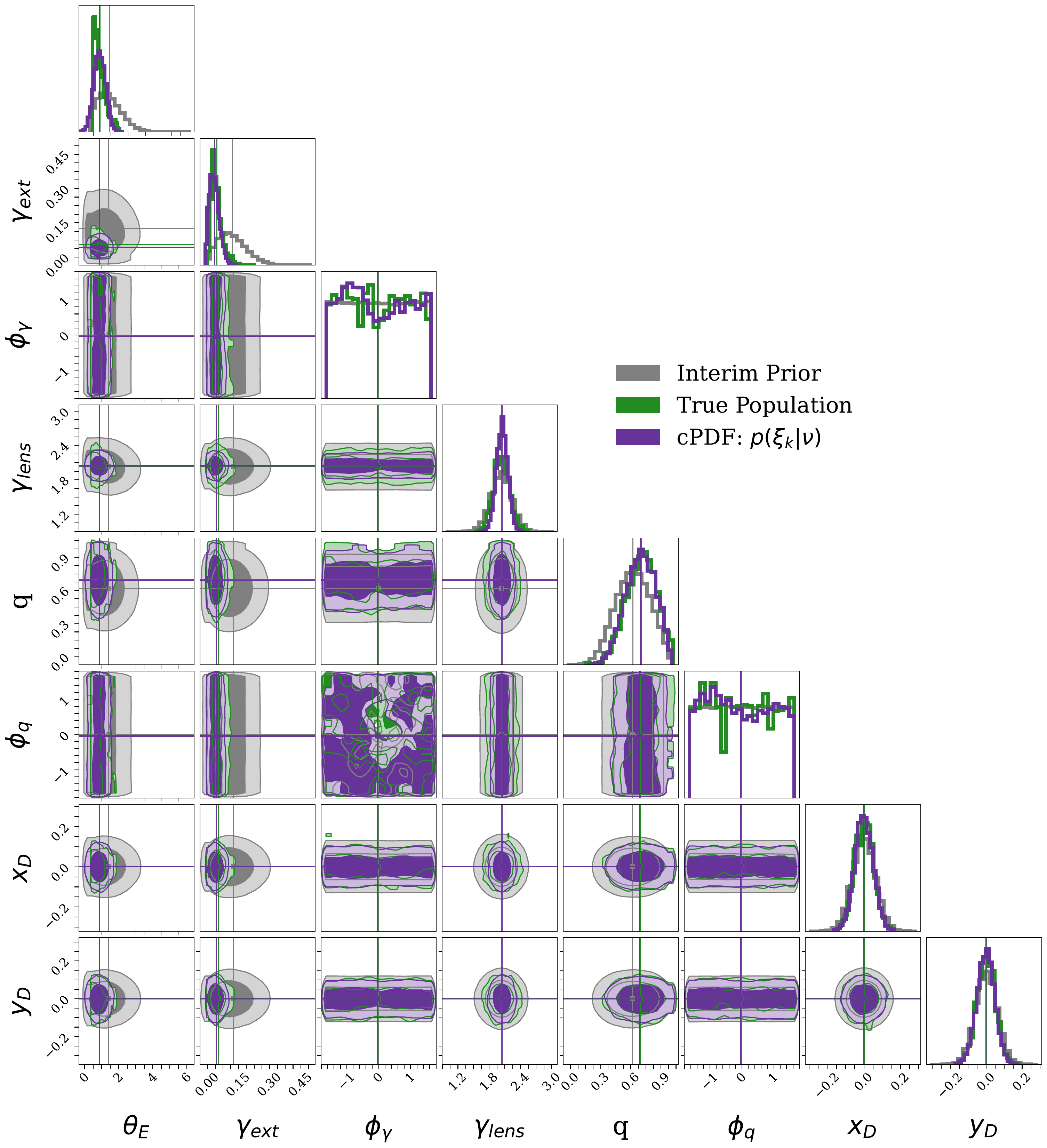}
    \caption{Here, we show the recovered conditional PDF $\rm p(\xi_k | \nu)$ where we infer $\nu$ hierarchically. We start by showing the network $5\times10^5$ combinations of lensing parameters sampled from the interim prior (gray contours) (Section \ref{method:train_data}). We apply the network to our realistic constructed sample of OM10 lenses (green contours) (Section \ref{data:om10_lens}) realized as LSST images [fiducial] (Section \ref{data:lsst}), infer a multivariate Gaussian posterior PDF of the lensing parameters from each image (Section \ref{method:NPE}). Then, we combine these posteriors in a hierarchical inference to learn the parameters characterizing the parent distribution they're sampled from (purple contours). We model the parent distribution as a \review{Gaussian with a diagonal covariance matrix} and thus infer the $M$ (mean) and $\Sigma$ (scatter) at the population-level (Section \ref{method:HBI}).}
    \label{fig:cPDF_fiducial}
\end{figure*}

\begin{figure*}
    \centering
    \includegraphics[width=0.8\textwidth]{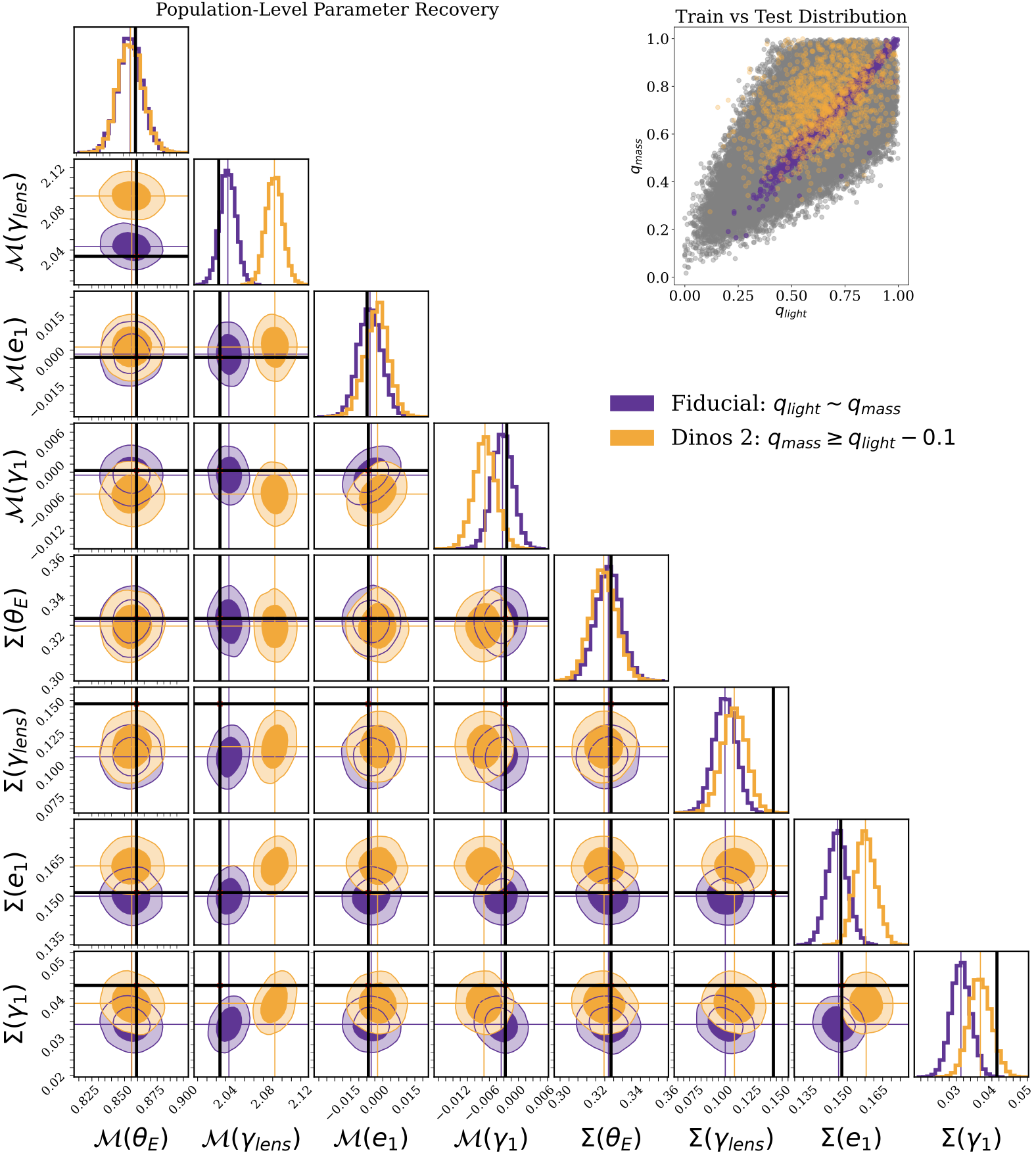}
    \caption{Recovery of population-level model parameters using 2 different test sets. In purple, we show our fiducial test set. In orange, we take the same test set and change the mass-light ellipticity relation to following $q_m \geq q_l - 0.1$ (taken from SLACS lenses \citep{bolton2008b, Tan_2024}). Inset panel: We show the two different mass-light ellipticity relations against the cloud of gray scatter points representing the mass-light ellipticity relationship in the training data. We demonstrate how sensitive the population-level parameter recovery is to such a shift, and suggest we need a more informative cPDF in order to work with such distribution shifts seen in real data.}
    \label{fig:lens_light_impact}
\end{figure*}

\section{Discussion}\label{sec:discussion}
After testing our automated modeling technique on a simulated LSST lensed AGN sample, we aim to understand our key performance drivers, and project how this technique will be adapted for future use.

\subsection{Impact of lens light subtraction on lens model recovery}\label{disc:lens_light_subtraction}
Lens light subtraction has been shown to benefit lens modeling generally \citep[as shown in e.g.][]{madireddy2019, lsst_euclid, Pearson_2021}, however, we do not find this to be true under all experiments. \reviewt{ We distill the impact of lens light subtraction to two factors: (i) encoding mass-light ellipticity correlations in the training data, and (ii) distribution shift in the latent parameters.}

\reviewt{Our fiducial experiment, in which mass and light ellipticity is correlated during training, is negatively impacted by lens light subtraction, as shown in Table \ref{Tab:optimize_data_and_method}. Under this experiment, the network uses light ellipticity to infer mass ellipticity, and removing this information has a negative impact.}

\reviewt{When we remove this mass-light ellipticity correlation in the training data and retrain, we find that lens light subtraction marginally improves the precision on $\theta_E$, $\rm e_{1/2}$ and $\rm \gamma_{1/2}$. It also marginally improves the scatter (MAE) on  $\theta_E$ and $\rm \gamma_{lens}$, as shown in Table \ref{Tab:optimize_data_and_method_full}.}

\reviewt{Regardless of the training data, we find that the bias on $\rm \gamma_{lens}$, in the presence of lens light subtraction, remains high. This is only decreased in the absence of any distribution shift between the train and test data, as seen in Figure \ref{fig:nolens_valid_interim}. This is discussed further in Section \ref{disc:gamma_lens_bias}.}

\subsection{$\gamma_{\rm lens}$ Bias}\label{disc:gamma_lens_bias}
We find consistent positive bias on NPE-modeled $\gamma_{\rm lens}$, on all experiments as seen in Table \ref{Tab:optimize_data_and_method}. This bias is propagated to \reviewt{population-level inference of mean mass density slope,} $\mathcal{M}(\gamma_{\rm lens})$, across all experiments as shown in Figure \ref{fig:pl_slope_pop}, and in Table \ref{Tab:pop_model}.  \reviewt{Measurements of the mass density slope from imaging data alone is subject to the mass sheet degeneracy (MSD). However, the appearance of lensed arcs from extended sources add to the constraining power of this parameter \citep{tdcosmo4, erickson2024}.}

We hypothesize several reasons for the $\rm \gamma_{lens}$ bias. 

\review{First, in the face of no data distribution shift, we recover $\gamma_{\rm lens}$ with 1\% bias on average, indicating that the distribution shape we infer could be changed.} Our network outputs a multivariate Gaussian PDF - a more malleable posterior functional form capable of capturing skewness and modality 
\review{might help in shifting the mean.}

We see that recovery of $\mathcal{M}(\gamma_{\rm lens})$ is influenced by the inferred scatter on ellipticity ($\Sigma(e_{1/2})$) and shear ($\Sigma{\gamma_{1/2}}$) as shown in Figure \ref{fig:HI_model}. $\gamma_{\rm lens}$ is degenerate with the axis ratio \reviewt{($q$)} and external shear \reviewt{($\rm \gamma_{ext}$) for a deflector galaxy}. The axis ratio $q$ is the polar characterization of Gaussian $e_1$ and $e_2$, thus it is a Rayleigh distribution characterized by $\Sigma(e_{1/2})$. In the fiducial case, an informative prior is \review{used} for these parameters, as we correlate lens mass and light ellipticity in the training. Even though we do not explicitly account for this in inferring the conditional PDF, learning ellipticity and shear more \reviewt{accurately} from each lens is captured in the full covariance matrix of the individual mass model posteriors that we feed to the HBI framework. \review{In Appendix \ref{app:gamma_lens_learn_all}, Figure \ref{fig:npe_to_post}, we show that there is an improvement in $\mathcal{M}(\gamma_{\rm lens})$ inference when we use all the parameter information from each lens.}
When lens light is subtracted, and lensed arcs and point source are visible more clearly, we identify that the individual mass model posterior results consistently show higher positive bias on $\gamma_{\rm lens}$ - both in case of LSST and emulated deconvolved data. \reviewt{We hypothesize that this is because the network confuses morphology of unresolved quasar images with highly magnified arcs resulting from a large extended AGN host galaxy being lensed by a deflector with a steep density profile.} \reviewt{Power-law slope inference is degenerate with source size and magnitude inference, \citep[effect shown in Appendix A of][]{marshall_2007}.} \review{We study this hypothesis by looking at the distribution shift in AGN host galaxy size between the training and test distribution:} In the fiducial case, the training set sees larger AGN host galaxies on average \review{($\rm R_{eff, src}$ peaking at 1.3 arcsec)}, and the test set has mostly smaller AGN host galaxies  \review{($\rm R_{eff, src}$ peaking at 0.7 arcsec)}. Thus, we hypothesize that the network believes that an AGN host galaxy, on average, must be bigger, and its compact appearance \review{in the test set images} is due to the deflector density slope. \reviewt{This is a manifestation of MSD.} To further test this, we study the lens light subtracted results in the \textit{absence} of train-test distribution shift across learned and latent parameters, including AGN galaxy host size (\reviewt{referred to as held-out test set} in Section \ref{res:dist_shift}). We show in Figure \ref{fig:nolens_valid_interim}, that \reviewt{the lens light subtracted held-out test set} yields better accuracy on $\rm \gamma_{lens}$. 

\reviewt{Finally, under the fiducial test set distribution shift, $\rm \gamma_{lens}$ is least biased when lens \textit{and} AGN light are subtracted. This reinforces that $\rm \gamma_{lens}$ information is inferred from appearance of the arcs, and that lens light subtraction alone is uniquely impacted by the distribution shift.}

From \citet{suyu_gamma, Treu_2016}, we know that mass distributions with steeper density profiles deflect light more and thus produce longer time delays, leading to measurements of shorter time-delay distances, and larger inferred values of \ho. Thus, our positive bias on the power-law density slope would translate directly to a positive bias on \ho.

\subsection{Realism of Training Data}\label{disc:training_data}

When we train on images with rough mass and light ellipticity correlation, and test on images that have the same relation, we see the best performance. Including loose correlations between mass and light in the training set should be done with caution. When we apply the network to test data that has any deviation from the correlation built into the training prior, we see that the calibration on ellipticity \review{becomes more under-confident}. This affects population-level parameter inference because we do not infer lens light parameters. In situations where we do infer the light as well, the inference pipeline will need to be provided with the information about the in-built correlation between mass and light in the modeling step.

We could also test this by changing mass-light alignment in the test set to follow SLACS lenses, where $\sigma_q = 0.12$ and $\sigma_\phi = 0.10^\circ$ \citep{bolton2008b}. This is adopted as prior in \cite{strides_schmidt} and \cite{Tan_2024} who present the following relationship for the axis ratio of mass and light: $q_m \geq q_l - 0.1$, a situation that could be produced in reality by a massive elliptical galaxy residing in a more spherical dark matter halo. We construct a test-set with the exact same properties as the original catalog presented in \ref{sec:data}, only changing the mass-light ellipticity relation to follow $q_m \geq q_l - 0.1$. \review{We apply the fiducial network with loose mass-light correlation to this modified-ellipticity test set. Combining individual lens models produced from this test set,} we present the population-level parameter inference in both test sets in Figure \ref{fig:lens_light_impact}. The subtle distribution shift between the training distribution mass-light ellipticity and the shifted mass-light ellipticity relation results in biased inference of $\Sigma(e_{1/2})$, impacting inference of $\mathcal{M}(\gamma_{\rm lens})$.


\subsection{Lensing Information in LSST Quality Images}\label{disc:lensing_info}
\review{We can extract the maximal lensing information by selecting for lenses with high SNR arcs, and/or by deconvolving the data.}

When we select high SNR arc lenses, we find that while precision on $\theta_E$ and $\gamma_{\rm lens}$ improves as described in Section \ref{res:bhg}, \review{this} imposes a stronger selection function (narrowing our sample to $\sim700$ lenses) and distribution shift that we do not fully recover from. Currently, adopting lens and AGN light subtraction could be a good way to extract signal from these high-SNR arcs. AGN light subtraction has not been attempted before, it could be possible with deconvolution.

\review{As for deconvolution, we find that, on average, parameter recovery is better from emulated deconvolution data (Section \ref{res:DECONV}).} However, we become much more sensitive to functional form mis-specifications, and our population level constraints, while much tighter, are slightly biased high on $\mathcal{M}(\gamma_{\rm lens})$.

\review{A caveat on the emulated deconvolved results:} We present results on the most optimistic threshold - we train and test on \textit{emulated} STARRED deconvolved data, and so there are no artifacts that confuse the network. If we actually apply STARRED to our simulated LSST \reviewt{coadd} images, we find that artifacts such as correlated noise can introduce a distribution shift in the simulation model and confuse the network. \reviewt{In practice, STARRED should be applied to a stack of exposures, which allows it to sample the noise better and prevent such correlated noise artifacts as detailed in Section \ref{data:3_preps}.}

\review{Finally, on the feasibility of deconvolution for training and testing data:} STARRED is currently optimized for light curve extraction of point sources. Extended sources (AGN host galaxies) are reconstructed on a higher resolution grid by combining dithered exposures and imposing a morphology prior through starlet regularization \citep{Starck2015}, which would not be feasible for generating STARRED-deconvolved training data. Other methods, such as Rubin SharPy (Kaehler et al. in prep) use a Recurrent Inference Machine (RIM) to deconvolve large datasets of images with extended sources at once. This could be a potential way to create a train and test dataset of deconvolved images.

\subsection{Timing}\label{disc:time}
A key benefit to machine learning \reviewt{based} lens modeling is its speed; here, we describe the timings of each piece of the lens modeling process. We generate $5\times10^5$ 33x33 pixel images for the training data which takes 3 hours on a CPU. We then generate 1344 images, one for each of the OM10 lens systems, which constitute the test data -- this takes 6 minutes on a CPU. Training for 200 epochs takes 1.5 hours on a GPU, and finally applying the trained network to the test set takes 5 minutes on a CPU. \review{Overall, for our sample of 1344 lenses, the modeling pipeline takes 12.5 seconds to model a single lens.} To \review{infer} the population model, the computational cost of MCMC was 1 - 2 CPU hours - primarily scaling with number of parameters inferred. For all experiments, we \review{infer the 16 hyperparameters} outlined in Section \ref{method:HBI}.

\subsection{Fiducial Test Set: Population Model Recovery}\label{disc:hypermodel}
There are two factors leading to the hypermodel recovery we see in Figure \ref{fig:HI_model} and this is attributed to the lack of information provided in construction of the conditional PDF (cPDF, shown in Figure \ref{fig:cPDF_fiducial}). Using an uninformative cPDF makes our results \reviewt{rely heavily on our data which is information-limited}.  The cPDF can be made more informative in two ways: (i) by building in the correlations between parameters that we expect to exist in real data, and using informative priors, and (ii) by allowing the cPDF to take on more flexible forms than a diagonal Gaussian that allow for correlations between parameters, and better characterize the distribution of parameters in the test set.

First, we use an uninformative uniform prior on the hypermodel parameters. The two parameters we are unable to learn in the hypermodel are $\Sigma(\gamma_{\rm lens})$ and $\Sigma(\gamma_{1/2})$. These are the two parameters that are hardest to \review{infer} with the data and most poorly constrained in the individual posterior level. We note that the recovery of external shear in individual mass model is still better than what has been recovered using galaxy-galaxy lenses in \cite{holismokes9} (Figure \ref{fig:all_interim_results}). We hypothesize that this is due to the additional image position information in the images from the bright background point source. However, the network is under-confident about these parameters, and the scatter of predictions around the truth is high. Thus, in the hypermodel inference, where we provide information-limited posteriors, and no additional information on existing correlations between parameters or tighter constraints on parameters that are hard to measure, we find that \textit{all the uncertainty is attributed to individual lenses} and the \textit{true scatter in the sample is underestimated}. On parameters like $\mathcal{M}(\theta_E)$ and $\mathcal{M}(e_{1/2})$, there is more information in the images, and our individual posteriors are better constrained. Thus, we find that despite $\theta_E$ following a skewed distribution\reviewt{,} we are able to accurately retrieve the mean and scatter of the underlying distribution for these parameters.

In future work, we can \review{use informative priors for the cPDF. We can use packages such as \texttt{SLSim} \citep{slsim} \reviewt{to} generate distributions with realistic parameter correlations, extending the sample with sources at $z>3.1$.} We can obtain informative priors on \thetaE from large galaxy-galaxy lens samples consisting of lenses with similar Einstein radii \citep{selection}. Also, we expect to be able to better measure parameters like \gammalens using the Gold sample and can use that to make a informative prior. The Gold sample can be used in general to obtain correlations between all lens mass model parameters. 

Second, our diagonal Gaussian cPDF does not represent the underlying test set as seen in Figure \ref{fig:cPDF_fiducial}. For example, $\theta_E$ is sampled from a non-Gaussian distribution in the fiducial test set, and in several empirical datasets \citep{selection}. The lensing selection results in non-Gaussianity in lens mass model parameters, due to observational effects (like projection effects that force more galaxies to look elliptical, and the seeing or PSF FWHM limiting the number of systems we can observe with $\rm \theta_{FWHM} > 2\times\theta_E$). Learning alternate functional forms, like a log-normal distribution instead of a diagonal Gaussian could be a good first step in this case.

\subsection{Applications of Silver Sample Beyond Cosmology}
We do not find any significant difference in accuracy of parameter recovery across the range of redshifts that we probe. This means that modeling LSST imaging will provide powerful insights into the size, evolution, and mass of galaxies up to redshift 2.



\subsection{Future Work}\label{disc:future}

In this section, we highlight the main work needed in the future of SBI-based lens modeling informed by our results and discussion. 

\textbf{Training Data:} There is limited information on certain parameters such as \gammalens available in ground-based images which makes the inference of these parameters prior-dominated. Although we can recover from ill-specified training priors on parameters like \thetaE, we do not see the same behavior on \gammalens. This means that we have to increase the realism of our training data in the following ways: include galaxy light-mass scaling relations, increase complexity of AGN host galaxy source structure to include features from disks, galaxy mergers etc. We also need to improve lens and environment model complexity and microlensing effects on AGN. \review{\texttt{SLSim} would be a good starting place to include these effects and simulate large sets of training data.}

\textbf{Normalizing flows:} We note that there is a small persistent bias in parameter recovery on the held-out test set. Thus, even when the training data are fully representative of the test set, we see that the approximate inference does not sufficiently capture the true posterior PDF. We suggest that using more flexible functional forms than multivariate Gaussian, and alternative methods to NPE (such as normalizing flows) to infer the individual mass model posteriors would be a useful next step, as demonstrated in \cite{poh_2025}.

\textbf{Multi-band imaging and modeling:} Multi-band imaging and modeling is required on many fronts. We can learn the AGN host galaxy structure better, we can model the deflector and subtract it out better and constrain the photometric redshifts of the systems better. From Holloway et al. 2025 (in prep), we can see that the presence of neighboring objects near the lens systems, such as nearby galaxies, degrade the precision of the predictions. This can also be mitigated by using multi-band imaging as shown in \cite{lsst_euclid}.

\section{Conclusions}\label{sec:conclusions}


We apply Neural Posterior Estimation (NPE) on simulated LSST imaging data of lensed AGN to construct lens mass models. We go on further to use hierarchical Bayesian inference (HBI) to learn the population from which our lenses were sampled. This work represents the first application of the NPE–HBI method to a large ground-based dataset of lensed AGN systems drawn from a realistically expected parameter distribution. We have demonstrated both the strengths and limitations of the approach, \review{with a specific focus on inferring population-level parameters, such as $\rm \mathcal{M}(\gamma_{lens})$, that could then affect cosmological parameter inference.} We show that there is useful information in the LSST lensed AGN sample for cosmology and outline clear next steps for further improving individual lens model parameter constraints and population-level parameter constraints.



Based on our investigations, we draw the following conclusions about the questions we have outlined in Section \ref{sec:intro}. All results quoted below are for the fiducial data (refer to Section \ref{res:LSST}). In the fiducial setup, we test on LSST lensed AGN constructed from the catalog detailed in Section \ref{sec:data}. We apply a network trained with mass-light ellipticity correlations matching the correlations in the test set.
\begin{enumerate}
    \item How accurately can we measure the key lens model parameters \thetaE (the Einstein Radius) and \gammalens (the density profile slope) from the LSST 5-year coadd imaging alone?
        \begin{itemize}
            \item We can recover \thetaE with less than 1\% bias on average, 6.5\% precision per lens and $\rm \gamma_{lens}$ with less than 3\% bias on average, 8\% precision per lens.
        \end{itemize}
    \item To what extent does this performance depend on the subtraction of the lens galaxy light, and the AGN point image light? How does deconvolution applied to the data prior to modeling improve the accuracy of measured lens models? 
    \begin{itemize}
        \item We find that lens light subtraction yields similar performance on \thetaE and increases bias on $\rm \gamma_{lens}$. We hypothesize this drop in performance is due to the network losing information about lens mass from lack of lens light, making the network more susceptible to distribution shifts.
        \item When subtracting both lens light and AGN light, leaving only the arc light from the AGN host galaxy, we see improved precision on \thetaE and $\rm \gamma_{lens}$. We hypothesize that most lensing information about \thetaE and $\rm \gamma_{lens}$ is inferred from the arcs.
        \item Emulated deconvolution improves the precision of inferred parameters by 30-60\%. It also lowers the bias on on parameters like $\rm \gamma_{lens}$. Thus we see that deconvolution prior to modeling will be beneficial in obtaining lens models from LSST images.
    \end{itemize}
    \item How do distribution shifts in learned and latent parameters impact the inference of lens mass model parameters?
    \begin{itemize}
        \item We find that removing distribution shift between the train and test data has a two-fold effect: (i) it reduces the bias on difficult to measure parameters like \gammalens (ii) the change in test distribution to a random sample of lenses, rather than the fiducial test set which is selected for detectability, reduces the average amount of information per lens, resulting in higher average uncertainty on parameters like \thetaE. These effects demonstrate the importance of testing under a realistic distributional shift.
    \end{itemize}
    \item How does the presence of mass-light ellipticity correlations in the training data impact the accuracy of inferred key lens model parameters, specifically \gammalens?
    \begin{itemize}
        \item In the fiducial case, we include mass-light ellipticity correlations in the training data and on average, the same mass-light ellipticity correlations in the test data. We find that removing this correlation from the training data biases \thetaE and \gammalens recovery, demonstrating the importance of a realistic prior on mass-light ellipticity.  
    \end{itemize}
    \item How accurately can we recover the population model for key lens mass model parameters?
    \begin{itemize}
        \item In the fiducial case, we \review{infer} the population level parameters for $\mathcal{M}(\theta_E)$ and $\Sigma(\theta_E)$ without bias. We \review{infer} $\rm \mathcal{M}(\gamma_{lens})$ with less than 1\% positive bias and $\rm \Sigma(\gamma_{lens})$ with 5\% negative bias. We find that distributional shifts in the mass/light ellipticity relation significantly impact the recovery of $\rm \mathcal{M}(\gamma_{lens})$. 
    \end{itemize}
\end{enumerate}

We demonstrate the feasibility of modeling LSST-quality images of lensed AGN with automated techniques, incorporating a thorough investigation of data preparations and distributional shifts. This work builds a foundation for utilizing all LSST imaging of lensed AGN for a dark energy measurement from time-delay cosmography.




\software{lenstronomy, paltas, emcee, corner \review{\citep{corner}}, matplotlib\review{\citep{matplotlib}}, numpy\review{\citep{numpy}}, pandas \review{\citep{pandas_pap, pandas_soft}}.}

\section*{Acknowledgments}
This paper has undergone internal review in the LSST Dark Energy Science Collaboration. 
P.V would like to thank internal reviewers Sreevani Jaragula and Ayan Mitra for their feedback. We additionally thank Anowar Shajib, Risa Wechsler, Sebastian Wagner-Carena, Gautham Narayan, Decker French, Margaret Verrico and the members of the DESC Time Domain Working Group and LSST Strong Lensing Science Collaboration for useful discussions.

P.V developed datasets, built on existing tools \texttt{paltas} and \texttt{lens-npe} to scale for large datasets, ran all analysis, produced all figures and wrote the main body of the text. S.E provided invaluable feedback on all components of this work: dataset development, analysis and writing and developed tools used for HBI analysis, and post-processing of NPE analysis. P.M helped define the lensed AGN sample, design the various experiments, design statistical framework to analyze results and provided invaluable feedback on all aspects. M.M provided interpretation and context for results and provided feedback on all aspects of analysis. P.H helped benchmark results on this dataset against other datasets, advised on implementation of tools for large-scale modeling, provided feedback on all components of analysis and writeup. S.B provided feedback on analysis choices, presentation of results, framing of methods and writeup. X.H provided feedback on analysis and writeup. N.K provided feedback on lens sample used for testing, simulation of training and test images. G.M provided input on initial construction of catalog, provided feedback on broad goals and scientific context of the project. A.R provided broad feedback on all components of this work. R.K provided data products used for analysis. S.D provided context for calibration of network, useful edits to writeup. A.M and S.J were internal reviewers for this work and provided valuable feedback on all components of the writeup. K.R’s work as a Rubin builder contributed to this work, especially through the realistic simulations we employ.


The work of PV was supported by KIPAC Post Baccalaureate Fellowship. SE acknowledges funding from NSF GRFP-2021313357 and the Stanford Data Science Scholars Program. MM acknowledges support by the SNSF (Swiss National Science Foundation) through mobility grant P500PT\_203114 and return CH grant P5R5PT\_225598. SB is supported by DoE Grant DE-SC0026113. PH acknowledges funding from the Science and Technology Facilities Council, Grant Code ST/W507726/1. This work was performed in part under DOE Contract DE-AC02-76SF00515.
The DESC acknowledges ongoing support from the Institut National de  Physique Nucl\'eaire et de Physique des Particules in France; the  Science \& Technology Facilities Council in the United Kingdom; and the Department of Energy, the National Science Foundation, and the LSST Discovery Alliance in the United States.  DESC uses resources of the IN2P3 Computing Center (CC-IN2P3--Lyon/Villeurbanne - France) funded by the  Centre National de la Recherche Scientifique; the National Energy 
Research Scientific Computing Center, a DOE Office of Science User 
Facility supported by the Office of Science of the U.S.\ Department of
Energy under Contract No.\ DE-AC02-05CH11231; STFC DiRAC HPC Facilities, 
funded by UK BEIS National E-infrastructure capital grants; and the UK 
particle physics grid, supported by the GridPP Collaboration.

\section*{Data and Code Availability}
Code developed for this work is publicly available in the  \texttt{lsst\_lensed\_agn\_modeling} \footnote{\href{https://github.com/padma18-vb/lsst_lensed_agn_modeling}{GitHub:https://github.com/padma18-vb/lsst\_lensed\_agn\_modeling}} repository. This research made use of \texttt{lenstronomy}, a multi-purpose gravitational lens modeling software package \citep{Birrer2021, Birrer_2018}. This work uses public software package \texttt{paltas}, a NPE tool for large-scale strong lens modeling \citep{paltas}. We used hierarchical inference scripts available in \texttt{lens-npe} \citep{lens-npe}. \review{Catalog-level data used in the fiducial set-up of this work, the ground truth catalogs, MCMC chains, network posteriors for all experiments are available on Zenodo \citep{venkatraman_zenodo}. Scripts to generate the imaging data using our training prior, or for the test catalog is available on GitHub.}



\newpage
\bibliography{main}{}

@string{apjl="ApJL"}

@string{apj="ApJ"}

@string{aj ="AJ"}

@string{mnras ="MNRAS"}

@string{aap="A\&A"}

@string{jcap="Journal of Comsology and Astroparticle Physics"}

@string{nature="Nature"}

@string{pasp="PASP"}

@article{corner,
      doi = {10.21105/joss.00024},
      url = {https://doi.org/10.21105/joss.00024},
      year  = {2016},
      publisher = {The Open Journal},
      volume = {1},
      number = {2},
      pages = {24},
      author = {Daniel Foreman-Mackey},
      title = {corner.py: Scatterplot matrices in Python},
      journal = {The Journal of Open Source Software}
    }

@ARTICLE{planck2020,
       author = {{Planck Collaboration} and {Aghanim}, N. and {Akrami}, Y. and {Ashdown}, M. and {Aumont}, J. and {Baccigalupi}, C. and {Ballardini}, M. and {Banday}, A.~J. and {Barreiro}, R.~B. and {Bartolo}, N. and {Basak}, S. and {Battye}, R. and {Benabed}, K. and {Bernard}, J.-P. and {Bersanelli}, M. and {Bielewicz}, P. and {Bock}, J.~J. and {Bond}, J.~R. and {Borrill}, J. and {Bouchet}, F.~R. and {Boulanger}, F. and {Bucher}, M. and {Burigana}, C. and {Butler}, R.~C. and {Calabrese}, E. and {Cardoso}, J.-F. and {Carron}, J. and {Challinor}, A. and {Chiang}, H.~C. and {Chluba}, J. and {Colombo}, L.~P.~L. and {Combet}, C. and {Contreras}, D. and {Crill}, B.~P. and {Cuttaia}, F. and {de Bernardis}, P. and {de Zotti}, G. and {Delabrouille}, J. and {Delouis}, J.-M. and {Di Valentino}, E. and {Diego}, J.~M. and {Dor{\'e}}, O. and {Douspis}, M. and {Ducout}, A. and {Dupac}, X. and {Dusini}, S. and {Efstathiou}, G. and {Elsner}, F. and {En{\ss}lin}, T.~A. and {Eriksen}, H.~K. and {Fantaye}, Y. and {Farhang}, M. and {Fergusson}, J. and {Fernandez-Cobos}, R. and {Finelli}, F. and {Forastieri}, F. and {Frailis}, M. and {Fraisse}, A.~A. and {Franceschi}, E. and {Frolov}, A. and {Galeotta}, S. and {Galli}, S. and {Ganga}, K. and {G{\'e}nova-Santos}, R.~T. and {Gerbino}, M. and {Ghosh}, T. and {Gonz{\'a}lez-Nuevo}, J. and {G{\'o}rski}, K.~M. and {Gratton}, S. and {Gruppuso}, A. and {Gudmundsson}, J.~E. and {Hamann}, J. and {Handley}, W. and {Hansen}, F.~K. and {Herranz}, D. and {Hildebrandt}, S.~R. and {Hivon}, E. and {Huang}, Z. and {Jaffe}, A.~H. and {Jones}, W.~C. and {Karakci}, A. and {Keih{\"a}nen}, E. and {Keskitalo}, R. and {Kiiveri}, K. and {Kim}, J. and {Kisner}, T.~S. and {Knox}, L. and {Krachmalnicoff}, N. and {Kunz}, M. and {Kurki-Suonio}, H. and {Lagache}, G. and {Lamarre}, J.-M. and {Lasenby}, A. and {Lattanzi}, M. and {Lawrence}, C.~R. and {Le Jeune}, M. and {Lemos}, P. and {Lesgourgues}, J. and {Levrier}, F. and {Lewis}, A. and {Liguori}, M. and {Lilje}, P.~B. and {Lilley}, M. and {Lindholm}, V. and {L{\'o}pez-Caniego}, M. and {Lubin}, P.~M. and {Ma}, Y.-Z. and {Mac{\'\i}as-P{\'e}rez}, J.~F. and {Maggio}, G. and {Maino}, D. and {Mandolesi}, N. and {Mangilli}, A. and {Marcos-Caballero}, A. and {Maris}, M. and {Martin}, P.~G. and {Martinelli}, M. and {Mart{\'\i}nez-Gonz{\'a}lez}, E. and {Matarrese}, S. and {Mauri}, N. and {McEwen}, J.~D. and {Meinhold}, P.~R. and {Melchiorri}, A. and {Mennella}, A. and {Migliaccio}, M. and {Millea}, M. and {Mitra}, S. and {Miville-Desch{\^e}nes}, M.-A. and {Molinari}, D. and {Montier}, L. and {Morgante}, G. and {Moss}, A. and {Natoli}, P. and {N{\o}rgaard-Nielsen}, H.~U. and {Pagano}, L. and {Paoletti}, D. and {Partridge}, B. and {Patanchon}, G. and {Peiris}, H.~V. and {Perrotta}, F. and {Pettorino}, V. and {Piacentini}, F. and {Polastri}, L. and {Polenta}, G. and {Puget}, J.-L. and {Rachen}, J.~P. and {Reinecke}, M. and {Remazeilles}, M. and {Renzi}, A. and {Rocha}, G. and {Rosset}, C. and {Roudier}, G. and {Rubi{\~n}o-Mart{\'\i}n}, J.~A. and {Ruiz-Granados}, B. and {Salvati}, L. and {Sandri}, M. and {Savelainen}, M. and {Scott}, D. and {Shellard}, E.~P.~S. and {Sirignano}, C. and {Sirri}, G. and {Spencer}, L.~D. and {Sunyaev}, R. and {Suur-Uski}, A.-S. and {Tauber}, J.~A. and {Tavagnacco}, D. and {Tenti}, M. and {Toffolatti}, L. and {Tomasi}, M. and {Trombetti}, T. and {Valenziano}, L. and {Valiviita}, J. and {Van Tent}, B. and {Vibert}, L. and {Vielva}, P. and {Villa}, F. and {Vittorio}, N. and {Wandelt}, B.~D. and {Wehus}, I.~K. and {White}, M. and {White}, S.~D.~M. and {Zacchei}, A. and {Zonca}, A.},
        title = "{Planck 2018 results. VI. Cosmological parameters}",
      journal = {\aap},
         year = 2020,
        month = sep,
       volume = {641},
          eid = {A6},
        pages = {A6},
          doi = {10.1051/0004-6361/201833910},
archivePrefix = {arXiv},
       eprint = {1807.06209},
 primaryClass = {astro-ph.CO},
       adsurl = {https://ui.adsabs.harvard.edu/abs/2020A&A...641A...6P}
}

@ARTICLE{des_combine_probes,
       author = {{DES Collaboration} and {Abbott}, T.~M.~C. and {Adamow}, M. and {Aguena}, M. and {Alarcon}, A. and {Allam}, S. and {Alves}, O. and {Amon}, A. and {Anbajagane}, D. and {Andrade-Oliveira}, F. and {Armstrong}, P. and {Avila}, S. and {Beas-Gonzalez}, J. and {Bechtol}, K. and {Becker}, M.~R. and {Bernstein}, G.~M. and {Bertin}, E. and {Blazek}, J. and {Bocquet}, S. and {Brooks}, D. and {Brout}, D. and {Burke}, D.~L. and {Camacho}, H. and {Camacho-Ciurana}, G. and {Camilleri}, R. and {Campailla}, G. and {Campos}, A. and {Carnero Rosell}, A. and {Carr}, A. and {Carretero}, J. and {Castander}, F.~J. and {Cawthon}, R. and {Chan}, K.~C. and {Chang}, C. and {Chen}, R. and {Coloma-Nadal}, J.~M. and {Conselice}, C. and {Costanzi}, M. and {Crocce}, M. and {d'Assignies}, W. and {da Costa}, L.~N. and {da Silva Pereira}, M.~E. and {Davis}, T.~M. and {De Vicente}, J. and {DePoy}, D.~L. and {DeRose}, J. and {Desai}, S. and {Diehl}, H.~T. and {Dodelson}, S. and {Doel}, P. and {Doux}, C. and {Drlica-Wagner}, A. and {Eifler}, T.~F. and {Elvin-Poole}, J. and {Everett}, S. and {Evrard}, A.~E. and {Ferrero}, I. and {Fert{\'e}}, A. and {Flaugher}, B. and {Fosalba}, P. and {de Souza}, D. Francis and {Frieman}, J. and {Galbany}, L. and {Garc{\'\i}a-Bellido}, J. and {Gatti}, M. and {Giannini}, G. and {Giles}, P. and {Glazebrook}, K. and {Gruen}, D. and {Gruendl}, R.~A. and {Gutierrez}, G. and {Harrison}, I. and {Hartley}, W.~G. and {Herner}, K. and {Hinton}, S.~R. and {Hollowood}, D.~L. and {Honscheid}, K. and {Huff}, E.~M. and {Huterer}, D. and {Jain}, B. and {James}, D.~J. and {Jarvis}, M. and {Jeffrey}, N. and {Jeltema}, T. and {Kent}, S. and {Kessler}, R. and {Kovacs}, A. and {Koyama}, K. and {Krause}, E. and {Kron}, R. and {Kuehn}, K. and {Lahav}, O. and {Lee}, J. and {Lee}, S. and {Legnani}, E. and {Li}, T.~S. and {Liddle}, A.~R. and {Lidman}, C. and {Lin}, H. and {Lin}, M. and {MacCrann}, N. and {Marshall}, J.~L. and {Mau}, S. and {McMahon}, R.~G. and {Mena-Fern{\'a}ndez}, J. and {Menanteau}, F. and {Miquel}, R. and {Mohr}, J.~J. and {Muir}, J. and {Myles}, J. and {M{\"o}ller}, A. and {Nichol}, R.~C. and {Ogando}, R.~L.~C. and {Percival}, W.~J. and {Petravick}, D. and {Pieres}, A. and {Plazas Malag{\'o}n}, A.~A. and {Popovic}, B. and {Porredon}, A. and {Prat}, J. and {Qu}, H. and {Raveri}, M. and {Rebou{\c{c}}as}, J. and {Riquelme}, W. and {Rodriguez-Monroy}, M. and {Rogozenski}, P. and {Romer}, A.~K. and {Roodman}, A. and {Rosenfeld}, R. and {Ross}, A.~J. and {Rykoff}, E.~S. and {Sako}, M. and {Samuroff}, S. and {S{\'a}nchez}, C. and {Sanchez}, E. and {Sanchez Cid}, D. and {Schutt}, T. and {Scolnic}, D. and {Sevilla-Noarbe}, I. and {Shah}, N. and {Shah}, P. and {Sheldon}, E. and {Smith}, M. and {Soares-Santos}, M. and {Suchyta}, E. and {Sullivan}, M. and {Swanson}, M.~E.~C. and {S{\'a}nchez}, B.~O. and {Tabbutt}, M. and {Tarle}, G. and {Taylor}, G. and {Thomas}, D. and {To}, C. and {Toribio San Cipriano}, L. and {Toy}, M. and {Troxel}, M.~A. and {Tucker}, D.~L. and {Vikram}, V. and {Vincenzi}, M. and {Weaverdyck}, N. and {Weller}, J. and {Whyley}, A. and {Wilkinson}, R.~D. and {Wiseman}, P. and {Yamamoto}, M. and {Yanny}, B. and {Yin}, B. and {Zhang}, Y. and {Zuntz}, J.},
        title = "{Constraints on Dynamical Dark Energy from Multiple Probes in the Full Dark Energy Survey}",
      journal = {arXiv e-prints},
         year = 2026,
        month = may,
          eid = {arXiv:2605.27221},
        pages = {arXiv:2605.27221},
          doi = {10.48550/arXiv.2605.27221},
archivePrefix = {arXiv},
       eprint = {2605.27221},
 primaryClass = {astro-ph.CO},
       adsurl = {https://ui.adsabs.harvard.edu/abs/2026arXiv260527221D}
}

@Article{matplotlib,
  Author    = {Hunter, J. D.},
  Title     = {Matplotlib: A 2D graphics environment},
  Journal   = {Computing in Science \& Engineering},
  Volume    = {9},
  Number    = {3},
  Pages     = {90--95},
  publisher = {IEEE COMPUTER SOC},
  doi       = {10.1109/MCSE.2007.55},
  year      = 2007
}

@Article{numpy,
 title         = {Array programming with {NumPy}},
 author        = {Charles R. Harris and K. Jarrod Millman and St{\'{e}}fan J.
                 van der Walt and Ralf Gommers and Pauli Virtanen and David
                 Cournapeau and Eric Wieser and Julian Taylor and Sebastian
                 Berg and Nathaniel J. Smith and Robert Kern and Matti Picus
                 and Stephan Hoyer and Marten H. van Kerkwijk and Matthew
                 Brett and Allan Haldane and Jaime Fern{\'{a}}ndez del
                 R{\'{i}}o and Mark Wiebe and Pearu Peterson and Pierre
                 G{\'{e}}rard-Marchant and Kevin Sheppard and Tyler Reddy and
                 Warren Weckesser and Hameer Abbasi and Christoph Gohlke and
                 Travis E. Oliphant},
 year          = {2020},
 month         = sep,
 journal       = {Nature},
 volume        = {585},
 number        = {7825},
 pages         = {357--362},
 doi           = {10.1038/s41586-020-2649-2},
 publisher     = {Springer Science and Business Media {LLC}},
 url           = {https://doi.org/10.1038/s41586-020-2649-2}
}

@software{pandas_soft,
    author       = {The pandas development team},
    title        = {pandas-dev/pandas: Pandas},
    month        = feb,
    year         = 2020,
    publisher    = {Zenodo},
    version      = {latest},
    doi          = {10.5281/zenodo.3509134},
    url          = {https://doi.org/10.5281/zenodo.3509134}
}

@InProceedings{pandas_pap,
  author    = { {W}es {M}c{K}inney },
  title     = { {D}ata {S}tructures for {S}tatistical {C}omputing in {P}ython },
  booktitle = { {P}roceedings of the 9th {P}ython in {S}cience {C}onference },
  pages     = { 56 - 61 },
  year      = { 2010 },
  editor    = { {S}t\'efan van der {W}alt and {J}arrod {M}illman },
  doi       = { 10.25080/Majora-92bf1922-00a }
}

@ARTICLE{emcee,
       author = {{Foreman-Mackey}, Daniel and {Hogg}, David W. and {Lang}, Dustin and {Goodman}, Jonathan},
        title = "{emcee: The MCMC Hammer}",
      journal = {\pasp},
         year = 2013,
        month = mar,
       volume = {125},
       number = {925},
        pages = {306},
          doi = {10.1086/670067},
archivePrefix = {arXiv},
       eprint = {1202.3665},
 primaryClass = {astro-ph.IM},
       adsurl = {https://ui.adsabs.harvard.edu/abs/2013PASP..125..306F}
}

@ARTICLE{slsim,
       author = {{Khadka}, Narayan and {Birrer}, Simon and {Best}, Henry and {Sharma}, Paras and {Abe}, Katsuya T. and {Tang}, Xianzhe and {Mistick}, Carly and {Urcelay}, Felipe and {Sonmez}, Emrecan M. and {Arendse}, Nikki and {Erickson}, Sydney and {Hjortlund}, Jacob O. and {Holloway}, Phil and {Huang}, Alan and {Karthik}, Rahul and {Lamontagne}, Mia and {Negi}, Vibhore and {Pierel}, Justin R. and {Sanchez}, Bruno and {Ece Saricaoglu}, Aysu and {Shajib}, Anowar and {Shao}, Yixuan and {Venkatraman}, Padma and {Wedig}, Bryce and {Agrawal}, Aadya and {Anguita}, Timo and {Bessa}, Pedro and {Bom}, Clecio R. and {Castillo}, Sofia and {Collett}, Thomas and {Daylan}, Tansu and {Dillmann}, Steven and {Grespan}, Margherita and {Hayes}, Erin E. and {Joseph}, Remy and {Kessler}, Richard and {Li}, Tian and {Marshall}, Phil and {More}, Anupreeta and {Motta}, Veronica and {Narayan}, Gautham and {O'Dowd}, Matt and {Oguri}, Masamune and {Smith}, Graham and {Verma}, Aprajita and {Vernardos}, Giorgos and {the Strong Lensing Science Collaboration} and {the LSST Dark Energy Science Collaboration}},
        title = "{SLSim: a strong lensing population simulation package}",
      journal = {arXiv e-prints},
         year = 2026,
        month = mar,
          eid = {arXiv:2603.17138},
        pages = {arXiv:2603.17138},
          doi = {10.48550/arXiv.2603.17138},
archivePrefix = {arXiv},
       eprint = {2603.17138},
 primaryClass = {astro-ph.CO},
       adsurl = {https://ui.adsabs.harvard.edu/abs/2026arXiv260317138K}
}

@dataset{venkatraman_zenodo,
  author       = {Venkatraman, Padmavathi},
  title        = {Catalog information, model predictions,
                   hierarchical inference chains for 'Lens Modeling
                   Accuracy in Expected LSST Lensed AGN Sample'
                  },
  month        = oct,
  year         = 2025,
  publisher    = {Zenodo},
  doi          = {10.5281/zenodo.17412880},
  url          = {https://doi.org/10.5281/zenodo.17412880},
}

@ARTICLE{investigating_de_constraint,
       author = {{Erickson}, Sydney and {Millon}, Martin and {Venkatraman}, Padmavathi and {Li}, Tian and {Holloway}, Philip and {Marshall}, Phil and {Shajib}, Anowar and {Birrer}, Simon and {Huang}, Xiang-Yu and {Anguita}, Timo and {Dillmann}, Steven and {Khadka}, Narayan and {Napier}, Kate and {Roodman}, Aaron and {The LSST Dark Energy Science Collaboration}},
        title = "{Investigating the Dark Energy Constraint from Strongly Lensed AGN at LSST-Scale}",
      journal = {arXiv e-prints},
         year = 2025,
        month = nov,
          eid = {arXiv:2511.13669},
        pages = {arXiv:2511.13669},
          doi = {10.48550/arXiv.2511.13669},
archivePrefix = {arXiv},
       eprint = {2511.13669},
 primaryClass = {astro-ph.CO},
       adsurl = {https://ui.adsabs.harvard.edu/abs/2025arXiv251113669E}
}

@misc{lens-npe,
  author       = {Erickson, Sydney},
  title        = {Model Weights, Predictions, and Chains: Lens
                   Modeling of STRIDES Strongly Lensed Quasars with
                   NPE
                  },
  month        = oct,
  year         = 2024,
  publisher    = {Zenodo},
  doi          = {10.5281/zenodo.13906030},
  url          = {https://doi.org/10.5281/zenodo.13906030},
}

@ARTICLE{fleury_shear_ellipticity,
       author = {{Fleury}, Pierre and {Larena}, Julien and {Uzan}, Jean-Philippe},
        title = "{Line-of-sight effects in strong gravitational lensing}",
      journal = {\jcap},
         year = 2021,
        month = aug,
       volume = {2021},
       number = {8},
          eid = {024},
        pages = {024},
          doi = {10.1088/1475-7516/2021/08/024},
archivePrefix = {arXiv},
       eprint = {2104.08883},
 primaryClass = {astro-ph.CO},
       adsurl = {https://ui.adsabs.harvard.edu/abs/2021JCAP...08..024F}
}

@article{shear_ellipticity,
   title={Foreground biases in strong gravitational lensing},
   volume={2024},
   ISSN={1475-7516},
   url={http://dx.doi.org/10.1088/1475-7516/2024/10/055},
   DOI={10.1088/1475-7516/2024/10/055},
   number={10},
   journal={Journal of Cosmology and Astroparticle Physics},
   publisher={IOP Publishing},
   author={Johnson, Daniel and Fleury, Pierre and Larena, Julien and Marchetti, Lucia},
   year={2024},
   month=oct, pages={055} }

@ARTICLE{tdcosmo_2025,
       author = {{TDCOSMO Collaboration} and {Birrer}, Simon and {Buckley-Geer}, Elizabeth J. and {Cappellari}, Michele and {Courbin}, Fr{\'e}d{\'e}ric and {Dux}, Fr{\'e}d{\'e}ric and {Fassnacht}, Christopher D. and {Frieman}, Joshua A. and {Galan}, Aymeric and {Gilman}, Daniel and {Huang}, Xiang-Yu and {Knabel}, Shawn and {Langeroodi}, Danial and {Lin}, Huan and {Millon}, Martin and {Morishita}, Takahiro and {Motta}, Veronica and {Mozumdar}, Pritom and {Paic}, Eric and {Shajib}, Anowar J. and {Sheu}, William and {Sluse}, Dominique and {Sonnenfeld}, Alessandro and {Spiniello}, Chiara and {Stiavelli}, Massimo and {Suyu}, Sherry H. and {Tan}, Chin Yi and {Treu}, Tommaso and {Van de Vyvere}, Lyne and {Wang}, Han and {Wells}, Patrick and {Williams}, Devon M. and {Wong}, Kenneth C.},
        title = "{TDCOSMO 2025: Cosmological constraints from strong lensing time delays}",
      journal = {arXiv e-prints},
         year = 2025,
        month = jun,
          eid = {arXiv:2506.03023},
        pages = {arXiv:2506.03023},
          doi = {10.48550/arXiv.2506.03023},
archivePrefix = {arXiv},
       eprint = {2506.03023},
 primaryClass = {astro-ph.CO},
       adsurl = {https://ui.adsabs.harvard.edu/abs/2025arXiv250603023T}
}

@ARTICLE{marshall_2007,
       author = {{Marshall}, Philip J. and {Treu}, Tommaso and {Melbourne}, Jason and {Gavazzi}, Rapha{\"e}l and {Bundy}, Kevin and {Ammons}, S. Mark and {Bolton}, Adam S. and {Burles}, Scott and {Larkin}, James E. and {Le Mignant}, David and {Koo}, David C. and {Koopmans}, L{\'e}on V.~E. and {Max}, Claire E. and {Moustakas}, Leonidas A. and {Steinbring}, Eric and {Wright}, Shelley A.},
        title = "{Superresolving Distant Galaxies with Gravitational Telescopes: Keck Laser Guide Star Adaptive Optics and Hubble Space Telescope Imaging of the Lens System SDSS J0737+3216}",
      journal = {\apj},
         year = 2007,
        month = dec,
       volume = {671},
       number = {2},
        pages = {1196-1211},
          doi = {10.1086/523091},
archivePrefix = {arXiv},
       eprint = {0710.0637},
 primaryClass = {astro-ph},
       adsurl = {https://ui.adsabs.harvard.edu/abs/2007ApJ...671.1196M}
}

@ARTICLE{foreman_mackey_exo,
       author = {{Foreman-Mackey}, Daniel and {Hogg}, David W. and {Morton}, Timothy D.},
        title = "{Exoplanet Population Inference and the Abundance of Earth Analogs from Noisy, Incomplete Catalogs}",
      journal = {\apj},
         year = 2014,
        month = nov,
       volume = {795},
       number = {1},
          eid = {64},
        pages = {64},
          doi = {10.1088/0004-637X/795/1/64},
archivePrefix = {arXiv},
       eprint = {1406.3020},
 primaryClass = {astro-ph.EP},
       adsurl = {https://ui.adsabs.harvard.edu/abs/2014ApJ...795...64F}
}

@misc{tensorflow,
title={ {TensorFlow}: Large-Scale Machine Learning on Heterogeneous Systems},
url={https://www.tensorflow.org/},
note={Software available from tensorflow.org},
author={
    Mart\'{i}n~Abadi and
    Ashish~Agarwal and
    Paul~Barham and
    Eugene~Brevdo and
    Zhifeng~Chen and
    Craig~Citro and
    Greg~S.~Corrado and
    Andy~Davis and
    Jeffrey~Dean and
    Matthieu~Devin and
    Sanjay~Ghemawat and
    Ian~Goodfellow and
    Andrew~Harp and
    Geoffrey~Irving and
    Michael~Isard and
    Yangqing Jia and
    Rafal~Jozefowicz and
    Lukasz~Kaiser and
    Manjunath~Kudlur and
    Josh~Levenberg and
    Dandelion~Man\'{e} and
    Rajat~Monga and
    Sherry~Moore and
    Derek~Murray and
    Chris~Olah and
    Mike~Schuster and
    Jonathon~Shlens and
    Benoit~Steiner and
    Ilya~Sutskever and
    Kunal~Talwar and
    Paul~Tucker and
    Vincent~Vanhoucke and
    Vijay~Vasudevan and
    Fernanda~Vi\'{e}gas and
    Oriol~Vinyals and
    Pete~Warden and
    Martin~Wattenberg and
    Martin~Wicke and
    Yuan~Yu and
    Xiaoqiang~Zheng},
  year={2015},
}

@misc{dolphin,
      title={dolphin: A fully automated forward modeling pipeline powered by artificial intelligence for galaxy-scale strong lenses}, 
      author={Anowar J. Shajib and Nafis Sadik Nihal and Chin Yi Tan and Vedant Sahu and Simon Birrer and Tommaso Treu and Joshua Frieman},
      year={2025},
      eprint={2503.22657},
      archivePrefix={arXiv},
      primaryClass={astro-ph.IM},
      url={https://arxiv.org/abs/2503.22657}, 
}

@ARTICLE{desi_2025,
       author = {{Abdul Karim}, M. and {Aguilar}, J. and {Ahlen}, S. and {Alam}, S. and {Allen}, L. and {Allende Prieto}, C. and {Alves}, O. and {Anand}, A. and {Andrade}, U. and {Armengaud}, E. and {Aviles}, A. and {Bailey}, S. and {Baltay}, C. and {Bansal}, P. and {Bault}, A. and {Behera}, J. and {BenZvi}, S. and {Bianchi}, D. and {Blake}, C. and {Brieden}, S. and {Brodzeller}, A. and {Brooks}, D. and {Buckley-Geer}, E. and {Burtin}, E. and {Calderon}, R. and {Canning}, R. and {Rosell}, A. Carnero and {Carrilho}, P. and {Casas}, L. and {Castander}, F.~J. and {Charles}, M. and {Chaussidon}, E. and {Chaves-Montero}, J. and {Chebat}, D. and {Chen}, X. and {Claybaugh}, T. and {Cole}, S. and {Cooper}, A.~P. and {Cuceu}, A. and {Dawson}, K.~S. and {de la Macorra}, A. and {de Mattia}, A. and {Deiosso}, N. and {Della Costa}, J. and {Demina}, R. and {Dey}, A. and {Dey}, B. and {Ding}, Z. and {Doel}, P. and {Edelstein}, J. and {Eisenstein}, D.~J. and {Elbers}, W. and {Fagrelius}, P. and {Fanning}, K. and {Fern{\'a}ndez-Garc{\'\i}a}, E. and {Ferraro}, S. and {Font-Ribera}, A. and {Forero-Romero}, J.~E. and {Frenk}, C.~S. and {Garcia-Quintero}, C. and {Garrison}, L.~H. and {Gazta{\~n}aga}, E. and {Gil-Mar{\'\i}n}, H. and {Gontcho A Gontcho}, S. and {Gonzalez}, D. and {Gonzalez-Morales}, A.~X. and {Gordon}, C. and {Green}, D. and {Gutierrez}, G. and {Guy}, J. and {Hadzhiyska}, B. and {Hahn}, C. and {He}, S. and {Herbold}, M. and {Herrera-Alcantar}, H.~K. and {Ho}, M.-F. and {Honscheid}, K. and {Howlett}, C. and {Huterer}, D. and {Ishak}, M. and {Juneau}, S. and {Kamble}, N.~V. and {Kara{\c{c}}ayl{\i}}, N.~G. and {Kehoe}, R. and {Kent}, S. and {Kim}, A.~G. and {Kirkby}, D. and {Kisner}, T. and {Koposov}, S.~E. and {Kremin}, A. and {Krolewski}, A. and {Lahav}, O. and {Lamman}, C. and {Landriau}, M. and {Lang}, D. and {Lasker}, J. and {Le Goff}, J.~M. and {Le Guillou}, L. and {Leauthaud}, A. and {Levi}, M.~E. and {Li}, Q. and {Li}, T.~S. and {Lodha}, K. and {Lokken}, M. and {Lozano-Rodr{\'\i}guez}, F. and {Magneville}, C. and {Manera}, M. and {Martini}, P. and {Matthewson}, W.~L. and {Meisner}, A. and {Mena-Fern{\'a}ndez}, J. and {Menegas}, A. and {Mergulh{\~a}o}, T. and {Miquel}, R. and {Moustakas}, J. and {Mu{\~n}oz-Guti{\'e}rrez}, A. and {Mu{\~n}oz-Santos}, D. and {Myers}, A.~D. and {Nadathur}, S. and {Naidoo}, K. and {Napolitano}, L. and {Newman}, J.~A. and {Niz}, G. and {Noriega}, H.~E. and {Paillas}, E. and {Palanque-Delabrouille}, N. and {Pan}, J. and {Peacock}, J.~A. and {Pellejero Ibanez}, M. and {Percival}, W.~J. and {P{\'e}rez-Fern{\'a}ndez}, A. and {P{\'e}rez-R{\`a}fols}, I. and {Pieri}, M.~M. and {Poppett}, C. and {Prada}, F. and {Rabinowitz}, D. and {Raichoor}, A. and {Ram{\'\i}rez-P{\'e}rez}, C. and {Rashkovetskyi}, M. and {Ravoux}, C. and {Rich}, J. and {Rocher}, A. and {Rockosi}, C. and {Rohlf}, J. and {Rom{\'a}n-Herrera}, J.~O. and {Ross}, A.~J. and {Rossi}, G. and {Ruggeri}, R. and {Ruhlmann-Kleider}, V. and {Samushia}, L. and {Sanchez}, E. and {Sanders}, N. and {Schlegel}, D. and {Schubnell}, M. and {Seo}, H. and {Shafieloo}, A. and {Sharples}, R. and {Silber}, J. and {Sinigaglia}, F. and {Sprayberry}, D. and {Tan}, T. and {Tarl{\'e}}, G. and {Taylor}, P. and {Turner}, W. and {Ure{\~n}a-L{\'o}pez}, L.~A. and {Vaisakh}, R. and {Valdes}, F. and {Valogiannis}, G. and {Vargas-Maga{\~n}a}, M. and {Verde}, L. and {Walther}, M. and {Weaver}, B.~A. and {Weinberg}, D.~H. and {White}, M. and {Wolfson}, M. and {Y{\`e}che}, C. and {Yu}, J. and {Zaborowski}, E.~A. and {Zarrouk}, P. and {Zhai}, Z. and {Zhang}, H. and {Zhao}, C. and {Zhao}, G.~B. and {Zhou}, R. and {Zou}, H. and {DESI Collaboration}},
        title = "{DESI DR2 results. II. Measurements of baryon acoustic oscillations and cosmological constraints}",
      journal = {\prd},
         year = 2025,
        month = oct,
       volume = {112},
       number = {8},
          eid = {083515},
        pages = {083515},
          doi = {10.1103/tr6y-kpc6},
archivePrefix = {arXiv},
       eprint = {2503.14738},
 primaryClass = {astro-ph.CO},
       adsurl = {https://ui.adsabs.harvard.edu/abs/2025PhRvD.112h3515A}
}

@ARTICLE{dist_shift,
       author = {{Filipp}, Andreas and {Hezaveh}, Yashar and {Perreault-Levasseur}, Laurence},
        title = "{Robustness of Neural Ratio and Posterior Estimators to Distributional Shifts for Population-level Dark Matter Analysis in Strong Gravitational Lensing}",
      journal = {\apj},
         year = 2025,
        month = aug,
       volume = {989},
       number = {2},
          eid = {226},
        pages = {226},
          doi = {10.3847/1538-4357/adee20},
archivePrefix = {arXiv},
       eprint = {2411.05905},
 primaryClass = {astro-ph.CO},
       adsurl = {https://ui.adsabs.harvard.edu/abs/2025ApJ...989..226F}
}

@ARTICLE{npe_uncertainties,
       author = {{Perreault Levasseur}, Laurence and {Hezaveh}, Yashar D. and {Wechsler}, Risa H.},
        title = "{Uncertainties in Parameters Estimated with Neural Networks: Application to Strong Gravitational Lensing}",
      journal = {\apjl},
         year = 2017,
        month = nov,
       volume = {850},
       number = {1},
          eid = {L7},
        pages = {L7},
          doi = {10.3847/2041-8213/aa9704},
archivePrefix = {arXiv},
       eprint = {1708.08843},
 primaryClass = {astro-ph.CO},
       adsurl = {https://ui.adsabs.harvard.edu/abs/2017ApJ...850L...7P}
}

@ARTICLE{auger,
       author = {{Auger}, M.~W. and {Treu}, T. and {Bolton}, A.~S. and {Gavazzi}, R. and {Koopmans}, L.~V.~E. and {Marshall}, P.~J. and {Moustakas}, L.~A. and {Burles}, S.},
        title = "{The Sloan Lens ACS Survey. X. Stellar, Dynamical, and Total Mass Correlations of Massive Early-type Galaxies}",
      journal = {\apj},
         year = 2010,
        month = nov,
       volume = {724},
       number = {1},
        pages = {511-525},
          doi = {10.1088/0004-637X/724/1/511},
archivePrefix = {arXiv},
       eprint = {1007.2880},
 primaryClass = {astro-ph.CO},
       adsurl = {https://ui.adsabs.harvard.edu/abs/2010ApJ...724..511A}
}

@ARTICLE{galan2022,
       author = {{Galan}, A. and {Vernardos}, G. and {Peel}, A. and {Courbin}, F. and {Starck}, J. -L.},
        title = "{Using wavelets to capture deviations from smoothness in galaxy-scale strong lenses}",
      journal = {\aap},
         year = 2022,
        month = dec,
       volume = {668},
          eid = {A155},
        pages = {A155},
          doi = {10.1051/0004-6361/202244464},
archivePrefix = {arXiv},
       eprint = {2207.05763},
 primaryClass = {astro-ph.IM},
       adsurl = {https://ui.adsabs.harvard.edu/abs/2022A&A...668A.155G}
}

@article{Tan_2024,
   title={Project Dinos I: A joint lensing–dynamics constraint on the deviation from the power law in the mass profile of massive ellipticals},
   volume={530},
   ISSN={1365-2966},
   url={http://dx.doi.org/10.1093/mnras/stae884},
   DOI={10.1093/mnras/stae884},
   number={2},
   journal={Monthly Notices of the Royal Astronomical Society},
   publisher={Oxford University Press (OUP)},
   author={Tan, Chin Yi and Shajib, Anowar J and Birrer, Simon and Sonnenfeld, Alessandro and Treu, Tommaso and Wells, Patrick and Williams, Devon and Buckley-Geer, Elizabeth J and Drlica-Wagner, Alex and Frieman, Joshua},
   year={2024},
   month=mar, pages={1474–1505} }

@misc{desi_foundry,
      title={DESI Strong Lens Foundry I: HST Observations and Modeling with GIGA-Lens}, 
      author={X. Huang and S. Baltasar and N. Ratier-Werbin and C. Storfer and W. Sheu and S. Agarwal and M. Tamargo-Arizmendi and D. J. Schlegel and J. Aguilar and S. Ahlen and G. Aldering and S. Banka and S. BenZvi and D. Bianchi and A. Bolton and D. Brooks and A. Cikota and T. Claybaugh and A. de la Macorra and A. Dey and P. Doel and J. Edelstein and A. Filipp and J. E. Forero-Romero and E. Gaztanaga and S. Gontcho A Gontcho and A. Gu and G. Gutierrez and K. Honscheid and E. Jullo and S. Juneau and R. Kehoe and D. Kirkby and T. Kisner and A. Kremin and K. J. Kwon and A. Lambert and M. Landriau and D. Lang and L. Le Guillou and J. Liu and A. Meisner and R. Miquel and J. Moustakas and A. D. Myers and S. Perlmutter and I. Perez-Rafols and F. Prada and G. Rossi and D. Rubin and E. Sanchez and M. Schubnell and Y. Shu and E. Silver and D. Sprayberry and N. Suzuki and G. Tarle and B. A. Weaver and H. Zou},
      year={2025},
      eprint={2502.03455},
      archivePrefix={arXiv},
      primaryClass={astro-ph.CO},
      url={https://arxiv.org/abs/2502.03455}, 
}

@article{strides_schmidt,
   title={STRIDES: automated uniform models for 30 quadruply imaged quasars},
   volume={518},
   ISSN={1365-2966},
   url={http://dx.doi.org/10.1093/mnras/stac2235},
   DOI={10.1093/mnras/stac2235},
   number={1},
   journal={Monthly Notices of the Royal Astronomical Society},
   publisher={Oxford University Press (OUP)},
   author={Schmidt, T and Treu, T and Birrer, S and Shajib, A J and Lemon, C and Millon, M and Sluse, D and Agnello, A and Anguita, T and Auger-Williams, M W and McMahon, R G and Motta, V and Schechter, P and Spiniello, C and Kayo, I and Courbin, F and Ertl, S and Fassnacht, C D and Frieman, J A and More, A and Schuldt, S and Suyu, S H and Aguena, M and Andrade-Oliveira, F and Annis, J and Bacon, D and Bertin, E and Brooks, D and Burke, D L and Carnero Rosell, A and Carrasco Kind, M and Carretero, J and Conselice, C and Costanzi, M and da Costa, L N and Pereira, M E S and De Vicente, J and Desai, S and Doel, P and Everett, S and Ferrero, I and Friedel, D and García-Bellido, J and Gaztanaga, E and Gruen, D and Gruendl, R A and Gschwend, J and Gutierrez, G and Hinton, S R and Hollowood, D L and Honscheid, K and James, D J and Kuehn, K and Lahav, O and Menanteau, F and Miquel, R and Palmese, A and Paz-Chinchón, F and Pieres, A and Plazas Malagón, A A and Prat, J and Rodriguez-Monroy, M and Romer, A K and Sanchez, E and Scarpine, V and Sevilla-Noarbe, I and Smith, M and Suchyta, E and Tarle, G and To, C and Varga, T N},
   year={2022},
   month=nov, pages={1260–1300} }

@ARTICLE{time,
       author = {{Ertl}, S. and {Schuldt}, S. and {Suyu}, S.~H. and {Schmidt}, T. and {Treu}, T. and {Birrer}, S. and {Shajib}, A.~J. and {Sluse}, D.},
        title = "{TDCOSMO. X. Automated modeling of nine strongly lensed quasars and comparison between lens-modeling software}",
      journal = {\aap},
         year = 2023,
        month = apr,
       volume = {672},
          eid = {A2},
        pages = {A2},
          doi = {10.1051/0004-6361/202244909},
archivePrefix = {arXiv},
       eprint = {2209.03094},
 primaryClass = {astro-ph.CO},
       adsurl = {https://ui.adsabs.harvard.edu/abs/2023A&A...672A...2E}
}

@Article{Verde2019,
author={Verde, Licia
and Treu, Tommaso
and Riess, Adam G.},
title={Tensions between the early and late Universe},
journal={Nature Astronomy},
year={2019},
month={Oct},
day={01},
volume={3},
number={10},
pages={891-895},
issn={2397-3366},
doi={10.1038/s41550-019-0902-0},
url={https://doi.org/10.1038/s41550-019-0902-0}
}

@article{Cranmer_2020,
   title={The frontier of simulation-based inference},
   volume={117},
   ISSN={1091-6490},
   url={http://dx.doi.org/10.1073/pnas.1912789117},
   DOI={10.1073/pnas.1912789117},
   number={48},
   journal={Proceedings of the National Academy of Sciences},
   publisher={Proceedings of the National Academy of Sciences},
   author={Cranmer, Kyle and Brehmer, Johann and Louppe, Gilles},
   year={2020},
   month=may, pages={30055–30062} }

@article{tdcosmo4,
   title={TDCOSMO: IV. Hierarchical time-delay cosmography – joint inference of the Hubble constant and galaxy density profiles},
   volume={643},
   ISSN={1432-0746},
   url={http://dx.doi.org/10.1051/0004-6361/202038861},
   DOI={10.1051/0004-6361/202038861},
   journal={Astronomy \& Astrophysics},
   publisher={EDP Sciences},
   author={Birrer, S. and Shajib, A. J. and Galan, A. and Millon, M. and Treu, T. and Agnello, A. and Auger, M. and Chen, G. C.-F. and Christensen, L. and Collett, T. and Courbin, F. and Fassnacht, C. D. and Koopmans, L. V. E. and Marshall, P. J. and Park, J.-W. and Rusu, C. E. and Sluse, D. and Spiniello, C. and Suyu, S. H. and Wagner-Carena, S. and Wong, K. C. and Barnabè, M. and Bolton, A. S. and Czoske, O. and Ding, X. and Frieman, J. A. and Van de Vyvere, L.},
   year={2020},
   month=nov, pages={A165} }

@ARTICLE{bolton2008b,
       author = {{Bolton}, Adam S. and {Treu}, Tommaso and {Koopmans}, L{\'e}on V.~E. and {Gavazzi}, Rapha{\"e}l and {Moustakas}, Leonidas A. and {Burles}, Scott and {Schlegel}, David J. and {Wayth}, Randall},
        title = "{The Sloan Lens ACS Survey. VII. Elliptical Galaxy Scaling Laws from Direct Observational Mass Measurements}",
      journal = {\apj},
         year = 2008,
        month = sep,
       volume = {684},
       number = {1},
        pages = {248-259},
          doi = {10.1086/589989},
archivePrefix = {arXiv},
       eprint = {0805.1932},
 primaryClass = {astro-ph},
       adsurl = {https://ui.adsabs.harvard.edu/abs/2008ApJ...684..248B}
}

@Article{Birrer2024,
author={Birrer, S.
and Millon, M.
and Sluse, D.
and Shajib, A. J.
and Courbin, F.
and Erickson, S.
and Koopmans, L. V. E.
and Suyu, S. H.
and Treu, T.},
title={Time-Delay Cosmography: Measuring the Hubble Constant and Other Cosmological Parameters with Strong Gravitational Lensing},
journal={Space Science Reviews},
year={2024},
month={Jun},
day={17},
volume={220},
number={5},
pages={48},
issn={1572-9672},
doi={10.1007/s11214-024-01079-w},
url={https://doi.org/10.1007/s11214-024-01079-w}
}

@article{selection,
   title={Strong lensing selection effects},
   volume={678},
   ISSN={1432-0746},
   url={http://dx.doi.org/10.1051/0004-6361/202346026},
   DOI={10.1051/0004-6361/202346026},
   journal={Astronomy \& Astrophysics},
   publisher={EDP Sciences},
   author={Sonnenfeld, Alessandro and Li, Shun-Sheng and Despali, Giulia and Gavazzi, Raphael and Shajib, Anowar J. and Taylor, Edward N.},
   year={2023},
   month=sep, pages={A4} }

@article{Pearson_2021,
   title={Strong lens modelling: comparing and combining Bayesian neural networks and parametric profile fitting},
   volume={505},
   ISSN={1365-2966},
   url={http://dx.doi.org/10.1093/mnras/stab1547},
   DOI={10.1093/mnras/stab1547},
   number={3},
   journal={Monthly Notices of the Royal Astronomical Society},
   publisher={Oxford University Press (OUP)},
   author={Pearson, James and Maresca, Jacob and Li, Nan and Dye, Simon},
   year={2021},
   month=may, pages={4362–4382} }

@article{holismokes9,
   title={HOLISMOKES: IX. Neural network inference of strong-lens parameters and uncertainties from ground-based images},
   volume={671},
   ISSN={1432-0746},
   url={http://dx.doi.org/10.1051/0004-6361/202244325},
   DOI={10.1051/0004-6361/202244325},
   journal={Astronomy \& Astrophysics},
   publisher={EDP Sciences},
   author={Schuldt, S. and Cañameras, R. and Shu, Y. and Suyu, S. H. and Taubenberger, S. and Meinhardt, T. and Leal-Taixé, L.},
   year={2023},
   month=mar, pages={A147} }

@ARTICLE{msd,
       author = {{Falco}, E.~E. and {Gorenstein}, M.~V. and {Shapiro}, I.~I.},
        title = "{On model-dependent bounds on H 0 from gravitational images : application to Q 0957+561 A, B.}",
      journal = {\apjl},
         year = 1985,
        month = feb,
       volume = {289},
        pages = {L1-L4},
          doi = {10.1086/184422},
       adsurl = {https://ui.adsabs.harvard.edu/abs/1985ApJ...289L...1F}
}

@article{taak_treu,
   title={Strong lensed QSOs with variability detectable by LSST: How many are there?},
   volume={524},
   ISSN={1365-2966},
   url={http://dx.doi.org/10.1093/mnras/stad2201},
   DOI={10.1093/mnras/stad2201},
   number={4},
   journal={Monthly Notices of the Royal Astronomical Society},
   publisher={Oxford University Press (OUP)},
   author={Taak, Yoon Chan and Treu, Tommaso},
   year={2023},
   month=jul, pages={5446–5453} }

@ARTICLE{npe_loss,
       author = {{Papamakarios}, George and {Murray}, Iain},
        title = "{Fast $\epsilon$-free Inference of Simulation Models with Bayesian Conditional Density Estimation}",
      journal = {arXiv e-prints},
         year = 2016,
        month = may,
          eid = {arXiv:1605.06376},
        pages = {arXiv:1605.06376},
          doi = {10.48550/arXiv.1605.06376},
archivePrefix = {arXiv},
       eprint = {1605.06376},
 primaryClass = {stat.ML},
       adsurl = {https://ui.adsabs.harvard.edu/abs/2016arXiv160506376P}
}

@ARTICLE{lemon,
       author = {{Gentile}, Fabrizio and {Tortora}, Crescenzo and {Covone}, Giovanni and {Koopmans}, L{\'e}on V.~E. and {Li}, Rui and {Leuzzi}, Laura and {Napolitano}, Nicola R.},
        title = "{LEMON: LEns MOdelling with Neural networks - I. Automated modelling of strong gravitational lenses with Bayesian Neural Networks}",
      journal = {\mnras},
         year = 2023,
        month = jul,
       volume = {522},
       number = {4},
        pages = {5442-5455},
          doi = {10.1093/mnras/stad1325},
archivePrefix = {arXiv},
       eprint = {2210.10793},
 primaryClass = {astro-ph.IM},
       adsurl = {https://ui.adsabs.harvard.edu/abs/2023MNRAS.522.5442G}
}

@article{gamma_lens_degeneracy,
   title={Gravitational Lensing Formalism in a Curved Arc Basis: A Continuous Description of Observables and Degeneracies from the Weak to the Strong Lensing Regime},
   volume={919},
   ISSN={1538-4357},
   url={http://dx.doi.org/10.3847/1538-4357/ac1108},
   DOI={10.3847/1538-4357/ac1108},
   number={1},
   journal={The Astrophysical Journal},
   publisher={American Astronomical Society},
   author={Birrer, Simon},
   year={2021},
   month=sep, pages={38} }

@article{sersic_massive_elliptical,
    author = "Caon, N. and Capaccioli, M. and D'Onofrio, M.",
    title = "{On the shape of the light profiles of early type galaxies}",
    eprint = "astro-ph/9309013",
    archivePrefix = "arXiv",
    doi = "10.1093/mnras/265.4.1013",
    journal = "Mon. Not. Roy. Astron. Soc.",
    volume = "265",
    pages = "1013",
    year = "1993"
}

@article{Treu_2022,
   title={Strong lensing time-delay cosmography in the 2020s},
   volume={30},
   ISSN={1432-0754},
   url={http://dx.doi.org/10.1007/s00159-022-00145-y},
   DOI={10.1007/s00159-022-00145-y},
   number={1},
   journal={The Astronomy and Astrophysics Review},
   publisher={Springer Science and Business Media LLC},
   author={Treu, Tommaso and Suyu, Sherry H. and Marshall, Philip J.},
   year={2022},
   month=nov }

@article{abdalla2022,
title = {Cosmology intertwined: A review of the particle physics, astrophysics, and cosmology associated with the cosmological tensions and anomalies},
journal = {Journal of High Energy Astrophysics},
volume = {34},
pages = {49-211},
year = {2022},
issn = {2214-4048},
doi = {https://doi.org/10.1016/j.jheap.2022.04.002},
url = {https://www.sciencedirect.com/science/article/pii/S2214404822000179},
author = {Elcio Abdalla and Guillermo Franco Abellán and Amin Aboubrahim and Adriano Agnello and Özgür Akarsu and Yashar Akrami and George Alestas and Daniel Aloni and Luca Amendola and Luis A. Anchordoqui and Richard I. Anderson and Nikki Arendse and Marika Asgari and Mario Ballardini and Vernon Barger and Spyros Basilakos and Ronaldo C. Batista and Elia S. Battistelli and Richard Battye and Micol Benetti and David Benisty and Asher Berlin and Paolo {de Bernardis} and Emanuele Berti and Bohdan Bidenko and Simon Birrer and John P. Blakeslee and Kimberly K. Boddy and Clecio R. Bom and Alexander Bonilla and Nicola Borghi and François R. Bouchet and Matteo Braglia and Thomas Buchert and Elizabeth Buckley-Geer and Erminia Calabrese and Robert R. Caldwell and David Camarena and Salvatore Capozziello and Stefano Casertano and Geoff C.-F. Chen and Jens Chluba and Angela Chen and Hsin-Yu Chen and Anton Chudaykin and Michele Cicoli and Craig J. Copi and Fred Courbin and Francis-Yan Cyr-Racine and Bożena Czerny and Maria Dainotti and Guido D'Amico and Anne-Christine Davis and Javier {de Cruz Pérez} and Jaume {de Haro} and Jacques Delabrouille and Peter B. Denton and Suhail Dhawan and Keith R. Dienes and Eleonora {Di Valentino} and Pu Du and Dominique Eckert and Celia Escamilla-Rivera and Agnès Ferté and Fabio Finelli and Pablo Fosalba and Wendy L. Freedman and Noemi Frusciante and Enrique Gaztañaga and William Giarè and Elena Giusarma and Adrià Gómez-Valent and Will Handley and Ian Harrison and Luke Hart and Dhiraj Kumar Hazra and Alan Heavens and Asta Heinesen and Hendrik Hildebrandt and J. Colin Hill and Natalie B. Hogg and Daniel E. Holz and Deanna C. Hooper and Nikoo Hosseininejad and Dragan Huterer and Mustapha Ishak and Mikhail M. Ivanov and Andrew H. Jaffe and In Sung Jang and Karsten Jedamzik and Raul Jimenez and Melissa Joseph and Shahab Joudaki and Marc Kamionkowski and Tanvi Karwal and Lavrentios Kazantzidis and Ryan E. Keeley and Michael Klasen and Eiichiro Komatsu and Léon V.E. Koopmans and Suresh Kumar and Luca Lamagna and Ruth Lazkoz and Chung-Chi Lee and Julien Lesgourgues and Jackson {Levi Said} and Tiffany R. Lewis and Benjamin L'Huillier and Matteo Lucca and Roy Maartens and Lucas M. Macri and Danny Marfatia and Valerio Marra and Carlos J.A.P. Martins and Silvia Masi and Sabino Matarrese and Arindam Mazumdar and Alessandro Melchiorri and Olga Mena and Laura Mersini-Houghton and James Mertens and Dinko Milaković and Yuto Minami and Vivian Miranda and Cristian Moreno-Pulido and Michele Moresco and David F. Mota and Emil Mottola and Simone Mozzon and Jessica Muir and Ankan Mukherjee and Suvodip Mukherjee and Pavel Naselsky and Pran Nath and Savvas Nesseris and Florian Niedermann and Alessio Notari and Rafael C. Nunes and Eoin {Ó Colgáin} and Kayla A. Owens and Emre Özülker and Francesco Pace and Andronikos Paliathanasis and Antonella Palmese and Supriya Pan and Daniela Paoletti and Santiago E. {Perez Bergliaffa} and Leandros Perivolaropoulos and Dominic W. Pesce and Valeria Pettorino and Oliver H.E. Philcox and Levon Pogosian and Vivian Poulin and Gaspard Poulot and Marco Raveri and Mark J. Reid and Fabrizio Renzi and Adam G. Riess and Vivian I. Sabla and Paolo Salucci and Vincenzo Salzano and Emmanuel N. Saridakis and Bangalore S. Sathyaprakash and Martin Schmaltz and Nils Schöneberg and Dan Scolnic and Anjan A. Sen and Neelima Sehgal and Arman Shafieloo and M.M. Sheikh-Jabbari and Joseph Silk and Alessandra Silvestri and Foteini Skara and Martin S. Sloth and Marcelle Soares-Santos and Joan {Solà Peracaula} and Yu-Yang Songsheng and Jorge F. Soriano and Denitsa Staicova and Glenn D. Starkman and István Szapudi and Elsa M. Teixeira and Brooks Thomas and Tommaso Treu and Emery Trott and Carsten {van de Bruck} and J. Alberto Vazquez and Licia Verde and Luca Visinelli and Deng Wang and Jian-Min Wang and Shao-Jiang Wang and Richard Watkins and Scott Watson and John K. Webb and Neal Weiner and Amanda Weltman and Samuel J. Witte and Radosław Wojtak and Anil Kumar Yadav and Weiqiang Yang and Gong-Bo Zhao and Miguel Zumalacárregui}
}

@ARTICLE{sersic,
       author = {{S{\'e}rsic}, J.~L.},
        title = "{Influence of the atmospheric and instrumental dispersion on the brightness distribution in a galaxy}",
      journal = {Boletin de la Asociacion Argentina de Astronomia La Plata Argentina},
         year = 1963,
        month = feb,
       volume = {6},
        pages = {41-43},
       adsurl = {https://ui.adsabs.harvard.edu/abs/1963BAAA....6...41S}
}

@article{Riess_2022,
doi = {10.3847/2041-8213/ac5c5b},
url = {https://dx.doi.org/10.3847/2041-8213/ac5c5b},
year = {2022},
month = {jul},
publisher = {The American Astronomical Society},
volume = {934},
number = {1},
pages = {L7},
author = {Adam G. Riess and Wenlong Yuan and Lucas M. Macri and Dan Scolnic and Dillon Brout and Stefano Casertano and David O. Jones and Yukei Murakami and Gagandeep S. Anand and Louise Breuval and Thomas G. Brink and Alexei V. Filippenko and Samantha Hoffmann and Saurabh W. Jha and W. D’arcy Kenworthy and John Mackenty and Benjamin E. Stahl and WeiKang Zheng},
title = {A Comprehensive Measurement of the Local Value of the Hubble Constant with 1 km s−1 Mpc−1 Uncertainty from the Hubble Space Telescope and the SH0ES Team},
journal = {The Astrophysical Journal Letters}
}

@ARTICLE{refsdal1964,
       author = {{Refsdal}, S.},
        title = "{On the possibility of determining Hubble's parameter and the masses of galaxies from the gravitational lens effect}",
      journal = {\mnras},
         year = 1964,
        month = jan,
       volume = {128},
        pages = {307},
          doi = {10.1093/mnras/128.4.307},
       adsurl = {https://ui.adsabs.harvard.edu/abs/1964MNRAS.128..307R}
}

@ARTICLE{lsst,
       author = {{Ivezi{\'c}}, {\v{Z}}eljko and {Kahn}, Steven M. and {Tyson}, J. Anthony and {Abel}, Bob and {Acosta}, Emily and {Allsman}, Robyn and {Alonso}, David and {AlSayyad}, Yusra and {Anderson}, Scott F. and {Andrew}, John and {Angel}, James Roger P. and {Angeli}, George Z. and {Ansari}, Reza and {Antilogus}, Pierre and {Araujo}, Constanza and {Armstrong}, Robert and {Arndt}, Kirk T. and {Astier}, Pierre and {Aubourg}, {\'E}ric and {Auza}, Nicole and {Axelrod}, Tim S. and {Bard}, Deborah J. and {Barr}, Jeff D. and {Barrau}, Aurelian and {Bartlett}, James G. and {Bauer}, Amanda E. and {Bauman}, Brian J. and {Baumont}, Sylvain and {Bechtol}, Ellen and {Bechtol}, Keith and {Becker}, Andrew C. and {Becla}, Jacek and {Beldica}, Cristina and {Bellavia}, Steve and {Bianco}, Federica B. and {Biswas}, Rahul and {Blanc}, Guillaume and {Blazek}, Jonathan and {Blandford}, Roger D. and {Bloom}, Josh S. and {Bogart}, Joanne and {Bond}, Tim W. and {Booth}, Michael T. and {Borgland}, Anders W. and {Borne}, Kirk and {Bosch}, James F. and {Boutigny}, Dominique and {Brackett}, Craig A. and {Bradshaw}, Andrew and {Brandt}, William Nielsen and {Brown}, Michael E. and {Bullock}, James S. and {Burchat}, Patricia and {Burke}, David L. and {Cagnoli}, Gianpietro and {Calabrese}, Daniel and {Callahan}, Shawn and {Callen}, Alice L. and {Carlin}, Jeffrey L. and {Carlson}, Erin L. and {Chandrasekharan}, Srinivasan and {Charles-Emerson}, Glenaver and {Chesley}, Steve and {Cheu}, Elliott C. and {Chiang}, Hsin-Fang and {Chiang}, James and {Chirino}, Carol and {Chow}, Derek and {Ciardi}, David R. and {Claver}, Charles F. and {Cohen-Tanugi}, Johann and {Cockrum}, Joseph J. and {Coles}, Rebecca and {Connolly}, Andrew J. and {Cook}, Kem H. and {Cooray}, Asantha and {Covey}, Kevin R. and {Cribbs}, Chris and {Cui}, Wei and {Cutri}, Roc and {Daly}, Philip N. and {Daniel}, Scott F. and {Daruich}, Felipe and {Daubard}, Guillaume and {Daues}, Greg and {Dawson}, William and {Delgado}, Francisco and {Dellapenna}, Alfred and {de Peyster}, Robert and {de Val-Borro}, Miguel and {Digel}, Seth W. and {Doherty}, Peter and {Dubois}, Richard and {Dubois-Felsmann}, Gregory P. and {Durech}, Josef and {Economou}, Frossie and {Eifler}, Tim and {Eracleous}, Michael and {Emmons}, Benjamin L. and {Fausti Neto}, Angelo and {Ferguson}, Henry and {Figueroa}, Enrique and {Fisher-Levine}, Merlin and {Focke}, Warren and {Foss}, Michael D. and {Frank}, James and {Freemon}, Michael D. and {Gangler}, Emmanuel and {Gawiser}, Eric and {Geary}, John C. and {Gee}, Perry and {Geha}, Marla and {Gessner}, Charles J.~B. and {Gibson}, Robert R. and {Gilmore}, D. Kirk and {Glanzman}, Thomas and {Glick}, William and {Goldina}, Tatiana and {Goldstein}, Daniel A. and {Goodenow}, Iain and {Graham}, Melissa L. and {Gressler}, William J. and {Gris}, Philippe and {Guy}, Leanne P. and {Guyonnet}, Augustin and {Haller}, Gunther and {Harris}, Ron and {Hascall}, Patrick A. and {Haupt}, Justine and {Hernandez}, Fabio and {Herrmann}, Sven and {Hileman}, Edward and {Hoblitt}, Joshua and {Hodgson}, John A. and {Hogan}, Craig and {Howard}, James D. and {Huang}, Dajun and {Huffer}, Michael E. and {Ingraham}, Patrick and {Innes}, Walter R. and {Jacoby}, Suzanne H. and {Jain}, Bhuvnesh and {Jammes}, Fabrice and {Jee}, M. James and {Jenness}, Tim and {Jernigan}, Garrett and {Jevremovi{\'c}}, Darko and {Johns}, Kenneth and {Johnson}, Anthony S. and {Johnson}, Margaret W.~G. and {Jones}, R. Lynne and {Juramy-Gilles}, Claire and {Juri{\'c}}, Mario and {Kalirai}, Jason S. and {Kallivayalil}, Nitya J. and {Kalmbach}, Bryce and {Kantor}, Jeffrey P. and {Karst}, Pierre and {Kasliwal}, Mansi M. and {Kelly}, Heather and {Kessler}, Richard and {Kinnison}, Veronica and {Kirkby}, David and {Knox}, Lloyd and {Kotov}, Ivan V. and {Krabbendam}, Victor L. and {Krughoff}, K. Simon and {Kub{\'a}nek}, Petr and {Kuczewski}, John and {Kulkarni}, Shri and {Ku}, John and {Kurita}, Nadine R. and {Lage}, Craig S. and {Lambert}, Ron and {Lange}, Travis and {Langton}, J. Brian and {Le Guillou}, Laurent and {Levine}, Deborah and {Liang}, Ming and {Lim}, Kian-Tat and {Lintott}, Chris J. and {Long}, Kevin E. and {Lopez}, Margaux and {Lotz}, Paul J. and {Lupton}, Robert H. and {Lust}, Nate B. and {MacArthur}, Lauren A. and {Mahabal}, Ashish and {Mandelbaum}, Rachel and {Markiewicz}, Thomas W. and {Marsh}, Darren S. and {Marshall}, Philip J. and {Marshall}, Stuart and {May}, Morgan and {McKercher}, Robert and {McQueen}, Michelle and {Meyers}, Joshua and {Migliore}, Myriam and {Miller}, Michelle and {Mills}, David J.},
        title = "{LSST: From Science Drivers to Reference Design and Anticipated Data Products}",
      journal = {\apj},
         year = 2019,
        month = mar,
       volume = {873},
       number = {2},
          eid = {111},
        pages = {111},
          doi = {10.3847/1538-4357/ab042c},
archivePrefix = {arXiv},
       eprint = {0805.2366},
 primaryClass = {astro-ph},
       adsurl = {https://ui.adsabs.harvard.edu/abs/2019ApJ...873..111I}
}

@ARTICLE{millon2020,
       author = {{Millon}, M. and {Galan}, A. and {Courbin}, F. and {Treu}, T. and {Suyu}, S.~H. and {Ding}, X. and {Birrer}, S. and {Chen}, G.~C. -F. and {Shajib}, A.~J. and {Sluse}, D. and {Wong}, K.~C. and {Agnello}, A. and {Auger}, M.~W. and {Buckley-Geer}, E.~J. and {Chan}, J.~H.~H. and {Collett}, T. and {Fassnacht}, C.~D. and {Hilbert}, S. and {Koopmans}, L.~V.~E. and {Motta}, V. and {Mukherjee}, S. and {Rusu}, C.~E. and {Sonnenfeld}, A. and {Spiniello}, C. and {Van de Vyvere}, L.},
        title = "{TDCOSMO. I. An exploration of systematic uncertainties in the inference of H$_{0}$ from time-delay cosmography}",
      journal = {\aap},
         year = 2020,
        month = jul,
       volume = {639},
          eid = {A101},
        pages = {A101},
          doi = {10.1051/0004-6361/201937351},
archivePrefix = {arXiv},
       eprint = {1912.08027},
 primaryClass = {astro-ph.CO},
       adsurl = {https://ui.adsabs.harvard.edu/abs/2020A&A...639A.101M}
}

@ARTICLE{wong2020,
       author = {{Wong}, Kenneth C. and {Suyu}, Sherry H. and {Chen}, Geoff C. -F. and {Rusu}, Cristian E. and {Millon}, Martin and {Sluse}, Dominique and {Bonvin}, Vivien and {Fassnacht}, Christopher D. and {Taubenberger}, Stefan and {Auger}, Matthew W. and {Birrer}, Simon and {Chan}, James H.~H. and {Courbin}, Frederic and {Hilbert}, Stefan and {Tihhonova}, Olga and {Treu}, Tommaso and {Agnello}, Adriano and {Ding}, Xuheng and {Jee}, Inh and {Komatsu}, Eiichiro and {Shajib}, Anowar J. and {Sonnenfeld}, Alessandro and {Blandford}, Roger D. and {Koopmans}, L{\'e}on V.~E. and {Marshall}, Philip J. and {Meylan}, Georges},
        title = "{H0LiCOW - XIII. A 2.4 per cent measurement of H$_{0}$ from lensed quasars: 5.3{\ensuremath{\sigma}} tension between early- and late-Universe probes}",
      journal = {\mnras},
         year = 2020,
        month = oct,
       volume = {498},
       number = {1},
        pages = {1420-1439},
          doi = {10.1093/mnras/stz3094},
archivePrefix = {arXiv},
       eprint = {1907.04869},
 primaryClass = {astro-ph.CO},
       adsurl = {https://ui.adsabs.harvard.edu/abs/2020MNRAS.498.1420W}
}

@ARTICLE{lemon_euclid,
       author = {{Euclid Collaboration} and {Busillo}, V. and {Tortora}, C. and {Metcalf}, R.~B. and {Nightingale}, J.~W. and {Meneghetti}, M. and {Gentile}, F. and {Gavazzi}, R. and {Zhong}, F. and {Li}, R. and {Cl{\'e}ment}, B. and {Covone}, G. and {Napolitano}, N.~R. and {Courbin}, F. and {Walmsley}, M. and {Jullo}, E. and {Pearson}, J. and {Scott}, D. and {Le Brun}, A.~M.~C. and {Leuzzi}, L. and {Aghanim}, N. and {Altieri}, B. and {Amara}, A. and {Andreon}, S. and {Aussel}, H. and {Baccigalupi}, C. and {Baldi}, M. and {Bardelli}, S. and {Battaglia}, P. and {Biviano}, A. and {Branchini}, E. and {Brescia}, M. and {Brinchmann}, J. and {Camera}, S. and {Ca{\~n}as-Herrera}, G. and {Capobianco}, V. and {Carbone}, C. and {Cardone}, V.~F. and {Carretero}, J. and {Casas}, S. and {Castellano}, M. and {Castignani}, G. and {Cavuoti}, S. and {Chambers}, K.~C. and {Cimatti}, A. and {Colodro-Conde}, C. and {Congedo}, G. and {Conselice}, C.~J. and {Conversi}, L. and {Copin}, Y. and {Courtois}, H.~M. and {Cropper}, M. and {Da Silva}, A. and {Degaudenzi}, H. and {de la Torre}, S. and {De Lucia}, G. and {Di Giorgio}, A.~M. and {Dinis}, J. and {Dole}, H. and {Dubath}, F. and {Dupac}, X. and {Dusini}, S. and {Escoffier}, S. and {Farina}, M. and {Farinelli}, R. and {Faustini}, F. and {Ferriol}, S. and {Finelli}, F. and {Fotopoulou}, S. and {Frailis}, M. and {Franceschi}, E. and {Galeotta}, S. and {George}, K. and {Gillard}, W. and {Gillis}, B. and {Giocoli}, C. and {Gracia-Carpio}, J. and {Granett}, B.~R. and {Grazian}, A. and {Grupp}, F. and {Haugan}, S.~V.~H. and {Holmes}, W. and {Hook}, I. and {Hormuth}, F. and {Hornstrup}, A. and {Hudelot}, P. and {Jahnke}, K. and {Jhabvala}, M. and {Joachimi}, B. and {Keih{\"a}nen}, E. and {Kermiche}, S. and {Kiessling}, A. and {Kubik}, B. and {K{\"u}mmel}, M. and {Kunz}, M. and {Kurki-Suonio}, H. and {Le Boulc'h}, Q. and {Ligori}, S. and {Lilje}, P.~B. and {Lindholm}, V. and {Lloro}, I. and {Mainetti}, G. and {Maino}, D. and {Maiorano}, E. and {Mansutti}, O. and {Marggraf}, O. and {Markovic}, K. and {Martinelli}, M. and {Martinet}, N. and {Marulli}, F. and {Massey}, R. and {Maurogordato}, S. and {Medinaceli}, E. and {Mei}, S. and {Mellier}, Y. and {Merlin}, E. and {Meylan}, G. and {Mora}, A. and {Moresco}, M. and {Moscardini}, L. and {Nakajima}, R. and {Neissner}, C. and {Niemi}, S. -M. and {Padilla}, C. and {Paltani}, S. and {Pasian}, F. and {Pedersen}, K. and {Pettorino}, V. and {Pires}, S. and {Polenta}, G. and {Poncet}, M. and {Popa}, L.~A. and {Pozzetti}, L. and {Raison}, F. and {Rebolo}, R. and {Renzi}, A. and {Rhodes}, J. and {Riccio}, G. and {Romelli}, E. and {Roncarelli}, M. and {Saglia}, R. and {Sakr}, Z. and {S{\'a}nchez}, A.~G. and {Sapone}, D. and {Sartoris}, B. and {Schewtschenko}, J.~A. and {Schirmer}, M. and {Schneider}, P. and {Schrabback}, T. and {Secroun}, A. and {Sefusatti}, E. and {Seidel}, G. and {Seiffert}, M. and {Serrano}, S. and {Simon}, P. and {Sirignano}, C. and {Sirri}, G. and {Smadja}, G. and {Stanco}, L. and {Steinwagner}, J. and {Tallada-Cresp{\'\i}}, P. and {Taylor}, A.~N. and {Tereno}, I. and {Toft}, S. and {Toledo-Moreo}, R. and {Torradeflot}, F. and {Tutusaus}, I. and {Valenziano}, L. and {Valiviita}, J. and {Vassallo}, T. and {Veropalumbo}, A. and {Wang}, Y. and {Weller}, J. and {Zamorani}, G. and {Zucca}, E. and {Allevato}, V. and {Ballardini}, M. and {Bolzonella}, M. and {Bozzo}, E. and {Burigana}, C. and {Cabanac}, R. and {Calabrese}, M. and {Di Ferdinando}, D. and {Escartin Vigo}, J.~A. and {Gabarra}, L. and {Huertas-Company}, M. and {Matthew}, S. and {Mauri}, N. and {Nucita}, A.~A. and {Pezzotta}, A. and {P{\"o}ntinen}, M. and {Porciani}, C. and {Scottez}, V. and {Tenti}, M. and {Viel}, M. and {Wiesmann}, M. and {Akrami}, Y. and {Alvi}, S. and {Andika}, I.~T. and {Anselmi}, S.},
        title = "{Euclid Quick Data Release (Q1). LEMON -- Lens Modelling with Neural networks. Automated and fast modelling of Euclid gravitational lenses with a singular isothermal ellipsoid mass profile}",
      journal = {arXiv e-prints},
         year = 2025,
        month = mar,
          eid = {arXiv:2503.15329},
        pages = {arXiv:2503.15329},
          doi = {10.48550/arXiv.2503.15329},
archivePrefix = {arXiv},
       eprint = {2503.15329},
 primaryClass = {astro-ph.CO},
       adsurl = {https://ui.adsabs.harvard.edu/abs/2025arXiv250315329E}
}

@article{holismokes10,
   title={HOLISMOKES: X. Comparison between neural network and semi-automated traditional modeling of strong lenses},
   volume={673},
   ISSN={1432-0746},
   url={http://dx.doi.org/10.1051/0004-6361/202244534},
   DOI={10.1051/0004-6361/202244534},
   journal={Astronomy \& Astrophysics},
   publisher={EDP Sciences},
   author={Schuldt, S. and Suyu, S. H. and Cañameras, R. and Shu, Y. and Taubenberger, S. and Ertl, S. and Halkola, A.},
   year={2023},
   month=may, pages={A33} }

@article{Park_2021,
   title={Large-scale Gravitational Lens Modeling with Bayesian Neural Networks for Accurate and Precise Inference of the Hubble Constant},
   volume={910},
   ISSN={1538-4357},
   url={http://dx.doi.org/10.3847/1538-4357/abdfc4},
   DOI={10.3847/1538-4357/abdfc4},
   number={1},
   journal={The Astrophysical Journal},
   publisher={American Astronomical Society},
   author={Park, Ji Won and Wagner-Carena, Sebastian and Birrer, Simon and Marshall, Philip J. and Lin, Joshua Yao-Yu and Roodman, Aaron},
   year={2021},
   month=mar, pages={39} }

@article{Wagner_Carena_2021,
   title={Hierarchical Inference with Bayesian Neural Networks: An Application to Strong Gravitational Lensing},
   volume={909},
   ISSN={1538-4357},
   url={http://dx.doi.org/10.3847/1538-4357/abdf59},
   DOI={10.3847/1538-4357/abdf59},
   number={2},
   journal={The Astrophysical Journal},
   publisher={American Astronomical Society},
   author={Wagner-Carena, Sebastian and Park, Ji Won and Birrer, Simon and Marshall, Philip J. and Roodman, Aaron and Wechsler, Risa H.},
   year={2021},
   month=mar, pages={187} }

@article{Treu_2016,
   title={Time delay cosmography},
   volume={24},
   ISSN={1432-0754},
   url={http://dx.doi.org/10.1007/s00159-016-0096-8},
   DOI={10.1007/s00159-016-0096-8},
   number={1},
   journal={The Astronomy and Astrophysics Review},
   publisher={Springer Science and Business Media LLC},
   author={Treu, Tommaso and Marshall, Philip J.},
   year={2016},
   month=jul }

@ARTICLE{suyu_gamma,
       author = {{Suyu}, S.~H. and {Auger}, M.~W. and {Hilbert}, S. and {Marshall}, P.~J. and {Tewes}, M. and {Treu}, T. and {Fassnacht}, C.~D. and {Koopmans}, L.~V.~E. and {Sluse}, D. and {Blandford}, R.~D. and {Courbin}, F. and {Meylan}, G.},
        title = "{Two Accurate Time-delay Distances from Strong Lensing: Implications for Cosmology}",
      journal = {\apj},
         year = 2013,
        month = apr,
       volume = {766},
       number = {2},
          eid = {70},
        pages = {70},
          doi = {10.1088/0004-637X/766/2/70},
archivePrefix = {arXiv},
       eprint = {1208.6010},
 primaryClass = {astro-ph.CO},
       adsurl = {https://ui.adsabs.harvard.edu/abs/2013ApJ...766...70S}
}

@ARTICLE{starred,
       author = {{Millon}, Martin and {Michalewicz}, Kevin and {Dux}, Fr{\'e}d{\'e}ric and {Courbin}, Fr{\'e}d{\'e}ric and {Marshall}, Philip J.},
        title = "{Image Deconvolution and Point-spread Function Reconstruction with STARRED: A Wavelet-based Two-channel Method Optimized for Light-curve Extraction}",
      journal = {\aj},
         year = 2024,
        month = aug,
       volume = {168},
       number = {2},
          eid = {55},
        pages = {55},
          doi = {10.3847/1538-3881/ad4da7},
archivePrefix = {arXiv},
       eprint = {2402.08725},
 primaryClass = {astro-ph.IM},
       adsurl = {https://ui.adsabs.harvard.edu/abs/2024AJ....168...55M}
}

@article{Barkana_1998,
   title={Fast Calculation of a Family of Elliptical Gravitational Lens Models},
   volume={502},
   ISSN={1538-4357},
   url={http://dx.doi.org/10.1086/305950},
   DOI={10.1086/305950},
   number={2},
   journal={The Astrophysical Journal},
   publisher={American Astronomical Society},
   author={Barkana, Rennan},
   year={1998},
   month=aug, pages={531–537} }

@article{paltas,
   title={From Images to Dark Matter: End-to-end Inference of Substructure from Hundreds of Strong Gravitational Lenses},
   volume={942},
   ISSN={1538-4357},
   url={http://dx.doi.org/10.3847/1538-4357/aca525},
   DOI={10.3847/1538-4357/aca525},
   number={2},
   journal={The Astrophysical Journal},
   publisher={American Astronomical Society},
   author={Wagner-Carena, Sebastian and Aalbers, Jelle and Birrer, Simon and Nadler, Ethan O. and Darragh-Ford, Elise and Marshall, Philip J. and Wechsler, Risa H.},
   year={2023},
   month=jan, pages={75} }

@ARTICLE{erickson2024,
       author = {{Erickson}, Sydney and {Wagner-Carena}, Sebastian and {Marshall}, Phil and {Millon}, Martin and {Birrer}, Simon and {Roodman}, Aaron and {Schmidt}, Thomas and {Treu}, Tommaso and {Schuldt}, Stefan and {Shajib}, Anowar J. and {Venkatraman}, Padma and {The LSST Dark Energy Science Collaboration}},
        title = "{Lens Modeling of STRIDES Strongly Lensed Quasars Using Neural Posterior Estimation}",
      journal = {\aj},
         year = 2025,
        month = jul,
       volume = {170},
       number = {1},
          eid = {44},
        pages = {44},
          doi = {10.3847/1538-3881/add99f},
archivePrefix = {arXiv},
       eprint = {2410.10123},
 primaryClass = {astro-ph.IM},
       adsurl = {https://ui.adsabs.harvard.edu/abs/2025AJ....170...44E}
}

@misc{gawade2024,
      title={Neural network prediction of model parameters for strong lensing samples from Hyper Suprime-Cam Survey}, 
      author={Priyanka Gawade and Anupreeta More and Surhud More and Akisato Kimura and Alessandro Sonnenfeld and Masamune Oguri and Naoki Yoshida},
      year={2024},
      eprint={2404.18897},
      archivePrefix={arXiv},
      primaryClass={astro-ph.CO},
      url={https://arxiv.org/abs/2404.18897}, 
}

@misc{poh2022,
      title={Strong Lensing Parameter Estimation on Ground-Based Imaging Data Using Simulation-Based Inference}, 
      author={Jason Poh and Ashwin Samudre and Aleksandra Ćiprijanović and Brian Nord and Gourav Khullar and Dimitrios Tanoglidis and Joshua A. Frieman},
      year={2022},
      eprint={2211.05836},
      archivePrefix={arXiv},
      primaryClass={astro-ph.IM},
      url={https://arxiv.org/abs/2211.05836}, 
}

@ARTICLE{poh_2025,
       author = {{Poh}, Jason and {Samudre}, Ashwin and {{\'C}iprijanovi{\'c}}, Aleksandra and {Frieman}, Joshua and {Khullar}, Gourav and {Nord}, Brian D.},
        title = "{Deep inference of simulated strong lenses in ground-based surveys}",
      journal = {\jcap},
         year = 2025,
        month = may,
       volume = {2025},
       number = {5},
          eid = {053},
        pages = {053},
          doi = {10.1088/1475-7516/2025/05/053},
archivePrefix = {arXiv},
       eprint = {2501.08524},
 primaryClass = {astro-ph.IM},
       adsurl = {https://ui.adsabs.harvard.edu/abs/2025JCAP...05..053P}
}

@misc{madireddy2019,
      title={A Modular Deep Learning Pipeline for Galaxy-Scale Strong Gravitational Lens Detection and Modeling}, 
      author={Sandeep Madireddy and Nesar Ramachandra and Nan Li and James Butler and Prasanna Balaprakash and Salman Habib and Katrin Heitmann and The LSST Dark Energy Science Collaboration},
      year={2022},
      eprint={1911.03867},
      archivePrefix={arXiv},
      primaryClass={astro-ph.IM},
      url={https://arxiv.org/abs/1911.03867}, 
}

@ARTICLE{lsst_euclid,
       author = {{Pearson}, James and {Li}, Nan and {Dye}, Simon},
        title = "{The use of convolutional neural networks for modelling large optically-selected strong galaxy-lens samples}",
      journal = {\mnras},
         year = 2019,
        month = sep,
       volume = {488},
       number = {1},
        pages = {991-1004},
          doi = {10.1093/mnras/stz1750},
archivePrefix = {arXiv},
       eprint = {1904.06199},
 primaryClass = {astro-ph.IM},
       adsurl = {https://ui.adsabs.harvard.edu/abs/2019MNRAS.488..991P}
}

@article{om10,
    author = {Oguri, Masamune and Marshall, Philip J.},
    title = "{Gravitationally lensed quasars and supernovae in future wide-field optical imaging surveys}",
    journal = {Monthly Notices of the Royal Astronomical Society},
    volume = {405},
    number = {4},
    pages = {2579-2593},
    year = {2010},
    month = {07},
    issn = {0035-8711},
    doi = {10.1111/j.1365-2966.2010.16639.x},
    url = {https://doi.org/10.1111/j.1365-2966.2010.16639.x},
    eprint = {https://academic.oup.com/mnras/article-pdf/405/4/2579/18441702/mnras0405-2579.pdf},
}

@ARTICLE{yue_2022,
       author = {{Yue}, Minghao and {Fan}, Xiaohui and {Yang}, Jinyi and {Wang}, Feige},
        title = "{A Mock Catalog of Gravitationally-lensed Quasars for the LSST Survey}",
      journal = {\aj},
         year = 2022,
        month = mar,
       volume = {163},
       number = {3},
          eid = {139},
        pages = {139},
          doi = {10.3847/1538-3881/ac4cb0},
archivePrefix = {arXiv},
       eprint = {2201.06761},
 primaryClass = {astro-ph.GA},
       adsurl = {https://ui.adsabs.harvard.edu/abs/2022AJ....163..139Y}
}

@ARTICLE{abe_2025,
       author = {{Abe}, Katsuya T. and {Oguri}, Masamune and {Birrer}, Simon and {Khadka}, Narayan and {Marshall}, Philip J. and {Lemon}, Cameron and {More}, Anupreeta and {LSST Dark Energy Science Collaboration}},
        title = "{A halo model approach for mock catalogs of time-variable strong gravitational lenses}",
      journal = {The Open Journal of Astrophysics},
         year = 2025,
        month = jan,
       volume = {8},
          eid = {8},
        pages = {8},
          doi = {10.33232/001c.128482},
archivePrefix = {arXiv},
       eprint = {2411.07509},
 primaryClass = {astro-ph.CO},
       adsurl = {https://ui.adsabs.harvard.edu/abs/2025OJAp....8E...8A}
}

@article{cosmodc2,
   title={CosmoDC2: A Synthetic Sky Catalog for Dark Energy Science with LSST},
   volume={245},
   ISSN={1538-4365},
   url={http://dx.doi.org/10.3847/1538-4365/ab510c},
   DOI={10.3847/1538-4365/ab510c},
   number={2},
   journal={The Astrophysical Journal Supplement Series},
   publisher={American Astronomical Society},
   author={Korytov, Danila and Hearin, Andrew and Kovacs, Eve and Larsen, Patricia and Rangel, Esteban and Hollowed, Joseph and Benson, Andrew J. and Heitmann, Katrin and Mao, Yao-Yuan and Bahmanyar, Anita and Chang, Chihway and Campbell, Duncan and DeRose, Joseph and Finkel, Hal and Frontiere, Nicholas and Gawiser, Eric and Habib, Salman and Joachimi, Benjamin and Lanusse, François and Li, Nan and Mandelbaum, Rachel and Morrison, Christopher and Newman, Jeffrey A. and Pope, Adrian and Rykoff, Eli and Simet, Melanie and To, Chun-Hao and Vikraman, Vinu and Wechsler, Risa H. and White, Martin},
   year={2019},
   month=dec, pages={26} }

@misc{adamopt,
      title={Adam: A Method for Stochastic Optimization}, 
      author={Diederik P. Kingma and Jimmy Ba},
      year={2017},
      eprint={1412.6980},
      archivePrefix={arXiv},
      primaryClass={cs.LG},
      url={https://arxiv.org/abs/1412.6980}, 
}

@misc{resnet34,
      title={Deep Residual Learning for Image Recognition}, 
      author={Kaiming He and Xiangyu Zhang and Shaoqing Ren and Jian Sun},
      year={2015},
      eprint={1512.03385},
      archivePrefix={arXiv},
      primaryClass={cs.CV},
      url={https://arxiv.org/abs/1512.03385}, 
}

@article{Birrer_2018,
   title={lenstronomy: Multi-purpose gravitational lens modelling software package},
   volume={22},
   ISSN={2212-6864},
   url={http://dx.doi.org/10.1016/j.dark.2018.11.002},
   DOI={10.1016/j.dark.2018.11.002},
   journal={Physics of the Dark Universe},
   publisher={Elsevier BV},
   author={Birrer, Simon and Amara, Adam},
   year={2018},
   month=dec, pages={189–201} }

@article{Birrer2021, doi = {10.21105/joss.03283}, url = {https://doi.org/10.21105/joss.03283}, year = {2021}, publisher = {The Open Journal}, volume = {6}, number = {62}, pages = {3283}, author = {Simon Birrer and Anowar J. Shajib and Daniel Gilman and Aymeric Galan and Jelle Aalbers and Martin Millon and Robert Morgan and Giulia Pagano and Ji Won Park and Luca Teodori and Nicolas Tessore and Madison Ueland and Lyne Van de Vyvere and Sebastian Wagner-Carena and Ewoud Wempe and Lilan Yang and Xuheng Ding and Thomas Schmidt and Dominique Sluse and Ming Zhang and Adam Amara}, title = {lenstronomy II: A gravitational lensing software ecosystem}, journal = {Journal of Open Source Software} }

@article{schuldt2021holismokes,
  title={HOLISMOKES-IV. Efficient mass modeling of strong lenses through deep learning},
  author={Schuldt, S and Suyu, SH and Meinhardt, T and Leal-Taix{\'e}, L and Ca{\~n}ameras, R and Taubenberger, S and Halkola, A},
  journal={Astronomy \& Astrophysics},
  volume={646},
  pages={A126},
  year={2021},
  publisher={EDP Sciences}
}

@ARTICLE{hezaveh2017,
       author = {{Hezaveh}, Yashar D. and {Perreault Levasseur}, Laurence and {Marshall}, Philip J.},
        title = "{Fast automated analysis of strong gravitational lenses with convolutional neural networks}",
      journal = {\nat},
         year = 2017,
        month = aug,
       volume = {548},
       number = {7669},
        pages = {555-557},
          doi = {10.1038/nature23463},
archivePrefix = {arXiv},
       eprint = {1708.08842},
 primaryClass = {astro-ph.IM},
       adsurl = {https://ui.adsabs.harvard.edu/abs/2017Natur.548..555H}
}

@article{ding2021time,
  title={Time delay lens modelling challenge},
  author={Ding, X and Treu, T and Birrer, S and Chen, G CF and Coles, J and Denzel, P and Frigo, M and Galan, A and Marshall, PJ and Millon, M and others},
  journal={Monthly Notices of the Royal Astronomical Society},
  volume={503},
  number={1},
  pages={1096--1123},
  year={2021},
  publisher={Oxford University Press}
}

@ARTICLE{shajib_2020,
       author = {{Shajib}, A.~J. and {Birrer}, S. and {Treu}, T. and {Agnello}, A. and {Buckley-Geer}, E.~J. and {Chan}, J.~H.~H. and {Christensen}, L. and {Lemon}, C. and {Lin}, H. and {Millon}, M. and {Poh}, J. and {Rusu}, C.~E. and {Sluse}, D. and {Spiniello}, C. and {Chen}, G.~C. -F. and {Collett}, T. and {Courbin}, F. and {Fassnacht}, C.~D. and {Frieman}, J. and {Galan}, A. and {Gilman}, D. and {More}, A. and {Anguita}, T. and {Auger}, M.~W. and {Bonvin}, V. and {McMahon}, R. and {Meylan}, G. and {Wong}, K.~C. and {Abbott}, T.~M.~C. and {Annis}, J. and {Avila}, S. and {Bechtol}, K. and {Brooks}, D. and {Brout}, D. and {Burke}, D.~L. and {Carnero Rosell}, A. and {Carrasco Kind}, M. and {Carretero}, J. and {Castander}, F.~J. and {Costanzi}, M. and {da Costa}, L.~N. and {De Vicente}, J. and {Desai}, S. and {Dietrich}, J.~P. and {Doel}, P. and {Drlica-Wagner}, A. and {Evrard}, A.~E. and {Finley}, D.~A. and {Flaugher}, B. and {Fosalba}, P. and {Garc{\'\i}a-Bellido}, J. and {Gerdes}, D.~W. and {Gruen}, D. and {Gruendl}, R.~A. and {Gschwend}, J. and {Gutierrez}, G. and {Hollowood}, D.~L. and {Honscheid}, K. and {Huterer}, D. and {James}, D.~J. and {Jeltema}, T. and {Krause}, E. and {Kuropatkin}, N. and {Li}, T.~S. and {Lima}, M. and {MacCrann}, N. and {Maia}, M.~A.~G. and {Marshall}, J.~L. and {Melchior}, P. and {Miquel}, R. and {Ogando}, R.~L.~C. and {Palmese}, A. and {Paz-Chinch{\'o}n}, F. and {Plazas}, A.~A. and {Romer}, A.~K. and {Roodman}, A. and {Sako}, M. and {Sanchez}, E. and {Santiago}, B. and {Scarpine}, V. and {Schubnell}, M. and {Scolnic}, D. and {Serrano}, S. and {Sevilla-Noarbe}, I. and {Smith}, M. and {Soares-Santos}, M. and {Suchyta}, E. and {Tarle}, G. and {Thomas}, D. and {Walker}, A.~R. and {Zhang}, Y.},
        title = "{STRIDES: a 3.9 per cent measurement of the Hubble constant from the strong lens system DES J0408-5354}",
      journal = {\mnras},
         year = 2020,
        month = jun,
       volume = {494},
       number = {4},
        pages = {6072-6102},
          doi = {10.1093/mnras/staa828},
archivePrefix = {arXiv},
       eprint = {1910.06306},
 primaryClass = {astro-ph.CO},
       adsurl = {https://ui.adsabs.harvard.edu/abs/2020MNRAS.494.6072S}
}

@article{MacLeod_2010,
   title={MODELING THE TIME VARIABILITY OF SDSS STRIPE 82 QUASARS AS A DAMPED RANDOM WALK},
   volume={721},
   ISSN={1538-4357},
   url={http://dx.doi.org/10.1088/0004-637X/721/2/1014},
   DOI={10.1088/0004-637x/721/2/1014},
   number={2},
   journal={The Astrophysical Journal},
   publisher={American Astronomical Society},
   author={MacLeod, C. L. and Ivezić, Z. and Kochanek, C. S. and Kozłowski, S. and Kelly, B. and Bullock, E. and Kimball, A. and Sesar, B. and Westman, D. and Brooks, K. and Gibson, R. and Becker, A. C. and de Vries, W. H.},
   year={2010},
   month=sep, pages={1014–1033} }

@book{Starck2015,
author="Starck, Jean-Luc
and Murtagh, Fionn
and Bertero, Mario",
editor="Scherzer, Otmar",
title="Starlet Transform in Astronomical Data Processing",
bookTitle="Handbook of Mathematical Methods in Imaging",
year="2015",
publisher="Springer New York",
address="New York, NY",
pages="2053--2098",
isbn="978-1-4939-0790-8",
doi="10.1007/978-1-4939-0790-8_34",
}
\bibliographystyle{aasjournal}

\appendix

\section{Relationship between accuracy of NPE posteriors and Population Mean}\label{app:gamma_lens_learn_all}

\begin{figure}[h!]
    \centering
    \includegraphics[width=0.5\textwidth]{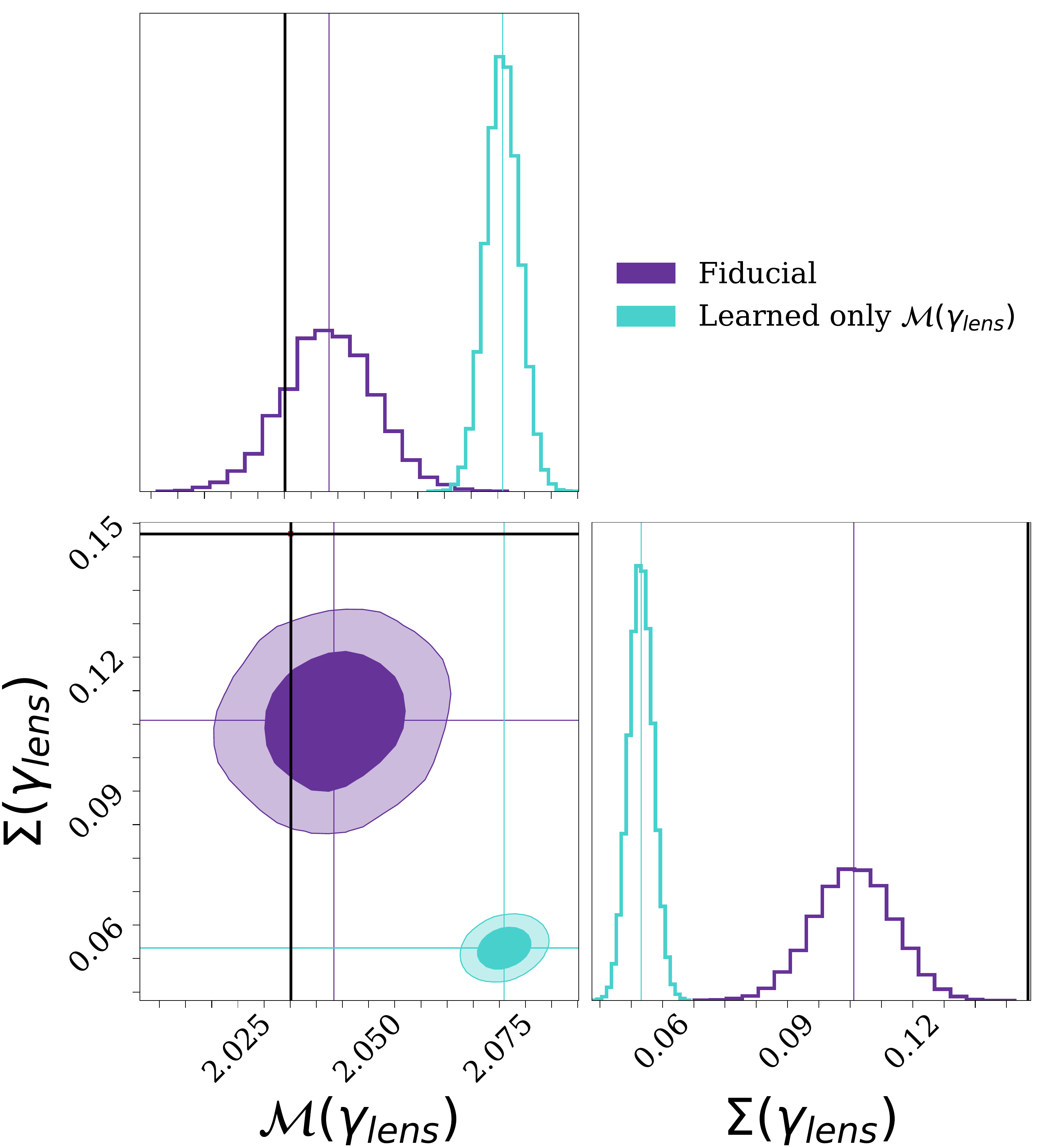}
    \caption{Black lines indicate the ground truth for the test set mean and scatter in $\rm \mathcal{M}(\gamma_{lens})$. Purple contour is reproduced from Figure \ref{fig:HI_model}, where we jointly learn all population model parameters for key 8 parameters. Turquoise contour is produced by only using \gammalens information from NPE posteriors to constrain population model for \gammalens. The mean error on \gammalens NPE posteriors is 0.04, and this is perfectly reproduced in the inferred population mean. }
    \label{fig:npe_to_post}
\end{figure}

In this section we elaborate on the relationship between the individual lens parameter posterior PDFs obtained using NPE and the inferred population parameters. We combine posterior PDFs from each individual lens using HBI described in Section \ref{method:HBI} to infer the population that the lenses are sampled from. In Figure \ref{fig:all_interim_results}, we show that the NPE model cannot predict \gammalens accurately from single lenses in the fiducial test set. The flat scatter of points around 2.03 of predicted values vs true values indicates that the network is just returning the mean of the \gammalens. However, since we infer each lens posterior PDF with a full covariance matrix, there are higher order terms that contain useful information. In Figure \ref{fig:npe_to_post}, we show that constraining the population model using only \gammalens sustains the bias we see in the interim posteriors. The average error on $\rm \gamma_{lens}$ in interim posteriors is 0.045, corresponding to the 0.042 error on $\rm \mathcal{M}(\gamma_{lens})$. However, jointly inferring all lens model parameters uses the higher-order information and de-biases the constraint $\rm \mathcal{M}(\gamma_{lens})$, bringing down the error to 0.01. We note that including information from all parameter dimensions also causes the uncertainty on inferred hyperparameters to inflate as the sampler now needs to accommodate a solution that fits more dimensions.

\section{Final Posteriors: Re-weighting of NPE posteriors}
The lens parameter posterior PDFs we infer from images $\rm d_k$, are conditioned on the prior that we implicitly introduce by training the NPE modeler on $\rm \xi_k; d_k$ pairs, where $\rm \xi_k$ is drawn from a broad distribution $\rm p(\xi | \nu_{int})$ (detailed in Section \ref{method:train_data}). We combine information gained from the full image dataset, $\rm \{d\}$, to infer the conditional PDF $p(\xi_k|\nu)$ (detailed in Section \ref{method:HBI}). We can now update the lens model inferred from images, $\rm p(\xi_k | d_k, \nu_{int})$, with information from all the images $\rm \{d\}$. Starting once again with Bayes' theorem,
\begin{equation}\label{eqn:ap1}
    \rm p(\xi_k | \{d\}) = \frac{p(\{d\}|\xi_k)}{p(\{d\})}.
\end{equation}

\begin{figure*}
    \includegraphics[width=0.45\textwidth]{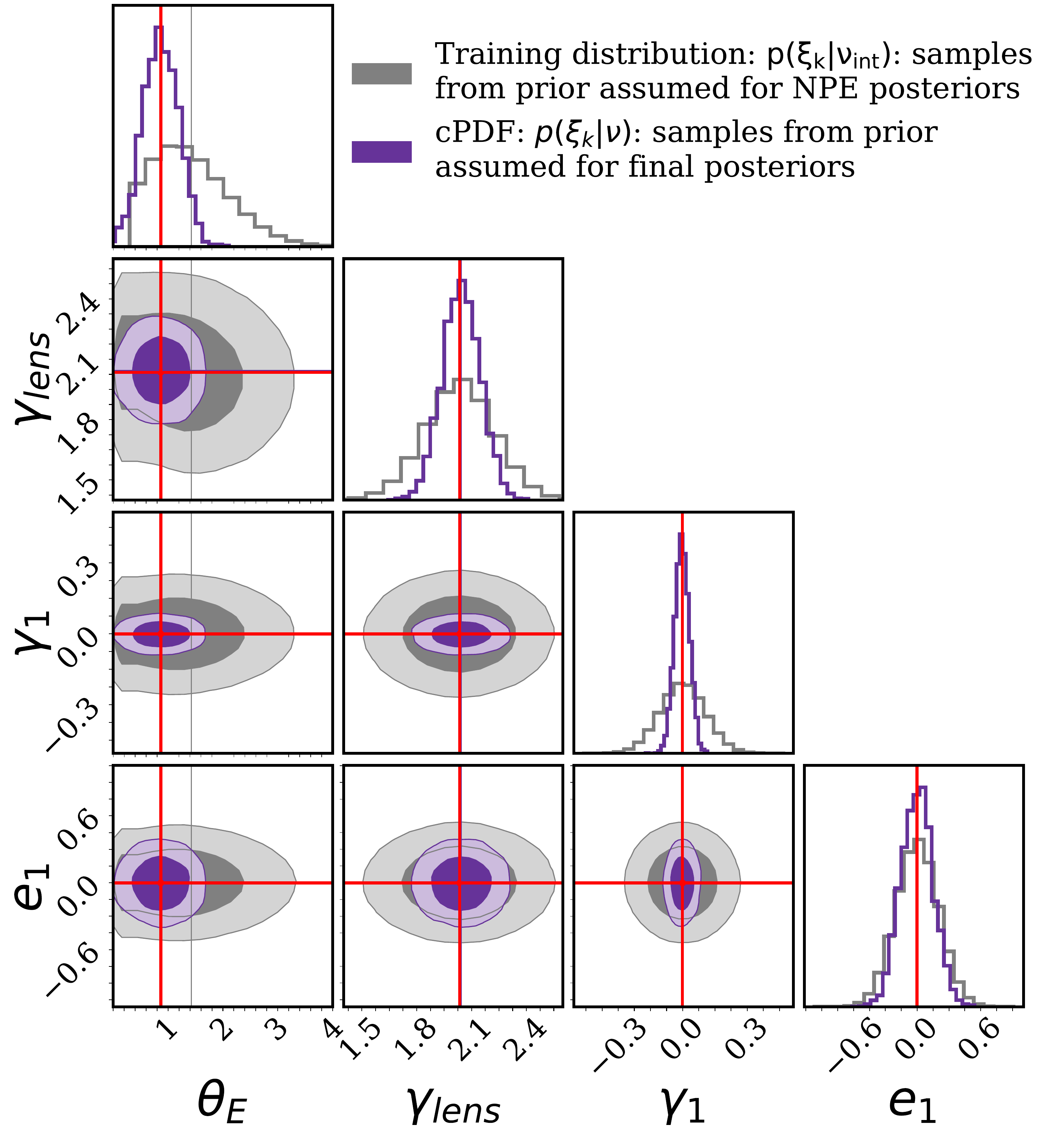}
    \includegraphics[width=0.58\textwidth]{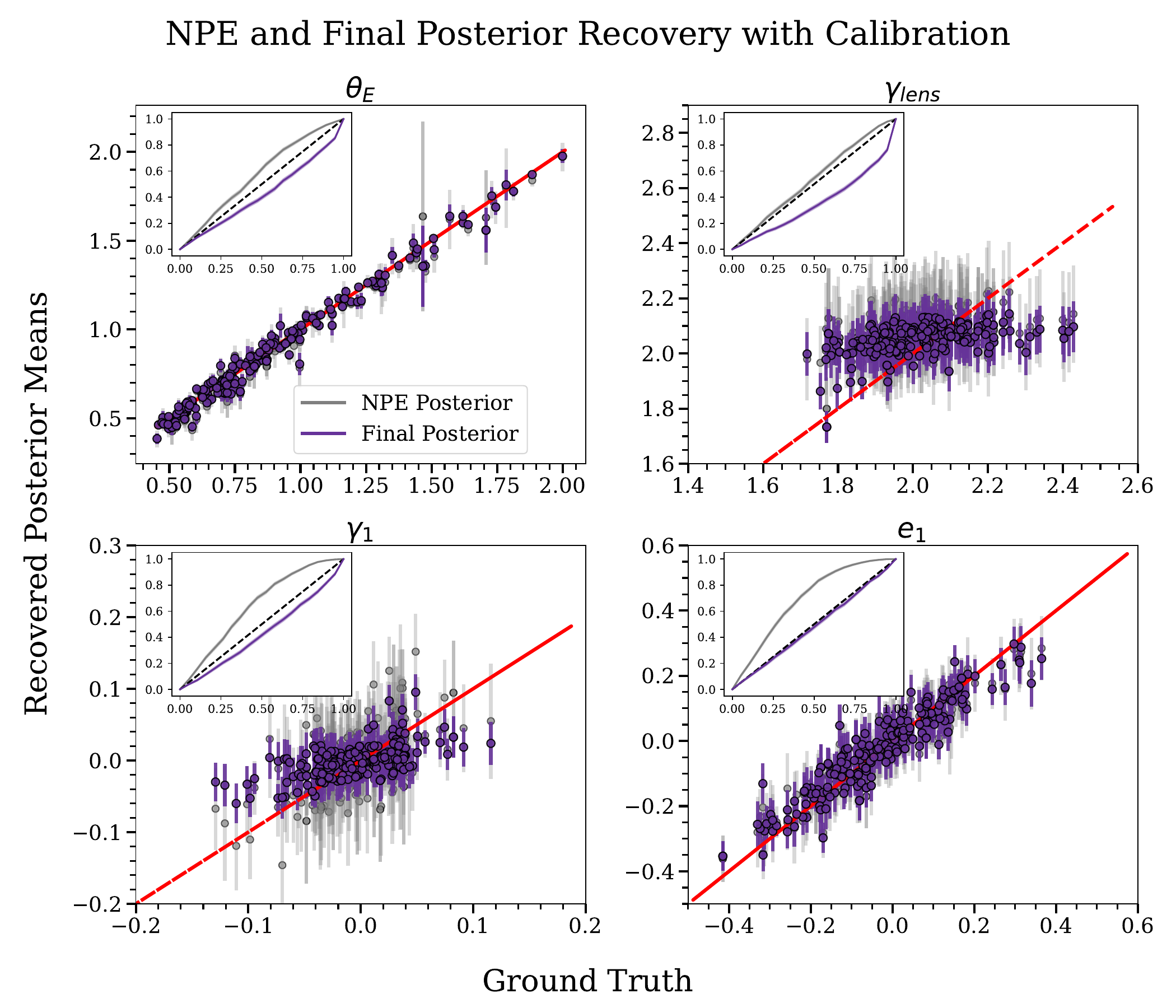}
    \caption{Here we show the impact of reweighting the NPE posteriors (Section \ref{method:NPE}) using the inferred cPDF (Section \ref{method:HBI}). Left: In gray, we show samples from the prior that the NPE posteriors are conditioned on, and in purple, we show samples from the prior that we learn and use for the final posteriors. Right: We show both NPE posteriors and final posteriors for a random sample of 200 objects. Final posteriors have smaller error-bars than the NPE posteriors, and their means are shifted towards the cPDF mean.}
    \label{fig:final_post_plus_calib}
\end{figure*}
\begin{figure*}
    \centering
    \includegraphics[width=\textwidth]{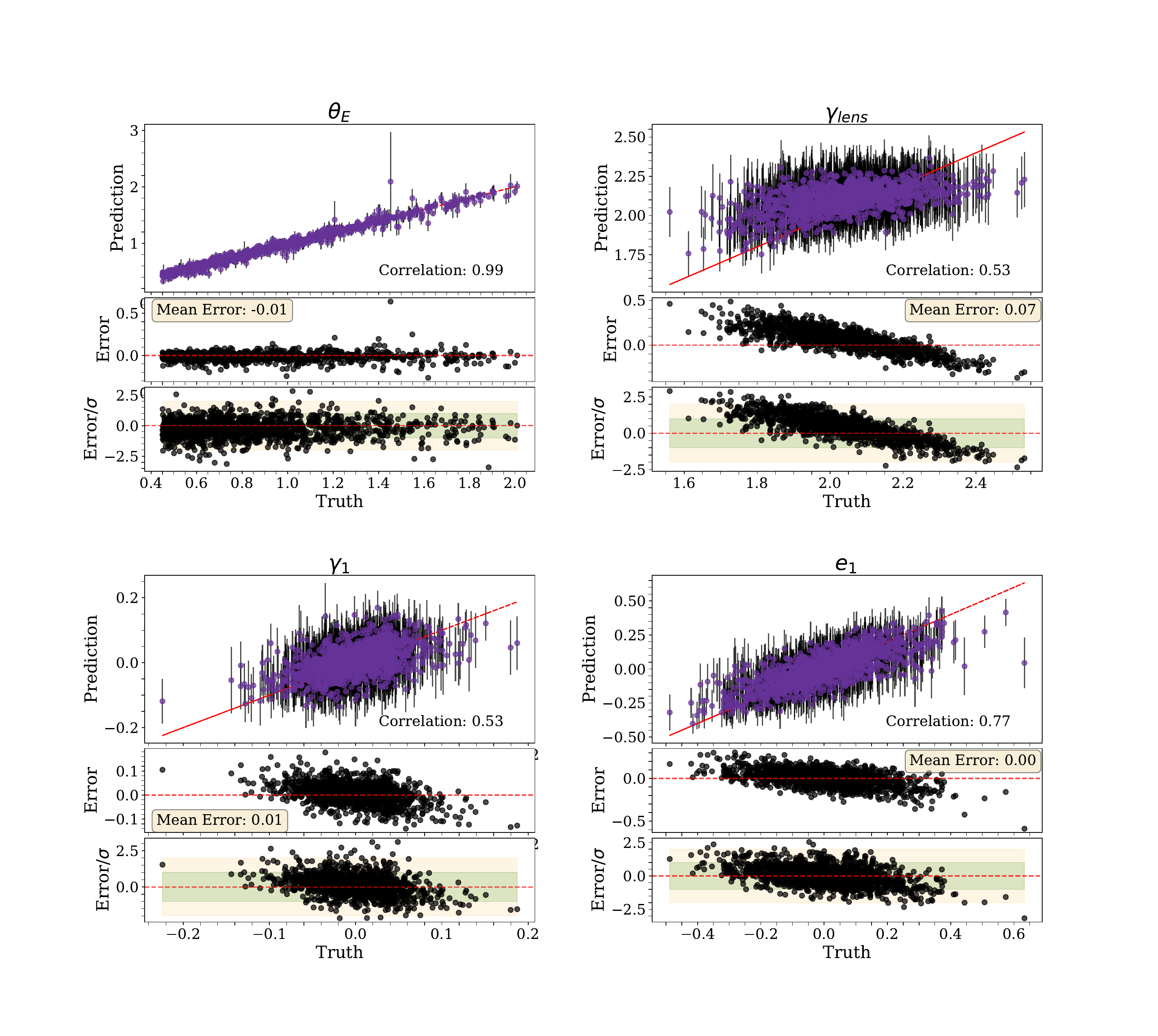}
    \caption{Here we reproduce Figure \ref{fig:all_interim_results}, but for the lens light subtracted preparation of data. Comparison of 1344     predicted Gaussian posteriors (obtained as described in Section \ref{method:NPE}) for lens light subtracted data (refer: Section \ref{data:3_preps}) against ground truth. Purple points correspond to the mean of the posteriors, and black lines correspond to the posterior $1 \,\sigma$ width or standard deviation. In the bottom most panel of each parameter subplot, the green shaded region represents the $1\,\sigma$ area and the yellow shaded region represents the $2\,\sigma$ region.}
    \label{fig:lens_light_subtracted_interim}
\end{figure*}

We can expand Equation \ref{eqn:ap1}, by including the dependency on $\nu$, and integrating over choice of $\nu$:

\begin{equation}\label{eqn:ap2}
    \rm p(\xi_k | \{d\}) = \int d\nu\frac{p(\{d\}|\xi_k, \nu)p(\xi_k|\nu)p(\nu)}{p(\{d\})}.
\end{equation}

Once again, since we have access to $\rm p(\xi_k|d_k, \nu)$ from the network, we can use the same expansion as Equation \ref{eqn:rewrite} in Section \ref{method:HBI}, to obtain:

\begin{equation}\label{eqn:ap2}
\begin{split}
    \rm p(\xi_k | \{d\}) = \int d\nu \,\,p(\xi_k|d_k, \nu_{int})p(d_k|\nu_{int})\frac{p(\xi_k|\nu)}{p(\xi_k|\nu_{int})}p(\nu|\{d\}) \\
    \rm
    \propto p(\xi_k|d_k, \nu_{int})\frac{1}{N}\sum_{\nu \sim p(\nu|\{d\})} \frac{p(\xi_k|\nu)}{p(\xi_k|\nu_{int})}
\end{split}
\end{equation}

In Equation \ref{eqn:ap2}, we obtain $\rm p(\nu|\{d\})$ from HBI. $\rm N$ refers to the number of samples that we pull to evaluate $\rm \frac{p(\xi_k|\nu)}{p(\xi_k | \nu_{int})}$. We can drop $\rm p(d_k | \nu_{int})$ which serves as a normalization constant.  An in-depth derivation of this reweighting scheme is presented in \cite{Wagner_Carena_2021}, Appendix C.

Overall, in cases where there is not much information in the image, network is not able to meaningfully hone in on the parameter value and the posterior is wide. This is further exacerbated by the fact that our multivariate gaussian function form is not malleable enough to capture skew/modality in the posterior PDF - so there will be instances where some posteriors are wider than they need to be. Under-confident NPE posteriors lead to \textit{under-estimated intrinsic scatter hyperparameters} and hence an overly-narrow conditional PDF, which then \textit{over-corrects the individual posteriors such that the final posteriors are over-confident}. We see this in the calibration curves in Figure \ref{fig:final_post_plus_calib} -- the gray curves indicating under-confidence, become over-confident post re-weighting.

\section{Lens Light Subtracted results}

We present NPE posteriors for the lens light subtracted case. Further discussion on results are presented in Sections \ref{res:LSST}, \ref{disc:lens_light_subtraction}.

\section{NPE posteriors for all parameters}\label{app:full}
We attach Table \ref{Tab:optimize_data_and_method_full} containing NPE posterior metrics for all parameters across all preparation. Additionally, we attach a table containing the population model inference for all parameters across all experiments in Table \ref{Tab:pop_model_full}.

\begin{deluxetable*}{|c|c||c|cccccc|}
\tablecaption{Summary of metrics for all experiment results detailed in Section \ref{sec:results}.}
\tablehead{
\colhead{Experiment} &
\colhead{Preparation} & \colhead{Metric} & \colhead{$\theta_E$} & \colhead{$\gamma_{\rm lens}$} & \colhead{$e_{1}$} & \colhead{$e_{2}$} & \colhead{$\gamma_{1}$} & \colhead{$\gamma_{2}$} \\
\hline
\cline{3-9}
 &  &  & (arcsec) &  &  &  &  &
}
\startdata
Fiducial & All Light Included & ME & -0.01 &  0.05 &  0.   & -0.   & -0.   & -0. \\  
& & MAE & 0.02 & 0.1  & 0.03 & 0.03 & 0.02 & 0.02 \\
& & Precision & 0.05 & 0.16 & 0.09 & 0.09 & 0.06 & 0.06 \\
\cline{3-9}
& Lens Light Subtracted & ME & -0.01 &  0.07 &  0.   & -0.   &  0.01 & -0. \\ 
& & MAE & 0.02 & 0.1  & 0.06 & 0.06 & 0.03 & 0.03 \\
& & Precision & 0.04 & 0.16 & 0.13 & 0.13 & 0.07 & 0.06 \\
\hline
\hline
\multicolumn{9}{c}{\textbf{Experiments optimizing the data}} \\
\hline
\hline
Bright Host Galaxies & All Light Included & ME & -0.   &  0.04 &  0.   & -0.   & -0.   &  0.   \\  
& & MAE & 0.02 & 0.1  & 0.03 & 0.03 & 0.02 & 0.02  \\
& & Precision & 0.04 & 0.16 & 0.09 & 0.09 & 0.06 & 0.06  \\
\cline{3-9}
& Lens Light Subtracted & ME & -0.01 &  0.06 &  0.01 & -0.   &  0.01 & -0.  \\  
& & MAE & 0.02 & 0.09 & 0.05 & 0.05 & 0.03 & 0.03  \\
& & Precision & 0.04 & 0.15 & 0.11 & 0.11 & 0.06 & 0.06 \\
\cline{3-9}
& Lens and AGN Light Subtracted & ME & -0.01 &  0.06 &  0.02 &  0.03 &  0.01 &  0.01  \\  
& & MAE &0.02 & 0.1  & 0.05 & 0.05 & 0.03 & 0.02 \\
& & Precision & 0.03 & 0.14 & 0.1  & 0.1  & 0.05 & 0.05 \\
\hline
Deconvolved & All Light Included & ME & 0.01 &  0.03 &  0.   &  0.   & -0.   &  0. \\  
& & MAE & 0.01 & 0.07 & 0.03 & 0.03 & 0.02 & 0.02 \\
& & Precision & 0.02 & 0.12 & 0.06 & 0.06 & 0.04 & 0.04 \\
\cline{3-9}
& Lens Light Subtracted & ME & -0.   &  0.05 & -0.   &  0.01 & -0.   &  0. \\
& & MAE & 0.01 & 0.05 & 0.03 & 0.03 & 0.02 & 0.02 \\
& & Precision & 0.02 & 0.08 & 0.07 & 0.07 & 0.04 & 0.04 \\
\cline{3-9}
& Lens and AGN Light Subtracted & ME & -0.01 &  0.02 & -0.01 & -0.   & -0.   & -0. \\  
& & MAE & 0.01 & 0.03 & 0.02 & 0.02 & 0.01 & 0.01 \\
& & Precision & 0.01 & 0.05 & 0.04 & 0.04 & 0.02 & 0.02 \\
\hline
\hline
\multicolumn{9}{c}{\textbf{Experiments modifying the modeling method}} \\
\hline
\hline
W/o mass/light correlations & All Light Included & ME & -0.02 &  0.06 &  0.01 & -0.   & -0.   &  0. \\  
& & MAE & 0.03 & 0.1  & 0.06 & 0.06 & 0.03 & 0.03 \\
& & Precision & 0.05 & 0.16 & 0.14 & 0.14 & 0.07 & 0.07 \\
\cline{3-9}
& Lens Light Subtracted & ME & -0.01 &  0.06 &  0.   &  0.01 &  0.   &  0.01 \\ 
& & MAE & 0.02 & 0.09 & 0.06 & 0.06 & 0.03 & 0.03 \\
& & Precision & 0.04 & 0.16 & 0.13 & 0.12 & 0.06 & 0.06 \\
\hline
No Distribution Shift & All Light Included & ME & 0.01 &  0.01 &  0.   & -0.01 & -0.   & -0. \\  
& & MAE &0.04 & 0.11 & 0.05 & 0.05 & 0.03 & 0.03 \\
& & Precision & 0.06 & 0.16 & 0.09 & 0.09 & 0.06 & 0.06 \\
\cline{3-9}
& Lens Light Subtracted & ME & -0.   &  0.01 &  0.   & -0.   &  0.   & -0. \\ 
& & MAE & 0.03 & 0.09 & 0.06 & 0.06 & 0.03 & 0.03 \\
& & Precision & 0.05 & 0.14 & 0.11 & 0.1  & 0.06 & 0.06 \\
\hline
\hline
\enddata
\tablecomments{We present tabulated results for all parameters other than lens position. We find that the network is almost always able to infer the lens position, and metrics do not provide useful insights on those parameters. Across all preparations, the bias on lens $x$ and $y$ position are less that 0.001'', and the median precision is less than 0.01'', thus, it errors and uncertainties are on a sub-pixel level for LSST. Details about every experiment is available in Section \ref{sec:results}. Details above every preparation is available in Section \ref{data:3_preps}. Metrics: ME = Mean error; MAE = Median absolute error; Precision - Median standard deviation of each posterior. Metrics are further defined in Section \ref{sec:results}.\label{Tab:optimize_data_and_method_full}}
\end{deluxetable*}

\renewcommand{\arraystretch}{1.15}
\begin{deluxetable*}{|c|c|c||cccccccc|}[b]
\tablecaption{Population-Level Model Parameter Recovery}
\tablehead{
\colhead{Experiment} &
\colhead{Preparation} &
\colhead{Metric} &
\colhead{$\mathcal{M}(\theta_E)$} & 
\colhead{$\Sigma(\theta_E)$} & 
\colhead{$\mathcal{M}(\gamma_{\rm lens})$} & 
\colhead{$\Sigma(\gamma_{\rm lens})$} & 
\colhead{$\mathcal{M}(e_{1\&2})$} & 
\colhead{$\Sigma(e_{1\&2})$} & 
\colhead{$\mathcal{M}(\gamma_{1\&2})$} & 
\colhead{$\Sigma(\gamma_{1\&2})$}}
\startdata
1 &  & Ground Truth & 0.87 & 0.34 & 2.04 & 0.15 & -0.00 & 0.04 & -0.00 & 0.15\\
\cline{3-11}
  & \multirow{2}{*}{ALL} & Inferred & $0.86_{-0.01}^{+0.01}$ & $0.33_{-0.01}^{+0.01}$ & $2.04_{-0.01}^{+0.01}$ & $0.11_{-0.01}^{+0.01}$ & $-0.00_{-0.002}^{+0.002}$ & $0.03_{-0.003}^{+0.003}$ & $-0.00_{-0.01}^{+0.01}$ & $0.15_{-0.004}^{+0.004}$ \\ 
  &     & Error & \textbf{-0.004} & \textbf{-0.001} & 0.01 & -0.041 & -0.001 & -0.01 &\textbf{ 0.001 }& \textbf{-0.001} \\ 
\cline{2-11}
  & \multirow{2}{*}{LLS} & Inferred & $0.85_{-0.01}^{+0.01}$ & $0.33_{-0.01}^{+0.01}$ & $2.13_{-0.01}^{+0.01}$ & $0.14_{-0.01}^{+0.01}$ & $0.01_{-0.003}^{+0.003}$ & $0.04_{-0.003}^{+0.003}$ & $0.01_{-0.01}^{+0.01}$ & $0.18_{-0.01}^{+0.01}$ \\
  &     & Error & -0.01 & -0.002 & 0.10 & -0.004 & 0.01 & 0.001 & 0.02 & 0.03 \\
\hline
\hline
2 &  & Ground Truth & 0.87 & 0.34 & 2.04 & 0.16 & -0.00 & 0.04 & 0.00 & 0.16 \\
\cline{3-11}
  & \multirow{2}{*}{ALL} & Inferred & $0.94_{-0.01}^{+0.01}$ & $0.34_{-0.01}^{+0.01}$ & $1.99_{-0.01}^{+0.01}$ & $0.11_{-0.01}^{+0.01}$ & $-0.00_{-0.003}^{+0.003}$ & $0.03_{-0.003}^{+0.003}$ & $-0.01_{-0.01}^{+0.01}$ & $0.14_{-0.01}^{+0.01}$ \\
  &     & Error & 0.07 & 0.01 & -0.05 & -0.05 & \textbf{0} & -0.01 & -0.01 & -0.01 \\ 
\cline{2-11}
  & \multirow{2}{*}{LLS} & Inferred & $0.93_{-0.02}^{+0.01}$ & $0.35_{-0.01}^{+0.01}$ & $2.08_{-0.01}^{+0.01}$ & $0.13_{-0.01}^{+0.01}$ & $0.01_{-0.003}^{+0.003}$ & $0.04_{-0.003}^{+0.003}$ & $0.01_{-0.01}^{+0.01}$ & $0.17_{-0.01}^{+0.01}$ \\
  &     & Error & 0.06 & 0.01 & 0.04 & -0.03 & 0.01 & \textbf{0} & 0.01 & 0.01 \\ 
\cline{2-11}
  & \multirow{2}{*}{LALS} & Inferred & $0.93_{-0.01}^{+0.01}$ & $0.04_{-0.003}^{+0.003}$ & $2.08_{-0.01}^{+0.01}$ & $0.15_{-0.01}^{+0.01}$ & $0.01_{-0.003}^{+0.003}$ & $0.09_{-0.004}^{+0.004}$ & $0.02_{-0.01}^{+0.01}$ & $0.04_{-0.002}^{+0.002}$ \\ 
  &     & Error & 0.05 & -0.3 & 0.04 & -0.01 & 0.01 & 0.05 & 0.02 & -0.11 \\ 
\hline
\hline
3 &  & Ground Truth & 0.87 & 0.34 & 2.04 & 0.15 & -0.00 & 0.04 & -0.00 & 0.15 \\
\cline{3-11}
  & \multirow{2}{*}{ALL} & Inferred & $0.87_{-0.01}^{+0.01}$ & $0.04_{-0.001}^{+0.002}$ & $2.05_{-0.004}^{+0.004}$ & $0.15_{-0.003}^{+0.004}$ & $-0.00_{-0.002}^{+0.002}$ & $0.11_{-0.003}^{+0.004}$ & $0.00_{-0.004}^{+0.01}$ & $0.05_{-0.001}^{+0.001}$ \\ 
  &     & Error & 0.01 & -0.27 & 0.02 & \textbf{0.002} & -0.002 & 0.07 & 0.01 & -0.11 \\ 
\cline{2-11}
  & \multirow{2}{*}{LLS} & Inferred & $0.87_{-0.01}^{+0.01}$ & $0.04_{-0.001}^{+0.001}$ & $2.07_{-0.004}^{+0.004}$ & $0.16_{-0.004}^{+0.004}$ & $-0.00_{-0.002}^{+0.002}$ & $0.13_{-0.003}^{+0.003}$ & $-0.01_{-0.01}^{+0.01}$ & $0.05_{-0.001}^{+0.001}$ \\
  &     & Error & -0.002 & -0.29 & 0.04 & 0.02 & -0.001 & 0.09 & -0.01 & -0.103 \\ 
\cline{2-11}
  & \multirow{2}{*}{LALS} & Inferred & $0.86_{-0.01}^{+0.01}$ & $0.04_{-0.001}^{+0.001}$ & $2.04_{-0.004}^{+0.004}$ & $0.15_{-0.003}^{+0.003}$ & $-0.00_{-0.001}^{+0.001}$ & $0.14_{-0.003}^{+0.003}$ & $-0.01_{-0.004}^{+0.004}$ & $0.05_{-0.001}^{+0.001}$ \\ 
  &     & Error & -0.01 & -0.3 & \textbf{0.01} & 0.01 & -0.002 & 0.1 & -0.01 & -0.1 \\ 
\hline
\hline
4 &  & Ground Truth & 0.86 & 0.33 & 2.03 & 0.15 & -0.00 & 0.04 & -0.00 & 0.15 \\
\cline{3-11}
  & \multirow{2}{*}{ALL} & Inferred & $0.84_{-0.01}^{+0.01}$ & $0.32_{-0.009}^{+0.02}$ & $2.14_{-0.01}^{+0.01}$ & $0.00_{-0.0}^{+0.08}$ & $0.01_{-0.01}^{+0.01}$ & $0.06_{-0.01}^{+0.01}$ & $0.02_{-0.014}^{+0.01}$ & $0.18_{-0.01}^{+0.01}$ \\
  &     & Error & -0.03 & -0.01 & 0.1 & -0.15 & 0.01 & 0.01 & 0.02 & 0.03 \\ 
\cline{2-11}
  & \multirow{2}{*}{LLS} & Inferred & $0.85_{-0.01}^{+0.01}$ & $0.32_{-0.01}^{+0.01}$ & $2.11_{-0.01}^{+0.01}$ & $0.13_{-0.01}^{+0.01}$ & $0.00_{-0.003}^{+0.003}$ & $0.05_{-0.003}^{+0.003}$ & $0.00_{-0.01}^{+0.01}$ & $0.17_{-0.01}^{+0.01}$ \\
  &     & Error & -0.01 & -0.02 & 0.08 & -0.02 & 0.01 & 0.002 & 0.01 & 0.02 \\ 
\hline
\enddata
\tablecomments{1 - Fiducial (\ref{res:LSST}); 2 - Bright Host Galaxies (\ref{res:bhg}); 3 - Emulated Deconvolved (\ref{res:DECONV}); 4 - Removing mass/light correlations in training prior (\ref{res:corr}). Note that we removed posteriors that were unconstrained (i.e.\ width of the posteriors was wider than the training prior) from the sample before computing the population parameters.}
\tablecomments{ ALL = All Light Included, 
LLS = Lens Light Subtracted, 
LALS = Lens and AGN Light Subtracted.}
\label{Tab:pop_model_full}
\end{deluxetable*}

\end{document}